\input epsf
\documentclass[12pt]{article}
\usepackage{latexsym,amsmath,amssymb,graphicx}
\renewcommand{\thefootnote}{\fnsymbol{footnote}}

\usepackage{amsmath}
\usepackage{amsfonts}

\numberwithin{equation}{section}

\def\doubleset#1#2{\bgroup%
\def\doit#1#2{%
\setbox\dblsetbox=\hbox{$\cstyle #1$}%
\raise#2\ht\dblsetbox\copy\dblsetbox%
\hskip-\wd\dblsetbox%
\raise-#2\ht\dblsetbox\box\dblsetbox}%
\mathchoice%
{\def\cstyle{\displaystyle}\doit#1#2}%
{\def\cstyle{\textstyle}\doit#1#2}%
{\def\cstyle{\scriptstyle}\doit#1#2}%
{\def\cstyle{\scriptscriptstyle}\doit#1#2}\egroup}
\newbox\dblsetbox

\newcommand{\plpl}{\doubleset{+}{.185}}
\newcommand{\mimi}{=}
\newcommand{\rd}{\mathrm{d}}
\newcommand{\intd}[1]{\int\mspace{-6mu}\rd#1\,}

\newlength{\extraspace}
\newlength{\extraspaces}
\newcommand{\be}{\begin{equation}
\addtolength{\abovedisplayskip}{\extraspaces}
\addtolength{\belowdisplayskip}{\extraspaces}
\addtolength{\abovedisplayshortskip}{\extraspace}
\addtolength{\belowdisplayshortskip}{\extraspace}}
\newcommand{\ee}{\end{equation}}

\newcommand{\ba}{\begin{eqnarray}
\addtolength{\abovedisplayskip}{\extraspaces}
\addtolength{\belowdisplayskip}{\extraspaces}
\addtolength{\abovedisplayshortskip}{\extraspace}
\addtolength{\belowdisplayshortskip}{\extraspace}}
\newcommand{\ea}{\end{eqnarray}}

\newcommand{\bd}{\begin{displaymath}
\addtolength{\abovedisplayskip}{\extraspaces}
\addtolength{\belowdisplayskip}{\extraspaces}
\addtolength{\abovedisplayshortskip}{\extraspace}
\addtolength{\belowdisplayshortskip}{\extraspace}}
\newcommand{\ed}{\end{displaymath}}

\newcounter{saveeqn}

\begin{document}
\addtolength{\baselineskip}{1.5mm}

\thispagestyle{empty}
\begin{flushright}
hep-th/0409196\\
\end{flushright}
\vbox{}
\vspace{2.0cm}

\begin{center}
{\LARGE{A Large $N$ Type IIB Duality via a $Spin(7)$ Geometric Transition as an F-theory Flop 
        }}\\[10mm]
{M.~C.~Tan\footnote{E-mail: g0306155@nus.edu.sg}}
\\[0mm]
{\it Department of Physics\\
National University of Singapore \\
Singapore 119260}\\[15mm]

\end{center}
\vspace{0.5 cm}

\centerline{\bf Abstract}\bigskip \noindent
The geometric features and toric descriptions of two different 8-dimensional $Spin(7)$ manifolds constructed via distinct resolutions of the cone over an $SU(3)/U(1)$ base, reveals that the geometry of the $Spin(7)$ conifold transition considered by Gukov et al. in \cite{gst}, is effected by a transition in its 6-dimensional submanifold which is isomorphic to a resolved or deformed Calabi-Yau 3-fold. This allows for a natural extension of the Gopakumar-Vafa large $N$ superstring duality of \cite{gv, v}; IIB superstring theory compactified on the $Spin(7)$ manifold with $N$ space-filling D5-branes wrapping an even-dimensional supersymmetric cycle, can be argued to undergo a large $N$ geometric transition at low energy to a $\it dual$ geometry with no branes but with certain units of 3-form fluxes through appropriate 3-cycles. For small $\it{or}$ large string coupling in a non-trivial axion field background, this large $N$ type IIB duality can be lifted to a purely geometric $\mathbb {RP}^5$ flop without D5-branes and 3-form fluxes via an F-theoretic description. The orientable, 10-dimensional, non-compact, Ricci-flat, $spin^c$ manifold undergoing the $\it{smooth}$ $\mathbb {RP}^5$ flop possesses an extended $SU(5) \odot {\mathbb Z_2}$ holonomy group, thus preserving $1/32$ of the maximal supersymmetry, consistent with the resulting $\mathcal N =(1,0)$ supersymmetric pure $SU(N)$ theory in $1+1$ dimensions. 

\newpage

\renewcommand{\thefootnote}{\arabic{footnote}}
\setcounter{footnote}{0}

\section{Introduction} 

In a seminal work by Witten \cite{cs}, $U(N)$ Chern-Simons theory on $S^3$ was given an equivalent description in terms of an open topologicial A string on the deformed conifold $T^*S^3$, with $N$ branes wrapping the Lagrangian submanifold $S^3$. The 't Hooft expansion of the $U(N)$ gauge theory, valid in the limit of large $N$, takes the form of an open string free energy expansion. Alternatively, it can also be recast into a form that resembles the free energy expansion of a closed string. Building upon this insight, it was conjectured by Gopakumar and Vafa in \cite{gv} that in the limit of large $N$, the deformed conifold $T^*S^3$ would undergo a geometric transition to a topologically distinct but nevertheless $\it dual$ geometry such that the $N$ branes which wrap the $S^3$ in the original open string description will disappear and be replace by $N$ units of RR 2-form fluxes through the $\mathbb P^1$ of the resolved conifold $\mathcal O(-1) \oplus \mathcal O(-1) \rightarrow \mathbb P^1$ in the corresponding dual $\it closed$ string description. Essentially, $SU(N)$ Chern-Simons theory on $S^3$ is conjectured to be exactly dual to a closed A-model topological string on the resolved conifold $\mathcal O(-1) \oplus \mathcal O(-1) \rightarrow \mathbb P^1$. This gauge/closed string duality conjecture has been verifed up to the level of the partition function on both sides for arbitrary 't Hooft coupling and to all orders in $1/N$ \cite{gv}. 

The important relevance of topological strings in the context of the physical superstring was discovered in \cite {topoamp1,topoamp2}, where it was shown that the open(closed) topological string amplitudes from a Calabi-Yau compactification of the topological string compute the $F$-term amplitudes of the resulting $\mathcal N= 1(2)$ theory in 4d from the corresponding superstring compactification on the $\it{same}$ Calabi-Yau 3-fold. Based on this connection, the large $N$ open/closed topological string duality was subsequently embedded in the superstring and re-expressed as a superstring duality in \cite{v} by Vafa, whereby the duality was shown to hold up to an equivalence of $F$-terms in the corresponding $\mathcal N= 1$ theory in 4d from the Calabi-Yau compactification of the superstring with D-branes and fluxes in the IR. In particular, type IIA string on $T^*S^3$ with $N$ space-filling D6-branes wrapping $S^3$, is shown to undergo a large $N$ geometric transition at low energy to the $\it{dual}$ $\mathcal O(-1) \oplus \mathcal O(-1) \rightarrow \mathbb P^1$ geometry with no branes and $N$ units of 2-form RR flux through the $\mathbb P^1$ cycle and additional 4-form RR flux through the dual 4-cycle. A mirror version of this type II string duality can also be considered. In this case, type IIB string on $\mathcal O(-1) \oplus \mathcal O(-1) \rightarrow \mathbb P^1$ with $N$ space-filling D5-branes wrapping $\mathbb P^1$, is shown to undergo a large $N$ geometric transition at low energy to the $\it{dual}$ $T^*S^3$ geometry with no branes but with $N$ units of 3-form RR flux accompanied by 3-form $NS$ flux through a compact $S^3$ and a dual non-compact 3-cycle of the deformed conifold respectively. The gauge/closed string duality that is implicit in this case was elucidated in a beautiful paper by Dijkgraaf and Vafa \cite{DV}, which exposes the deep connection between matrix models in the planar limit and closed topological B strings on CY manifolds.   

The large $N$ type IIA open/closed string duality has been generalized to a large class of Calabi-Yau 3-folds \cite{17}, which has been further studied in \cite{18}, leading to the development of powerful methods to compute all-loop A-model topological string amplitudes \cite{19,20}. Likewise, the corresponding mirror type IIB duality has also been generalized and used to derive highly non-trivial results and relationships amongst various 4d supersymmetric Yang-Mills theories resulting from string compactifications on different Calabi-Yau 3-folds \cite{10,11,12,13,14,15}.

By considering M-theory compactification on 7-dimensional spaces with/without certain singularities, it is possible to arrive at a purely geometric and equivalent  description of the type IIA configuration on either side of the geometric transition \cite{acharya}. Specifically, the large $N$ type IIA duality obtained via the deformed/resolved conifold transition in the presence of branes/fluxes, has a lift to a purely geometric $S^3$ flop in the equivalent M-theory on a $G_2$ manifold without branes and fluxes \cite{amv}. In a similar spirit, type IIA compactification on a 7-dimensional $G_2$ manifold undergoing a flop due to a phase transition with D6-branes/RR 2-form fluxes was conjectured by Gukov et al. in \cite{gst}. This flop was then argued to have a lift to an 8-dimensional $Spin(7)$ conifold transition in the equivalent M-theory resulting from a phase transition involving the condensation of $M5$ branes and the emergence of $G$-fluxes. The resulting $Spin(7)$ manifold preserves $1/16$ of the maximal supersymmetry, consistent with the effective $\mathcal{N}=1$ supersymmetry in $d=3$. Subsequently, a toric description of this $Spin(7)$ transition was provided, and its corresponding type IIA/B compactification in the presence of branes and fluxes considered in \cite{bel1,bel2}. 

In this paper, we will first reconsider the toric $\it and$ geometric descriptions of the $Spin(7)$ conifold transition conjectured by Gukov et al. \cite{gst}. We then find that there exists a natural extension of the original Gopakumar-Vafa large $N$, CY 3-fold, type IIB duality of \cite{gv, v}, such that type IIB superstring compactified on the $Spin(7)$ manifold  with $N$ space-filling D5-branes wrapping an even-dimensional supersymmetric cycle can be argued to undergo a large $N$ geometric transition at low energy to a $\it dual$ geometry with no branes but with certain units of 3-form fluxes through appropriate 3-cycles. In addition, arguments from a purely gauge-theoretic point of view of the effective theory in 1+1 dimensions allow for an alternative verification of this IIB $Spin(7)$ duality in further support of its physical validity. We next show that for small $\it{or}$ large values of string coupling, in a non-trivial albeit finite and smoothly varying axion field background, this large $N$, $Spin(7)$, type IIB duality can be lifted to a purely geometric $\mathbb {RP}^5$ flop without D5-branes and 3-form fluxes in the corresponding F-theoretic description. The non-compact, 10-dimensional Ricci-flat manifold undergoing the smooth $\mathbb {RP}^5$ flop is found to have an extended $SU(5) \odot {\mathbb Z_2}$ holonomy group, which preserves $1/32$ of the maximal supersymmetry, in agreement with the effective $\mathcal N =(1,0)$ supersymmetric pure $SU(N)$ theory in $d=1+1$ from the IIB compactification on the $Spin(7)$ with D5-branes/fluxes. The flop manifold lacks the usual $spin$ or $pin$ structures. However, one can show that it has a $spin^c$ structure, thus implying that it is nevertheless capable of supporting a covariantly constant spinor as required of a supersymmetric compactification.

To elucidate the above-mentioned statements and findings, the paper will be organized as follows: in $\S$2, we will review and provide the toric/geometric descriptions of the $Spin(7)$ transition and discuss its relation to the CY geometric transition of \cite{gv,v}. In $\S$3, we will demonstrate the natural extension of the large $N$ type IIB duality relating the resolved/deformed conifold to our case involving the $Spin(7)$ geometric transition and present the arguments which support its validity. In $\S$4, we will first geometrically lift the IIB duality to an equivalent background without D5-branes and 3-form fluxes. In addition, a gauge-theoretic interpretation of this duality in the lifted background will be provided as a consistency check and for completeness. Next, we will discuss the corresponding results for large string coupling via an application of a IIB S-duality transformation. Following that, we will review the F-theoretic description of a general IIB vacua/background. Finally, from the above equivalent IIB background at small or large string coupling, we will demonstrate the lift of the large $N$, $Spin(7)$, type IIB duality to a purely geometric $\mathbb {RP}^5$ flop without D5-branes and 3-form fluxes via its F-theoretic description. Details of the solution of the metric on the 10-dimensional flop manifold are also provided. We conclude the paper in $\S$5 with a summary of the results. In order to make the paper self-contained, a pedagogical construction of the $\mathcal N =(1,0)$ supersymmetric pure $SU(N)$ theory in $1+1$ dimensions is also furnished in Appendix A.

\section{Toric/geometric description of the $Spin(7)$ transition} 

The two 8-dimensional, asymptotically conical, Ricci-flat and non-compact $Spin(7)$ manifolds related via a geometric transition as considered by Gukov et al. in \cite{gst}, have been constructed in \cite{sparks} out of two different resolutions of the cone over the weak $G_2$ holonomy Aloff-Wallach space $SU(3)/U(1)$. Numerical evidence for the existence of these solutions was given in \cite{gibbons}. The first resolution of the cone results in the universal quotient bundle of $\mathbb {CP}^2$ endowed with a $Spin(7)$ holonomy. As shown in \cite{sparks}, this new $Spin(7)$ manifold $\mathcal Q$ is a chiral spin bundle\footnote{Notice that the $\mathbb {CP}^2$ base is not $spin$. However, it is $spin^c$ \cite{michel}. More precisely, this allows one to define the chiral spin bundle as a $ spin^c$ bundle over $\mathbb {CP}^2$, which is in turn isomorphic to a trivial $\mathbb R^4$ bundle.} and is therefore isomorphic to a trivial $\mathbb R^4$ bundle over $\mathbb {CP}^2$. In other words, its topology is given by
\be
\mathcal Q \cong \mathbb R^4 \times \mathbb {CP}^2.
\label{res1}
\ee
The second resolution of the cone results in a $Spin(7)$ manifold $\mathcal X$ which is a trivial $H^0(2)/U(1)$ bundle over $S^5$ \cite{gst}, whereby the 4-dimensional $H^0(2)$ space is a trivial (i.e. $c_1 =0$) complex line bundle over $S^2$ with a $U(1)$ action along the $\mathbb C$ fibre and $H^0(2)/U(1) = \mathbb R^3$ via a hyperk\"ahler moment map of the $U(1)$ action \cite{aw}. Hence, its topology is given by 
\be
\mathcal X \cong \mathbb R^3 \times S^5.
\label{res2}
\ee
The isometry group of both $\mathcal Q$ and $\mathcal X$ is given by $U(3)$. There is a single modulus $a > 0$ that corresponds to the size of the $\mathbb {CP}^2$ or $S^5$ of $\mathcal Q$ or $\mathcal X$ respectively. 

Recall that the resolved and deformed conifold of the 6-dimensional CY 3-fold considered in \cite{gv,v} is given by $\mathcal O(-1) \oplus \mathcal O(-1) \rightarrow \mathbb {CP}^1$ and $T^*S^3$ respectively. The resolved conifold $\mathcal O(-1) \oplus \mathcal O(-1) \rightarrow \mathbb {CP}^1$ consists of two copies of the spinor bundle over $\mathbb {CP}^1$. Indeed, it has been shown by Atiyah in \cite{atiyah} that it is isomorphic to a trivial $\mathbb R^4$ bundle over $\mathbb {CP}^1$ as one might have expected. In other words, its topology is given by 
\be
Q \equiv \mathcal O(-1) \oplus \mathcal O(-1) \rightarrow \mathbb {CP}^1 \cong \mathbb R^4 \times \mathbb {CP}^1.
\label{resolved}
\ee
On the other hand, $T^*S^3$ is isomorphic to a trivial $\mathbb R^3$ bundle over $S^3$. Hence, its topology is given by 
\be
X \equiv T^*S^3 \cong \mathbb R^3 \times S^3.
\label{deformed}
\ee  

The resolved and deformed conifold $Q$ and $X$ are related to each other via the Gopakumar-Vafa geometric transition $Q \leftrightarrow X$ \cite{gv,v}. Consequently, the two smooth $Spin(7)$ resolutions $\mathcal Q \cong \mathbb R^4 \times \mathbb {CP}^2$ and $\mathcal X \cong \mathbb R^3 \times S^5$, which are related to each other via the geometric transition $\mathcal Q \leftrightarrow \mathcal X$ of Gukov et al. \cite{gst}, have been termed the `resolution' and `deformation' of the $Spin(7)$ conifold because of their close analogy with $Q \cong \mathbb R^4 \times \mathbb {CP}^1$ and $X \cong \mathbb R^3 \times S^3$. In fact, it can be shown that the 6-dimensional subspace of $\mathcal Q$ or $\mathcal X$, which we will henceforth denote as $\widetilde Q$ or $\widetilde X$, is actually isomorphic to $Q$ or $X$, such that the $Spin(7)$ geometric transition $\mathcal Q \rightarrow \mathcal X$, which is found to be solely effected by the geometric transition ${\widetilde Q \rightarrow \widetilde X}$ of the 6-dimensional subspace, can be implied from the Gopakumar-Vafa transition $Q\rightarrow X$! In other words, $\mathcal Q \rightarrow \mathcal X$ can be regarded as an 8-dimensional extension of the transition $Q \rightarrow X$ which preserves $1/16$ of the maximal supersymmetry. To this end, we will first review the toric description of $\mathbb {CP}^n$ and its relationship to odd-dimensional spheres $S^{2n+1}$. We will then proceed to explore the features of the 6-dimensional subspaces of the $Spin(7)$ manifolds $\widetilde Q$ and $\widetilde X$ in order to elucidate their relationship with the resolved and deformed conifolds $Q$ and $X$ such that we can see how the $Spin(7)$ geometric transition $\mathcal Q \rightarrow \mathcal X$ can be regarded as an 8-dimensional extension of $Q \rightarrow X$.

\subsection{Toric description of complex projective spaces $\mathbb {CP}^n$} 

Toric geometry involves the viewing of a manifold as a torus fibration of some base space. A toric manifold is therefore expressed as an  n-complex dimensional space which is a $T^n$ fibration of an n-real dimensional base with boundaries \cite{leung}. The fibration is non-trivial due to the fact that certain cycles of the tori can shrink or degenerate along some loci of the base manifold. The nature of each of the $n$ independent $U(1)$ toric actions along the $n$ cycles of the $T^n$ fibre allows one to encode the combinatoric data of ``which cycles shrink where" in an n-real dimensional polytope $\triangle_n$ which is really the base space itself. Essentially, one associates the $(n-m)$-dimensional faces of $\triangle_n$ with the fixed points of the $U(1)^m$ toric action along the corresponding $m$ cycles in the $T^n$ fibre. Let us first illustrate this general concept with $\mathbb C^{n+1}$ since the essential example of $\mathbb {CP}^n$ is just a restricted case.   

Let the complex coordinates of $\mathbb C^{n+1}$ be $(z_1, z_2,...z_{n+1})$, where $z_k = |z_k|e^{i \theta_k}$ for $k=1,2,...n+1$. Alternatively, we can parameterize $\mathbb C^{n+1}$ in terms of the $2(n+1)$ real coordinates $(({|z_1|}^2, \theta_1), ({|z_2|}^2, \theta_2),...({|z_{n+1}|}^2, \theta_{n+1}))$, where $|z_k|^2 \geq 0$ and $\theta_k \sim {\theta_k + 2\pi i}$. As such, notice that the $n+1$ positive ${|z_k|^2}s$ will span the positive octant $\mathbb R^{n+1}_{\ge 0}$ while the ${\theta_k}s$ will parameterize the $n+1$ circles which naturally represent the cycles of a $T^{n+1}$ space. We can therefore view $\mathbb C^{n+1}$ as a non-trivial $T^{n+1}$ fibration of $\mathbb R^{n+1}_{\ge 0}$. Notice also that for any $|z_k|^2=0$, the circle parameterized by ${\theta_k}$ (i.e. $k^{th}$ cycle of $T^{n+1}$) degenerates; since $z_k=|z_k|e^{i\theta_k}=0$, the toric $U(1)$ action which acts on ${z_k}\in \mathbb C^{n+1}$ via shifts in $\theta_k$ along the circle is fixed, i.e. it spans a circle of zero size. Likewise, $m$ circles/cycles will degenerate where there are $m$ values of $k$ for which $|z_k|^2=0$.      

Now consider the $n$-dimensional projective space $\mathbb {CP}^n$. It is defined as follows:
\be
\mathbb {CP}^{n} = {[\mathbb C^{n+1}\setminus (0,0,0...)]} / {\mathbb{C}^*},
\label{cpn}
\ee
where the $\mathbb C^*$ action acts on the complex coordinates $(z_1,...,z_{n+1})$ parameterising $\mathbb C^{n+1}$ and is given by $\mathbb C^* : (z_1,...,z_{n+1}) \rightarrow (e^{i\theta}z_1,...,e^{i\theta}z_{n+1})$ for real $\theta$ and therefore $e^{i\theta}\in {\mathbb C^*}$. The complex homogenous coordinates of $\mathbb {CP}^n$ are the same coordinates that parameterize $\mathbb C^{n+1}$, the space in which it is embedded in. As was done in the $\mathbb C^{n+1}$ case, we can re-express these $n+1$ complex coordinates in terms of 2(n+1) real coordinates $(({|z_1|}^2, \theta_1), ({|z_2|}^2, \theta_2),...({|z_{n+1}|}^2, \theta_{n+1}))$. Naively, one would then expect $\mathbb {CP}^n$ to be expressed as a $T^{n+1}$ fibration of $\mathbb R^{n+1}_{\ge 0}$. However, this is $\it not$ the case.\footnote{As is commonly known, $\mathbb {CP}^n$ is an $n$-dimensional complex manifold. Hence, it should be given by a $T^n$ fibration of an $n$-real dimensional base space.} Notice that the modding of the $C^*$ action in (\ref{cpn}) results in the identification $(z_1,...,z_{n+1}) \sim (e^{i\theta}z_1,...,e^{i\theta}z_{n+1})$ in $\mathbb {CP}^n$. This can be used to trivialize and hence eliminate one of the $U(1)$ actions, thus resulting in an overall effective $U(1)^{n}$ toric action along $n$ cycles of a $T^n$ and $\it{not}$ $T^{n+1}$ space. Moreover, one can make a rescaling of the ${z_k}s$ such that they obey the constraint $|z_1|^2 + |z_2|^2 + ...|z_{n+1}|^2 = r$. This means that the ${|z_k|^2}s$ neccessarily parameterize an $n$-dimensional polytope $\triangle_n$ embedded in $\mathbb R^{n+1}_{\ge 0}$. Hence, the toric description of $\mathbb {CP}^n$ is really given by a $T^n$ fibration of $\triangle_n$. Let us now specialize this general result to the relevant cases of $\mathbb {CP}^1\cong S^2$ and $\mathbb {CP}^2$ respectively. 
\vspace{0.2cm}
\newline         
(i) $\mathbb {CP}^1$ projective space:
\newline
based on the above discussion concerning $\mathbb {CP}^n$ for any $n$, the topology or toric description of $\mathbb {CP}^1\cong S^2$ should be given by a non-trivial $T^1 \cong S^1$ fibration of $\triangle_1$, where $\triangle_1$ is a line segment (i.e. 1-simplex). Indeed the single independent $U(1)$ toric action along an $S^1$ cycle acts as $z \rightarrow e^{i\theta}z$, where $z$ is either $z_1$ or $z_2$, the coordinates of $\mathbb C^2$ in which $\mathbb{CP}^1$ is embedded in. The constraint in this case imposed via a rescaling of the ${z_k}s$ is given by $|z_1|^2 + |z_2|^2 = r$. This necessarily means that $|z_1|^2$ and $|z_2|^2$ together parameterize a 1-dimensional polytope of finite length embedded in $\mathbb R^{2}_{\ge 0}$ giving $\triangle_1$. The $S^1$ cycle collapses where $|z_1|^2=0$ or $|z_2|^2=0$ and notice that $|z_1|^2=0$ or $|z_2|^2=0$ at either of the ``faces" (i.e. ends) of the finite line segment. In other words, as shown in Fig.1, $\mathbb {CP}^1$ can be viewed as the line segment with a circle fibre on top, such that the circle fibre collapses to zero size at the end points which describe the north and south poles of the $S^2$.  

\bigskip
\centerline{\epsfxsize 2.truein \epsfysize 2.truein\epsfbox{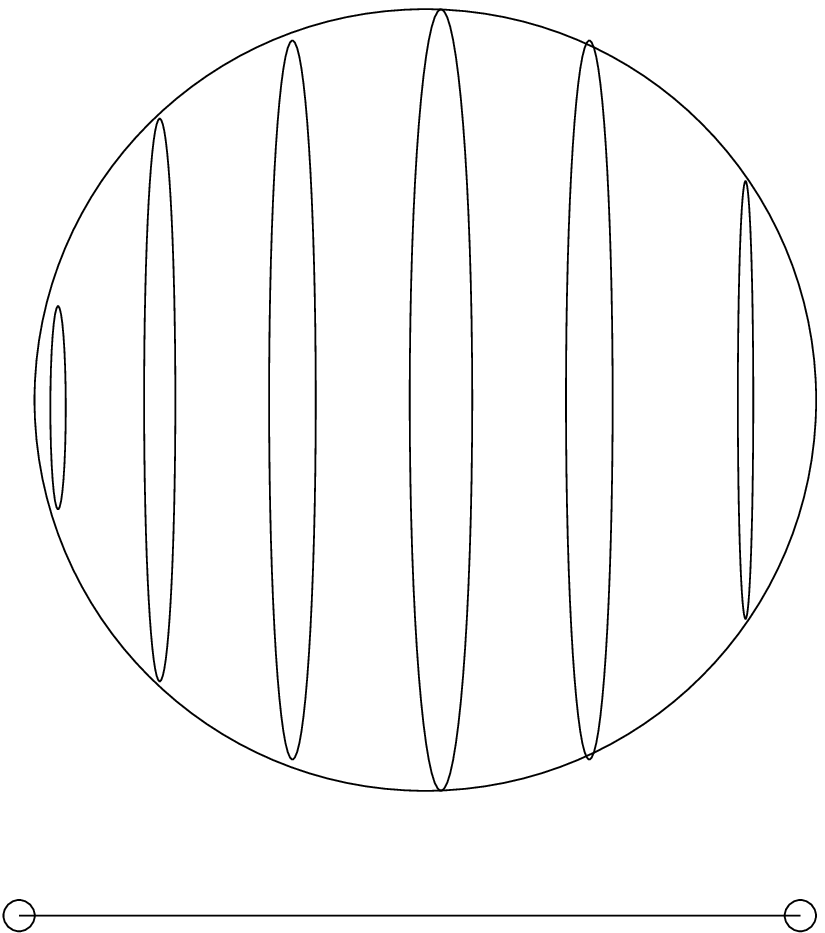}}
\noindent{\baselineskip=8pt {\bf Fig.1}: {\rm
The 2-sphere can be viewed as an interval with a circle
on top, where the circle shrinks to zero size at the two ends.}}
\bigskip
\newline
(ii) $\mathbb {CP}^2$ projective space:
\newline
specializing the above discussion on $\mathbb {CP}^n$ to $n=2$, the topology or toric description of $\mathbb {CP}^2$ should be given by a non-trivial $T^2$ fibration of $\triangle_2$, where $\triangle_2$ is a 2-simplex. Indeed the two independent $U(1)$ toric action along the two $S^1$ cycles of $T^2$ (which we name $a$ and $b$ for convenience) can be chosen to act as $(z_1, z_2, z_3) \rightarrow (e^{i\theta_1}z_1, e^{i\theta_2}z_2, z_3)$, where $(z_1, z_2, z_3)$ are the coordinates of the $\mathbb C^3$ space in which $\mathbb {CP}^2$ is embedded in. The constraint in this case imposed via a rescaling of the ${z_k}s$ is given by $|z_1|^2 + |z_2|^2 + |z_3|^2 = r$. This necessarily means that $|z_1|^2$, $|z_2|^2$ and $|z_3|^2$ together parameterize a 2-dimensional polytope of finite area embedded in $\mathbb R^{3}_{\ge 0}$ giving $\triangle_2$. An $S^1$ cycle collapses where $z_i = |z_i|^2=0$ for any $i=1,2,3$,\footnote{If we consider the diagonal $U(1)$ toric action such that $\theta_1=\theta_2=\phi$, we can, via a $\mathbb C^*$ equivalence rescaling, define a circle toric action on $z_3$ such that $(z_1, z_2, z_3) \rightarrow (z_1, z_2, e^{-i\phi}z_3)$. We thus see that a fixed point of this toric action exists such that the cycle generated by $e^{-i\phi}\sim e^{-i(\theta_1/ 2)} e^{-i (\theta_2/ 2)}$ degenerates at $z_3 = 0$. As noted above, the cycles $a$ and $b$ of the $T^2$ fibre are generated by $e^{i\theta_1}$ and $e^{i\theta_2}$ respectively. This means that there will also be a degenerating cycle at $z_3=0$, which can be viewed as the cycle $a+b$.} while the $T^2$ collapses to a point where $z_i =z_j = 0$ (i.e. $|z_i|^2 = |z_j|^2 = 0$) for any $i,j=1,2,3$. Notice that $z_i=0$ for $i=1,2,3$ at either one of the three faces (i.e. edges) of the 2-simplex, while $z_i=z_j=0$ at the vertices (i.e. points where the edges intersect) of the 2-simplex. In other words, as shown in Fig.2, $\mathbb {CP}^2$ can be viewed as a 2-simplex with a $T^2$ fibre on top, such that an $S^1$ cycle of the $T^2$ fibre collapses to zero size at each of the three faces while the entire $T^2$ fibre shrinks to zero size at each of the three vertices. In accordance with our discussions in (i) above, the remaining non-trivial $S^1$ fibration of each of the three finite length edges (such that the fibre degenerates at the edge ends), will individually result in a $\mathbb {CP}^1\cong S^2$ subspace. In other words, each edge of the 2-simplex $\triangle_2$ will torically represent a $\mathbb {CP}^1 \cong S^2$. 


\bigskip

\centerline{\epsfxsize 2.truein \epsfysize 2.truein\epsfbox{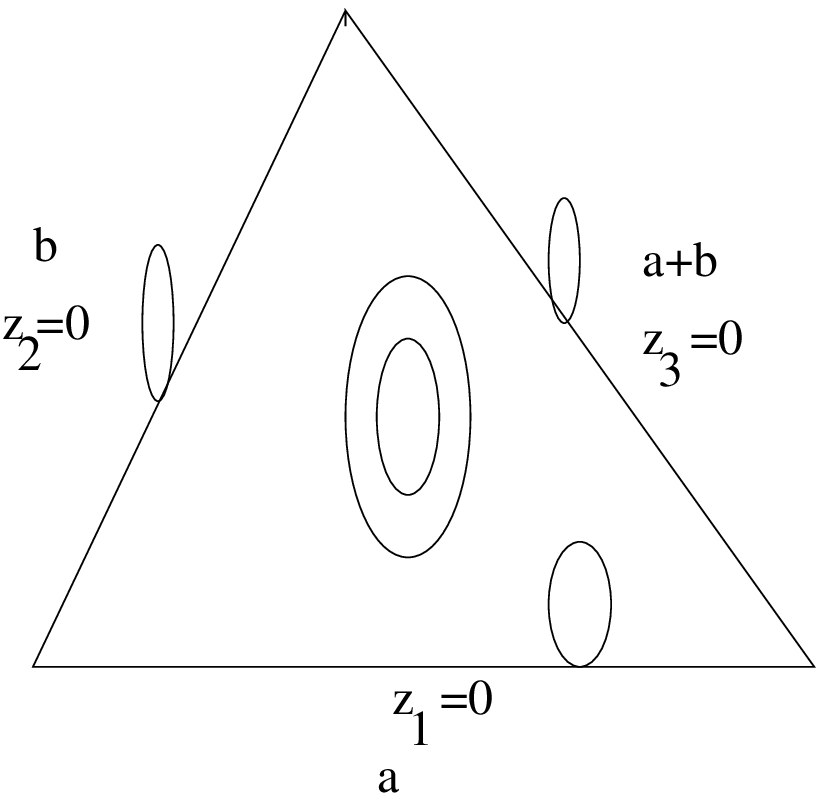}}
\noindent{\baselineskip=8pt {\bf Fig.2}: {\rm The toric
description of $\mathbb {CP}^2$ involving a 2-simplex over which 
there is a 2-torus at each point which shrinks to a circle
at each edge and further to a point at each vertex. Each edge of the 2-simplex with the circle fibre above it represents
a $\mathbb {CP}^1$.}}


\subsection{$\mathbb {CP}^n$ and odd-dimensional spheres $S^{2n+1}$} 

Due to its definition in (\ref{cpn}), $\mathbb{CP}^n$ must satisfy a constraint relation obtained via a rescaling of the ${z_k}s$ as mentioned in the previous subsection. This relation is given by 
\be
\sum_{k=1}^{n+1}|z_k|^2\ =\ r,  \qquad r \in \mathbb R_{> 0}.
\label{s^n}
\ee 
Notice that this relation actually parameterizes an odd-dimensional sphere $S^{2n+1}$ of non-zero size. Recall from the definition in (\ref{cpn}) that the ${z_k}$ coordinates also satisfy the projective identification $(z_1,...,z_{n+1}) \sim (e^{i\theta}z_1,...,e^{i\theta}z_{n+1})$, whereby $e^{i\theta}\in U(1)$ and $\theta \in \mathbb R$. 

From the above, we can see that there is an alternative description of $\mathbb {CP}^n$ as an $S^{2n+1}$ sphere modulo a $U(1)$ identification of the complex coordinates parameterizing the $\mathbb C^n$ space in which the sphere is embedded in. In other words, noting that $U(1) \cong S^1$, we can simply write the relation between $\mathbb {CP}^n$ and $S^{2n+1}$ as 
\be
{\mathbb{CP}^n} \cong {S^{2n+1}}/{S^1}\,.
\label{cpn,symplectic}
\ee 
Hence, $S^{2n+1}$ is isomorphic to a non-trivial $S^1$ fibration of $\mathbb {CP}^n$:
\be
{S}^1 \hookrightarrow {S}^{2n+1} \to {{\mathbb {CP}^n}},
\label{sfibration}
\ee
i.e. $S^{2n+1}$ is given by a Hopf fibration of $\mathbb{CP}^n$. A well-known example would be $S^3$, which is a Hopf fibration of $\mathbb {CP}^1 \cong S^2$ given by ${S}^1 \hookrightarrow {S}^3 \to {S}^2$.

\subsection{The subspaces $\widetilde Q$, $\widetilde X$, the Gopakumar/Vafa geometric transition and the $Spin(7)$ geometric transition $\mathcal Q \rightarrow \mathcal X$}                                  

Note that the toric description of a manifold gives us its $\it{topology}$. In the case of $\mathbb {CP}^1$, we have shown that it is torically described by a circle fibre over a finite length interval such that the circle degenerates at the endpoints. Indeed this gives us the topology of an $S^2 \cong \mathbb {CP}^1$. Likewise for $\mathbb {CP}^2$, we find that it is torically described by a non-trivial $T^2$ fibration of a 2-simplex base $\triangle_2$, such that the cycle/s degenerate at the edges/vertices. Hence, from Fig. 2, we find that the $\mathbb {CP}^2$ manifold is given by the $\it{non}$-$\it{trivial}$ product of a 2-dimensional space having the topology of three $S^2$s (each represented by an edge of $\triangle_2$) intersecting each other (at the vertices where the edges meet in $\triangle_2$), and a 2-dimensional space having the topology of a non-trivial circle fibration of the `inner' dimension of $\triangle_2$. We will henceforth label the latter as $\widetilde {\mathcal M}_2$ in anticipation of its further reference in the paper. 

Note here that one has the result that $b_l(\mathbb{CP}^n)=0$ for $l$ odd and $b_l(\mathbb{CP}^n)=1$ for $l$ even and $l \leq 2n$, whereby $b_l$ is the $l^{th}$ betti number \cite{hou}. This means that the second betti number of $\mathbb{CP}^2$ is given by $b_2(\mathbb {CP}^2) = 1$, i.e. there is a $\it{single}$ 2-cycle within $\mathbb {CP}^2$ given by a $\mathbb{CP}^1\in H_2(\mathbb{CP}^2)$. This means that the 2-dimensional space spanned by the three intersecting $S^2$s at the edges of $\triangle_2$ must be homeomorphic (i.e. topologically equivalent) to a single $S^2$. In other words, $\mathbb{CP}^2$ is given by a non-trivial $\widetilde {\mathcal M}_2$ fibre over an $S^2$ base, i.e. $\widetilde M_2 \hookrightarrow \mathbb{CP}^2 \to S^2$.    

Let us now recall the relationship between $\mathbb{CP}^n$ and $S^{2n+1}$ in eqn.(\ref{sfibration}) here. In particular, we have the non-trivial Hopf fibration $S^1\hookrightarrow S^5 \to {\mathbb {CP}^2}$ which says that $S^5$ is isomorphic to a non-trivial $S^1$ fibration of $\mathbb {CP}^2$. In light of the structure of $\mathbb {CP}^2$ explored above, which reveals the presence of an $S^2 \subset \mathbb {CP}^2$, the non-trivial Hopf fibration $S^1\hookrightarrow S^3 \to S^2$ then implies that there should exist a 3-dimensional subspace in $S^5$ which is isomorphic to an $S^3$ and is normal to an $\widetilde {\mathcal M}_2$ fibre. In other words, $S^5$ can also be viewed as a non-trivial $\widetilde {\mathcal M}_2$ fibre over $S^3$, i.e. $\widetilde M_2 \hookrightarrow S^5 \to S^3$. Indeed, it is known that the geometry of $SU(3)$ is a product of a 3-sphere and a 5-sphere such that $S^5 \cong SU(3) / SU(2)$, whereby $SU(2) \cong S^3$ \cite{hou}. One also has the following decomposition of the $SU(3)$ group manifold \cite{byrd}: an arbitrary element of $SU(3)$ denoted by $D^{(3)}$ can be expressed as $D^{(3)}(\alpha,\beta,\gamma,\theta,a,b,c,\phi) = D^{(2)}(\alpha,\beta,\gamma)e^{(i\lambda_5 \theta)}D^{(2)}(a,b,c)e^{(i\lambda_8 \phi)}$, where $D^{(2)}$ is an arbitrary element of $SU(2) \subset SU(3)$, $(\lambda_1, \lambda_2, ..., \lambda_8)$ are the 8 generators of $SU(3)$, and $(\alpha,\beta,\gamma,\theta,a,b,c,\phi)$ are real parameters. Hence, one can see that in modding $D^{(3)}$ by a single $D^{(2)}$ to arrive at $S^5$, one is left with an $SU(2) \cong S^3$ and a 2-dimensional submanifold spanned by $e^{(i\lambda_5 \theta)} e^{(i\lambda_8 \phi)}$ which corresponds to the $\widetilde M_2$ fibre.    
        
Via the description of $\mathbb {CP}^2$ above, we can hence view the `resolved' $Spin(7)$ conifold $\mathcal Q \cong \mathbb R^4 \times \mathbb {CP}^2$ as being isomorphic to a non-trivial $\widetilde {\mathcal M}_2$ fibre over a 6-dimensional base given by a trivial $\mathbb R^4$ fibration of $S^2$ or $\mathbb R^4 \times S^2$. Thus, we find that the 6-dimensional subspace of $\mathcal Q$ (i.e. the base of the $\widetilde {\mathcal M}_2$ fibration), which we will denote as $\widetilde {Q}$, is such that $\widetilde{Q} \cong {\mathbb R^4 \times S^2} \cong {\mathcal O(-1) \oplus \mathcal O(-1) \rightarrow \mathbb {CP}^1}$. Hence, we have $\widetilde M_2 \hookrightarrow \mathcal Q \to ({\widetilde Q \cong {\mathcal O(-1) \oplus \mathcal O(-1) \rightarrow \mathbb {CP}^1}})$.   

Via the description of $S^5$ above, we can likewise view the `deformed' $Spin(7)$ conifold $\mathcal X \cong \mathbb R^3 \times S^5$ as being isomorphic to a non-trivial $\widetilde {\mathcal M}_2$ fibre over a 6-dimensional base given by a trivial $\mathbb R^3$ fibration of $S^3$ or $\mathbb R^3 \times S^3$. Thus, we find that the 6-dimensional subspace of $\mathcal X$ (i.e. the base of the $\widetilde {\mathcal M}_2$ fibration), which we will denote as $\widetilde {X}$, is such that $\widetilde{X} \cong {\mathbb R^3 \times S^3} \cong T^*S^3$. Hence, we have $\widetilde M_2 \hookrightarrow \mathcal X \to ({\widetilde X \cong T^*S^3})$.   
 
In fact, more can be said about the isomorphism between $\widetilde X$ and $T^*S^3$ as follows. First, let us note that $T^*S^3$ has topology $S^3 \times S^2 \times \mathbb R_{\geq 0}$ at infinity, whereby $S^3$ here is the projection of the compact 3-sphere at the origin to infinity. The $(S^2 \times \mathbb R_{\geq 0})$ space normal to the $S^3$ is the unbounded and thus non-compact\footnote{Note that a compact manifold must be closed (has a boundary) $\it{and}$ bounded at the same time. Thus, being closed alone as in the case of $(S^2 \times \mathbb R_{\geq 0})$ is not a sufficient condition for compactness.} 3-cycle with the topology of a 3-ball. It resides in the cotangent bundle (isomorphic to $\mathbb R^3$) and is dual to $S^3$ \cite{la ossa}. Recall from the earlier description of $\mathcal X$ that the $\mathbb R^3$ fibre of $\mathcal X \cong \mathbb R^3 \times S^5$ and therefore $\widetilde X \cong \mathbb R^3 \times S^3$ is actually an $H^0(2)/U(1)$ space, whereby the 4-dimensional $H^0(2)$ space is a $\it{trivial}$ $\mathbb C$ bundle over $S^2$ with a $U(1)$ action along the $\mathbb C = \mathbb R^2$ fibre \cite{aw}. The presence of a $U(1)$ action along the $\mathbb R^2$ fibre of $H^0(2)$ suggests that the $\mathbb R^2$ fibre can be viewed as a trivial product of two decompactified circles such that the $U(1)$ action acts along one of them. This then implies that the quotient space $H^0(2)/U(1) \cong H^0(2)/S^1$ is a trivial product of a decompactified circle fibre (i.e. the real line $\mathbb R$) with an $S^2$. Notice then that $S^2 \times \mathbb R = S^2 \times(\mathbb R_{<0} \cup \mathbb R_{\geq 0})=(S^2 \times\mathbb R_{<0}) \cup (S^2 \times \mathbb R_{\geq 0})$ and $(S^2 \times \mathbb R_{\geq 0})$ is nothing but an unbounded and hence non-compact 3-cycle with the topology of a 3-ball.  As one might have expected, just as in the case of the deformed conifold $T^*S^3$, there exists a dual $\it{non}$-$\it{compact}$ 3-cycle with the topology of a 3-ball in the $\mathbb R^3$ bundle of $\widetilde X \cong \mathbb R^3 \times S^3$.    

The Gopakumar-Vafa geometric transition of \cite{gv,v} given by ${\mathcal O_{\mathbb {CP}^1}(-1) \oplus \mathcal O_{\mathbb {CP}^1}(-1)}\rightarrow T^*S^3$ implies that it is natural to consider the transition $\widetilde Q \rightarrow \widetilde X$ in light of the isomorphisms $\widetilde Q \cong {\mathcal O(-1) \oplus \mathcal O(-1)\rightarrow {\mathbb {CP}^1}}$ and $\widetilde X \cong T^*S^3$. In fact by doing so, we arrive at the $Spin(7)$ geometric transition $\mathcal Q \rightarrow \mathcal X$ of Gukov et al. \cite{gst}; since $\mathcal Q$ and $\mathcal X$ are isomorphic to a non-trivial $\widetilde {\mathcal M}_2$ fibration of $\widetilde Q$ and $\widetilde X$ which we will henceforth denote as $\widetilde M_2 \hookrightarrow \mathcal Q \to \widetilde Q$ and $\widetilde M_2 \hookrightarrow \mathcal X \to \widetilde X$ respectively, $\mathcal Q \rightarrow \mathcal X$ is simply effected by the geometric transition of the base $\widetilde  Q \rightarrow \widetilde X$! The $\widetilde M_2$ fibres of $\mathcal Q$ and $\mathcal X$ are in fact  `spectator' subspaces in the $Spin(7)$ geometric transition $\mathcal Q \rightarrow \mathcal X$. Let us look at this more closely. 

The Gopakumar-Vafa geometric transition is effected by a blow-down of a $\mathbb {CP}^1$ in $\mathcal O(-1) \oplus \mathcal O(-1) \rightarrow \mathbb P^1 \cong \widetilde Q$ followed by a blow-up of an $S^3$ in $T^*S^3 \cong \widetilde X$. Likewise, the geometric transition ${\widetilde Q} \rightarrow {\widetilde X}$ is effected by a blow-down and and blow-up of a $\mathbb {CP}^1$ and $S^3$ respectively. The $\mathbb {CP}^1$ in $\widetilde Q$ is a subspace of the $\mathbb {CP}^2$ in $\mathcal Q$ such that $\widetilde M_2 \hookrightarrow \mathbb {CP}^2 \to \mathbb {CP}^1$ (recall that $\widetilde M_2 \hookrightarrow (\mathcal Q \cong \mathbb R^4 \times \mathbb {CP}^2) \to(\widetilde Q \cong \mathbb R^4 \times \mathbb {CP}^1)$). This means that $vol (\mathbb {CP}^2) = vol(\widetilde M_2) \cdot vol (\mathbb {CP}^1)$. Hence, a blown-down (i.e. zero volume) $\mathbb {CP}^1$ will imply a blown-down $\mathbb {CP}^2$. The $S^3$ of $\widetilde X$ is a subspace of the $S^5$ in $\mathcal X$ such that $\widetilde M_2 \hookrightarrow S^5 \to S^3$ (recall that $\widetilde M_2 \hookrightarrow (\mathcal X \cong \mathbb R^3 \times S^5) \to(\widetilde X \cong \mathbb R^3 \times S^3)$). Hence, $vol (S^5) = vol(\widetilde M_2) \cdot vol (S^3)$. In other words, a blown-up (i.e. non-zero volume) $S^3$ implies a blown-up $S^5$. Thus, we see that the geometric transition of the 6-dimensional bases $\widetilde Q \rightarrow \widetilde X$ indeed results in an 8-dimensional geometric transition from the `resolved' to `deformed' $Spin(7)$ conifold $\mathcal Q \rightarrow \mathcal X$ given respectively by a blow-down of a $\mathbb {CP}^2$ in $\mathcal Q$ and a blow-up of an $S^5$ in $\mathcal X$! Thus, the $Spin(7)$ geometric transition can be regarded as an 8-dimensional extension of the conifold transition ${\mathcal O(-1) \oplus \mathcal O(-1)} \rightarrow T^*S^3$, which now preserves $1/16$ of the maximal supersymmetry.

\section{Large $N$ type IIB duality via a $Spin(7)$ geometric transition}  

In this section, we will demonstrate a large $N$ type IIB duality via a $Spin(7)$ geometric transition. To do so, we will first argue for a physically consistent $Spin(7)$ extension of the resolved/deformed conifold duality of Gopakumar-Vafa. We will next discuss the relevant aspects of the resulting $\mathcal N = (1,0)$ pure $SU(N)$ theory in 1+1 dimensions. Finally, we will present a purely gauge-theoretic description of the $Spin(7)$ geometric transition as an alternative verification of this type IIB, $Spin(7)$ duality in further support of our earlier arguments.

\subsection{A $Spin(7)$ extension of the resolved/deformed conifold superstring duality} 

\vspace{0.3cm}

Since the $Spin(7)$ geometric transition $\mathcal Q \rightarrow \mathcal X$ (whereby a $\mathbb{CP}^2$ of $\mathcal Q$ blows down and an $S^5$ of $\mathcal X$ blows up) is simply effected by the geometric transition of its base $\widetilde Q \rightarrow \widetilde X$, and can thus be regarded as an 8-dimensional extension of the CY conifold transition, it is only natural to believe that there should exist a $Spin(7)$ extension of the Gopakumar-Vafa large $N$ geometric transition superstring duality, such that IIB string theory on $\mathcal Q$ with $N$ wrapped D-branes, where $N$ is large, will undergo a geometric transition at low energy to a $\it{dual}$ IIB background on $\mathcal X$ $\it{without}$ these D-branes but with fluxes instead.  In fact, to describe the IIB compactification on $\mathcal Q$ in a background of $N$ space-filling D5-branes wrapping the $\mathbb {CP}^2 \subset \mathcal Q$ at $\it{low}$ energy, where $N$ is large, first notice that in light of the fibration $\widetilde M_2 \hookrightarrow \mathcal Q \to \widetilde Q$, we can simply view the compactification of the IIB string (from $d=9+1$ to $d=1+1$) on the 8-dimensional `resolved' $Spin(7)$ conifold $\mathcal Q$ as an initial compactification on a CY 3-fold $\widetilde Q \cong {\mathcal O(-1) \oplus \mathcal O(-1)\rightarrow {\mathbb {CP}^1}}$ (down to $d=3+1$) followed by an additional compactification on a non-trivial $\widetilde M_2$ fibre (down to $d=1+1$). Also recall from $\S$2.3 that $\mathbb {CP}^2$ is given by an $\widetilde M_2$ fibre over a $\mathbb {CP}^1$. Hence, the configuration of $N$ space-filling D5-branes wrapping the 4-cycle $\mathbb {CP}^2$ of $\mathcal Q \cong \mathbb R^4 \times \mathbb {CP}^2$ in the IIB compactification on $\mathcal Q$ down to $d=1+1$ can be viewed as a configuration of space-filling D5-branes with two of its spatial dimensions first wrapping the $\mathbb {CP}^1$ base of $\widetilde Q \cong {\mathcal O(-1) \oplus \mathcal O(-1)\rightarrow {\mathbb {CP}^1}}$ in an initial IIB compactification on this resolved conifold down to $d=3+1$, followed by an additional wrapping of its remaining two spatial dimensions on an $\widetilde M_2$ fibre in a further compactification of the IIB theory on this non-trivial $\widetilde M_2$ fibre space down to $d=1+1$. Thus, our IIB background on $\mathcal Q$ with $N$ space-filling D5-branes wrapping the $\mathbb{CP}^2 \subset \mathcal Q$ at low energy can be equivalently described as an initial IIB compactification on $\widetilde Q \cong {\mathcal O(-1) \oplus \mathcal O(-1)\rightarrow {\mathbb {CP}^1}}$ with $N$ space-filling D5-branes wrapping the $\mathbb {CP}^1 \subset \widetilde Q$ at $\it{low}$ energy, followed by a further compactification on the non-trivial $\widetilde M_2$ fibre. According to the Gopakumar-Vafa superstring duality, the IIB string on  $[\mathcal O(-1) \oplus \mathcal O(-1)\rightarrow {\mathbb {CP}^1}] \cong \widetilde Q$ with $N$ space-filling D5-branes wrapping the $\mathbb {CP}^1$ at low energy, where $N$ is large, is $\it{equivalent}$ to the IIB string on $T^*S^3 \cong \widetilde X$ with no D-branes but with $N$ units of 3-form $H_{RR}$ flux through the compact $S^3$ in $\widetilde X$ accompanied by 3-form $H_{NS}$ flux through the dual non-compact 3-cycle (with topology of a 3-ball) in the $\mathbb R^3$ bundle of $\widetilde X$. From a further compactification on the non-trivial $\widetilde M_2$ fibre, we indeed arrive at a background which is given by a IIB compactification on $\mathcal X$ with no D-branes but with $N$ units of 3-form $H_{RR}$ flux through an $S^3 \subset \widetilde X \subset \mathcal X$ and 3-form $H_{NS}$ flux through the non-compact 3-cycle in the $\mathbb R^3$ bundle of $\widetilde X \subset \mathcal X$. Hence, by this token of a direct application of the Gopakumar-Vafa superstring duality in the intermediate $d=3+1$ theory, one indeed has a $Spin(7)$ extension of the large $N$ superstring duality. Let us now look at things in greater detail so as to verify the physical consistency and validity of this extension.   
\newpage
\noindent{\it The IIB Background on ${\mathcal Q}$}
\vspace{0.2cm}
   
Firstly, observe that $\mathcal Q$ is neccessarily the physically consistent choice for a IIB $Spin(7)$ compactification with wrapped D-branes as follows: since we are dealing with a IIB string theory, we can have Dp-branes where p is odd. As we are investigating the extension of the original CY conifold duality which involves $N$ D5-branes, it is natural to consider a background of $N$ D5-branes. Recall that D-branes are BPS objects, i.e. they have minimal mass. This means that one must wrap the $N$ D5-branes around minimal (supersymmetric) cycles of a $Spin(7)$ manifold if a supersymmetric worldvolume theory is desired. Note that a $Spin(7)$ manifold comes equipped with a closed 4-form $\Psi$ and a corresponding 4-dimensional $\Psi$-submanifold called the Cayley 4-fold. They are known as the calibration and calibrated submanifold respectively. All calibrated submanifolds are minimal because they are volume-minimising in their homology class, i.e. they are supersymmetric cycles \cite{joyce}. In the case of the `resolved' or `deformed' $Spin(7)$ conifold, the $\mathbb {CP}^2$ submanifold is a supersymmetric cycle while the $S^5$ submanifold is not. Hence, a physically consistent choice would indeed be to consider a background in which the $N$ space-filling D5-branes wrap the 4-dimensional supersymmetric $\mathbb {CP}^2$ cycle of $\mathcal Q \cong \mathbb R^4 \times \mathbb {CP}^2$, so as to result in a supersymmetric theory in the uncompactified 1+1 dimensions.                    

It is prudent at this point to mention that there is a topological obstruction to the wrapping of D-branes on a submanifold of 10-dimensional spacetime $Y$ due to a global Freed-Witten anomaly in the worldsheet path integral of the IIB string theory \cite{baryons,freed-witten}. In particular, the configuration of $N$ space-filling D5-branes wrapping $\mathbb {CP}^2$ with worldvolume $\mathcal W = \mathbb R^{1,1} \times \mathbb {CP}^2$, is anomalous unless the following topological condition is satisfied:
\be
\zeta|_{\mathcal W} = W_3(\mathcal W)= W_3(\mathbb {CP}^2),
\label{zeta}
\ee
where $\zeta = H/{2\pi}$ and $H=dB$ is the curvature of the NS B-field. For any given manifold $M$, $W_3(M) \in H^3(M, \mathbb Z)$ such that $W_3(M) = \beta(w_2(M))$, whereby $w_2(M) \in H^2 (M, \mathbb Z_2)$ is the second Stiefel-Whitney class of $M$ and $\beta$ is given by the ``Bockstein" map $\beta : H^2(M; {\mathbb Z}_2) \rightarrow H^3(M;\mathbb Z)$. Note that the second equality in (\ref{zeta}) arises from the fact that the space $\mathbb R^{1,1}$ is topologically trivial and hence orientable and spin, i.e. the first and second Stiefel-Whitney classes of this space vanish. Hence, from the Whitney sum formula \cite{hou}, we have $(1+ w_1(\mathcal W) + w_2(\mathcal W) + ...) = (1+ 0 + 0 +...)(1+ w_1(\mathbb {CP}^2) + w_2(\mathbb{CP}^2) + ...)$, i.e. $w_i (\mathcal W) = w_i (\mathbb {CP}^2)$ and therefore, $W_3 (\mathcal W) = W_3(\mathbb {CP}^2)$.     

Returning to the case at hand, it is known that $W_3(M) = 0$ if and only if $w_2(M) =c$ mod 2, where $c \in H^2(M,\mathbb Z)$ \cite{michel}. As $\mathbb {CP}^2$ is $spin^c$, it has $w_2(\mathbb {CP}^2) =c$ mod 2 \cite{spin spaces}. Thus, we have $W_3 (\mathbb {CP}^2) = 0$. In other words, from (\ref{zeta}), our background must satisfy the following condition for the theory to be anomaly-free:
\be      
H|_{\mathbb R^{1,1} \times \mathbb {CP}^2} = 0,
\label{dB=0}
\ee
i.e. the NS 3-form H-flux must be zero when restricted to $\mathbb R^{1,1} \times \mathbb {CP}^2$. This condition is indeed compatible with our background of vanishing NS 3-form H-flux when there are D5-branes.\footnote{Recall that in the case of the original Gopakumar-Vafa duality, 3-form NS and RR fluxes only need to be turned on after the wrapped D5-branes disappear under a geometric transition of the CY conifold. This is also true of a consistent $Spin(7)$ extension of this duality as we will see shortly.} Hence, we shall not worry about this any further.

The $Spin(7)$ manifold $\mathcal Q$ preseves 1/16 of the 32 conserved supercharges of the IIB theory. The BPS D5-branes preseve 1/2 of these leftover conserved supercharges on its worldvolume theory along the uncompactified directions. This means that we effectively have 1 conserved supercharge in the resulting $d=1+1$ worldvolume theory or $\mathcal N=(1,0)$ supersymmetry. As will be shown later, the final result for large string coupling will be equivalent to that of small string coupling under an S-duality when we do a geometric lift to a IIB background without D5-branes and fluxes, consistent with the fact that the duality is to hold for all values of 't Hooft and hence string coupling, as had been emphasized in \cite{v} for the case of the type II CY conifold duality. Hence, we will just need to consider the limit of small string coupling here. In the limit of small string coupling $g_s$, the D-brane excitations decouple from the closed string modes (i.e. gravity in the bulk). Thus, the effective theory of the IIB string compactified on $\mathcal Q \cong \mathbb R^4 \times \mathbb {CP}^2$ with $N$ space-filling D5-branes wrapping the $\mathbb{CP}^2$ 4-cycle is an $\mathcal N=(1,0)$ supersymmetric pure $SU(N)$ theory in $d=1+1$.\footnote{The worlvolume theory of $N$ coincident D-branes is actually a $U(N)$ gauge theory. However, we can neglect the overall $U(1)$ in $U(N)=SU(N) \times U(1)$ for large $N$.} One could of course consider and derive a worlvolume theory with $N_f$ massive flavours by adding $N_f$ space-filling D5-branes that wrap a 4-cycle $\subset {\mathcal Q}$ which doesn't coincide with the $\mathbb {CP}^2$ such that the open strings which connect the $N_f$ and $N$ D5-branes result in light states which form the matter multiplets of the $d=1+1$ supersymmetric worldvolume theory. The mass of these states will then be proportional to the length of the connecting open strings. The IIB duality has also been shown to hold even in the presence of massive hypermultiplets for the case of the resolved/deformed CY conifold in \cite{10}. However, we will be content with the simplest case of a pure SYM worldvolume theory in this paper.  
\newpage
\noindent{\it The Dual IIB Background on ${\mathcal X}$ and the Conjectured $Spin(7)$ Duality }
\vspace{0.2cm}
   
In a supposed geometric transition $\mathcal Q \rightarrow \mathcal X$ such that there is a blow-down of the $\mathbb {CP}^2 \subset \mathcal Q$ and a blow-up of $S^5 \subset \mathcal X$, the D5-branes wrapping the blown-down $\mathbb {CP}^2 \subset \mathcal Q$ will therefore disappear as one transitions to the $\it {dual}$ $\mathcal X$ geometry. By Gauss's law, there should be fluxes to replace the vanishing D5-branes if the $\it{non}$-$\it{compact}$ $\mathcal X$ is to be a dual compactification. Indeed if such a duality is to hold in 10-dimensional spacetime, the $N$ D5-branes that are present in the `resolved' $Spin(7)$ conifold background $\mathcal Q$ must act, after vanishing in a geometric transition, as the magnetic source for $N$ units of (Hodge dual) 3-form $dB_{RR} = H_{RR}$ flux through a compact 3-cycle with the topology of a 3-sphere in $\mathcal X$, which is embedded in the 4-dimensional space normal to the $N$ coincident D5-branes that it surrounds.\footnote{Recall that in a $d$-dimensional space of a $(d+1)$-dimensional spacetime, the dual flux due to the charges of an $m$-dimensional solitonic object must flow through a $(d-m-1)$-sphere, which is embedded in the $(d-m)$-dimensional space normal to the object that it surrounds. A trivial example would be a 0-brane (i.e. point charge) in the 3-dimensional space of 4-dimensional spacetime with 2-form flux through a 2-sphere, which is embedded in the 3-dimensional space normal to the 0-brane that it surrounds.} This compact 3-cycle must be the $S^3$ base of the $\widetilde X$ subspace of $\mathcal X$ as follows: the $N$ D5-branes wrap the $\mathbb {CP}^2\subset \mathcal Q$ and fill spacetime. This means that the compact 3-cycle which surrounds the D5-branes must be normal to $\mathbb{CP}^2$ $\it{and}$ the uncompactified directions of spacetime. The former implies that the 3-cycle must be normal to the $\mathbb{CP}^1\subset \mathbb{CP}^2 \subset \mathcal Q$ $\it{and}$ the fibre $\widetilde M_2 \subset \mathbb {CP}^2 \subset \mathcal Q$. The latter implies that the 3-cycle must be a subspace of $\mathcal X$. Recall that since $\widetilde X \cong T^*S^3$ and $\widetilde Q \cong {\mathcal O(-1) \oplus \mathcal O(-1) \rightarrow \mathbb {CP}^1}$, and that $T^*S^3$ and ${\mathcal O(-1) \oplus \mathcal O(-1) \rightarrow \mathbb {CP}^1}$ are the deformation and resolution of a cone over an $S^3 \times S^2$ base respectively \cite{la ossa}, the $S^3 \subset \widetilde X$ must be normal to the $\mathbb{CP}^1\subset \widetilde Q$, whereby $S^3 \subset S^5 \subset \mathcal X$ and $\mathbb{CP}^1 \subset \mathbb{CP}^2 \subset \mathcal Q$. Moreover, from the fibration $\widetilde M_2 \hookrightarrow (S^5\subset \mathcal X) \to S^3$, we see that the $S^3$ is normal to the $\widetilde M_2$ fibre of $S^5$, whereby $\widetilde M_2 \subset \mathbb {CP}^2 \subset \mathcal Q$ as well (recall that the $\widetilde M_2$ fibre is a spectator space in the geometric transition $\mathcal Q \rightarrow \mathcal X$). Thus, the compact 3-cycle with the topology of a 3-sphere which surrounds the $N$ D5-branes must given by $S^3 \subset \widetilde X \subset \mathcal X$. 

In addition, due to the changing volume of the $S^5 \subset \mathcal X$ induced by the $Spin(7)$ geometric transition $\mathcal Q \rightarrow \mathcal X$, there must also be 3-form $dB_{NS}= H_{NS}$ flux through a corresponding 3-cycle in $\mathcal X$; we have shown in $\S$2.3 that the $\widetilde M_2$ fibre is just a spectator subspace during a $Spin(7)$ geometric transition $\mathcal Q \rightarrow \mathcal X$ such that a blow-up of an $S^5$ in $\mathcal X$ is purely effected by a blow-up of an $S^3$ in $\widetilde X \cong T^*S^3$. Noting that $\widetilde M_2 \hookrightarrow \mathcal X \to \widetilde X$, we can view the IIB string on $\mathcal X$ as an initial IIB compactification on a CY 3-fold $\widetilde X \cong T^*S^3$ followed by a further compactification on the $\widetilde M_2$ fibre. From the perspective of the initial IIB compactification on $\widetilde X$, a blow-up of an $S^3 \subset \widetilde X$ means that $\delta \int_{S^3} \Omega \neq 0$, whereby $\Omega$ is the holomorphic volume (3,0)-form defined on $\widetilde X$, and $\delta \int_{S^3} \Omega = \int_{\delta S^3} \Omega$. However, one can $\it{equivalently}$ view the variation as $ \delta \int_{S^3} \Omega = \int_{S^3} \delta \Omega \neq 0$, i.e. $\delta \Omega \neq 0$, such that the $S^3$ is non-varying or $\delta S^3 = 0$ instead. Since $\delta \Omega \neq 0$, one will have $\bar {\partial} \Omega \neq 0$ and thus $d\Omega \neq 0$. The non-closure of $\Omega$ in a supersymmetric type IIB compactification on a 6-dimensional manifold endowed with an $SU(3)$ structure such as $\widetilde X$ (with non-varying volume) is measure by the following relation \cite{flux}:
\be
d \Omega = \bar{W}_5 \wedge \Omega,
\label{flux}
\ee
whereby $W_5$ is a complex (1,0)-form representing a torsion class whose degrees of freedom are the $3 \oplus \bar{3}$ of $SU(3)$, and a non-zero $\bar W_5$ is induced by a unit of 3-form $NS$ flux through a 3-cycle in $\widetilde X$ normal to the tangent directions of $S^3\subset \widetilde X$ along which $\Omega$ is non-closed. In other words, we must turn on a unit of 3-form $dB_{NS}= H_{NS}$ flux through the non-compact 3-cycle (normal to the $S^3$, with topology of a 3-ball) in the $\mathbb R^3$ bundle of $\widetilde X$.

Note that the presence of $RR$ and $NS$ fluxes in the $\mathcal X$ background will effectively break supersymmetry in the resulting d=1+1 theory; recall that the `deformed' $Spin(7)$ conifold $\mathcal X$ should only preseve 1/16 of the maximal supersymmetry while a CY space such as $\widetilde X \cong T^*S^3$ should only preserve 1/4 of the maximal supersymmetry. This then implies that a further compactification of the IIB theory on $\widetilde M_2$ should preserve 1/4 of the remaining conserved supercharges from its initial compactification on $\widetilde X$. Hence, since $H_{RR}$ and $H_{NS}$ only reside within $\widetilde X$, we can view the flux compactification of the IIB string on the 8-dimensional $\mathcal X$ down to $d=1+1$ as a flux compactification on a CY 3-fold $\widetilde X \cong T^*S^3$ down to $d=3+1$, with $N$ units of $H_{RR}$ flux through the compact $S^3 \subset \widetilde X$ and a unit of $H_{NS}$ flux through the non-compact 3-cycle of $\widetilde X$, followed by an additional compactification on a non-trivial $\widetilde M_2$ fibre down to $d=1+1$, which preserves 1/4 of the remaining conserved supercharges from the flux compactification on $\widetilde X$. Looking at the initial IIB compactification on $\widetilde X \cong T^*S^3$, the presence of $H_{RR}$ and $H_{NS}$ fluxes in $\widetilde X$ will result in the introduction of a superpotential term to the existing Lagrangian of the $\mathcal N= 2$ theory in $d=3+1$ from a IIB compactification on $\widetilde X$ $\it{without}$ fluxes or space-filling D-branes. This term is given by \cite{gw,tv,mayr2}
\be     
W=\int_{\widetilde X} {(H_{RR} + \tau H_{NS}) \wedge \Omega},
\label{W}
\ee      
where $\tau = {\theta_{4d} \over 2\pi} + {4\pi i \over {g_{4d}}^2}$ is the complexified IIB coupling, of which $g_{4d}$ and $\theta_{4d}$ are the bare gauge coupling constant and $\theta$-angle of the $d=3+1$ theory respectively. We can choose a basis in $H_3(\widetilde X, \mathbb Z)$ given by $(A,B)$, where $A$ and $B$ are the compact $S^3$ and non-compact 3-cycle of $\widetilde X$ respectively. Since we require $N$ units of $H_{RR}$ flux through $A \equiv S^3$ and a unit of $H_{NS}$ flux through the dual non-compact 3-cycle $B$, we will have 
\be
N = \int_{A} H_{RR}, \qquad 1 = \int_{B}H_{NS},
\label{basis}
\ee
and
\be
\int_{A} H_{NS} = \int_{B} H_{RR}=0.
\ee   
Using the Riemann billinear identity
\be
\int_{\widetilde X} H \wedge \Omega = \int_{A} H \int_{B}\Omega - \int_{B} H \int_{A} \Omega,
\label{riemann}
\ee
where $H = (H_{RR} + \tau H_{NS})$, and the special geometry relations
\be
\int_{A} \Omega = 2\pi i S, \qquad \int_{B} \Omega = {\partial {\mathcal F_0} \over \partial {S}},
\label{special}
\ee
noting that $S \sim vol (S^3)$ and $\mathcal F_0$ is the free energy at genus 0 of the closed topological B string on $\widetilde X \cong T^*S^3$ given by $\mathcal F_0 = {1\over 2}{S}^2 log S + P_2(S)$, whereby $P_2(S)$ is a certain degree 2 polynomial in $S$, we obtain 
\be
W = N {\partial \mathcal F_0 \over \partial S} - 2\pi i \tau S = NS log S -2\pi i \tau S.
\label{glueball}
\ee   
This is the standard Veneziano-Yankielowicz glueball superpotential for $\mathcal N =1$ supersymmetric pure $SU(N)$ in $d=3+1$ if we identify $S$ as the glueball field \cite{ven}. In other words, the compactification of the IIB string on $\widetilde X$ in the presence of the $H_{RR}$ and $H_{NS}$ fluxes results in an $\mathcal N=1$ ($\it{not}$ $\mathcal N=2$) theory in $d=3+1$. In fact, by extremizing the superpotential $W$ in (\ref{glueball}), we obtain the vev of the glueball condensate as 
\be
{\langle S \rangle} \sim e^{{2\pi i \tau} \over N}.
\label{Nvacua}
\ee
The non-zero vev of the glueball condensate $\langle S \rangle$ breaks a global $\mathbb Z_{2N}$ chiral symmetry down to its $\mathbb Z_2$ subgroup, thus resulting in the $N$ inequivalent vacua typical of an effective $\mathcal N =1$ supersymmetric pure $SU(N)$ theory in $d=3+1$ at low energy. This phenomenon is implicit in (\ref{Nvacua}), where one finds that by making the substitution $\theta_{4d} \rightarrow \theta_{4d} + 2\pi m$, contrary to the original high energy regime in which the global $\mathbb Z_{2N}$ chiral symmetry is intact and the vacuum remains invariant under this shift in $\theta_{4d}$,\footnote{When the global $\mathbb Z_N$ symmetry is unbroken, the relevant phase factor is given by $e^{2\pi i \tau} = e^{(i\theta_{4d} - 8 \pi^2/ g^2_{4d})}$. Hence, a shift such as $\theta_{4d} \rightarrow \theta_{4d} + 2\pi m$ leaves this phase factor and consequently the vacuum invariant.} there is now a choice of $N$ distinct inequivalent vacua labelled by the phases ${2 \pi m} \over N$, whereby $m =$ 0, 1, 2..., $N-1$. Hence, a further compactification on the $\widetilde M_2$ fibre down to $d=1+1$ will preserve 1/4 of the four conserved supercharges from the $\mathcal N=1$ SUSY algebra in $d=3+1$, which effectively results in one conserved supercharge in $d=1+1$. Therefore, the presence of $RR$ and $NS$ fluxes on $\mathcal X$ effectively breaks supersymmetry in the $d=1+1$ theory by preserving only one out of the two conserved supercharges from a IIB compactification on $\mathcal X$ without fluxes. Note also that due to small string coupling $g_s$ and the non-compactness of $\mathcal X \cong \mathbb R^3 \times S^5$, there is a decoupling of gravitational and stringy effects.  Thus, as expected of a dual background, we indeed have an $\mathcal N = (1,0)$ globally supersymmetric gauge field theory in $d=1+1$ from the IIB flux compactification on the `deformed' $Spin(7)$ conifold $\mathcal X$ with no D-branes, in agreement with the effective supersymmetric gauge field worldvolume theory in $d=1+1$ from the IIB string compactification on the `resolved' $Spin(7)$ conifold $\mathcal Q$ with space-filling D5-branes and no fluxes before the geometric transition. Moreover, since there are no D-branes in the `deformed' $Spin(7)$ conifold closed string background, there is an absence of the Freed-Witten global anomaly (which is due to open string worldsheets ending on D-branes) discussed earlier. Hence, there is $\it{no}$ restriction on $H_{NS}$ in this case (recall that $H_{NS}|_{\mathbb R^{1,1} \times \mathbb {CP}^2} = 0$ must be imposed to cancel the anomaly in the case when there are space-filling D5-branes wrapping $\mathbb{CP}^2 \subset \mathcal Q$). Therefore, our dual IIB background on the `deformed' $Spin(7)$ conifold $\mathcal X$ with no D-branes is hence anomaly-free even in the presence of $\it{non}$-$\it{vanishing}$ 3-form $H_{NS}$ flux.   

Thus, we indeed have a physically consistent and anomaly-free $Spin(7)$ extension of the original Gopakumar-Vafa type IIB CY conifold duality; the IIB background on $\mathcal Q$ with $N$ space-filling D5-branes wrapping the $\mathbb{CP}^2 \subset \mathcal Q$, where $N$ is large, can be viewed to undergo a $Spin(7)$ geometric transition at $\it{low}$ energy to a $\it{dual}$ IIB background on $\mathcal X$ with no D-branes but with $N$ units of 3-form $H_{RR}$ flux through $S^3 \subset \mathcal X$ and a unit of 3-form $H_{NS}$ flux through a non-compact 3-cycle in $\mathcal X$!

\subsection{Relevant aspects of the effective $\mathcal N = (1,0)$ supersymmetric pure $SU(N)$ theory in 1+1 dimensions}       

Let us now investigate the relevant aspects of the effective theory in $d=1+1$ before we proceed to furnish a gauge theoretic interpretation of the $Spin(7)$ geometric transition with D5-branes/fluxes which will then allow for an alternative verification of the IIB $Spin(7)$ duality in further support of its physical validity. The effective $\mathcal N = (1,0)$ supersymmetric pure $SU(N)$ theory in $d=1+1$ from the IIB compactification on the non-compact $Spin(7)$ manifold $\mathcal Q$ ($\mathcal X$) in the presence of $N$ space-filling D5-branes (3-form fluxes) for small $g_s$, large $N$ and hence finite 't Hooft coupling given by\footnote{The 't Hooft coupling is actually given by $\lambda_s N$, whereby $\lambda_s$ is the $\it{topological}$ string coupling. However, the topological string amplitude at each genus $g$ (associated with the factor $\lambda_s^{2g-2}$) corresponds to the F-term correction in the superstring theory coming from the same genus $g$ amplitude (associated with the factor $g_s^{2g-2}$). Thus, we can identify the topological string coupling $\lambda_s$ with the superstring coupling $g_s$. Hence, the 't Hooft coupling is equivalently given by $g_s N$.} $t = g_s N$, has an action (see eqn. (\ref{finalaction}) of Appendix A)
\begin{align}
S_{SYM}= {{a} \over {g^2}} \intd{^2x}Tr(F_{\plpl\mimi}^{\it k} F_{\plpl\mimi}^{\it k}   - i  \chi_-^{\it k} {\nabla}_{\plpl}\chi_-^{\it k}) + \ b\ \theta \intd{^2x} Tr(F_{\plpl\mimi}^{\it k}),  
\label{SYM action} 
\end{align}
whereby $g$ and $\theta$ are the usual unrenormalized gauge coupling and theta angle respectively, $a$ and $b$ are dimensionless constants of proportionality, while $x^\plpl$ and $x^\mimi$ are bosonic light-cone/null coordinates which can be expressed in terms of the $d=1+1$ coordinates $x$ and $t$ as $(x^\plpl,x^\mimi) \equiv (x+t,x-t)$. ${\nabla}_{\plpl}$ here is a $\it spacetime$ gauge covariant derivative compatible with the spinor gaugino fields $\chi^{\it k}_-(x,t)$. The trace has been taken over the matrix components of the generators (labelled by $k)$ in the corresponding representation of the gauge group, which have been suppressed here for simplicity in expression. As the Yang-Mills multiplets are in the adjoint representation of the $SU(N)$ gauge group, the sum over $\it k$ in \eqref{SYM action} is taken over the values ${\it k}=1,2,3$.....dim $\mathfrak {su(n)}$, where $\mathfrak {su(n)}$ is the Lie algebra of $SU(N)$ and dim $\mathfrak {su(n)}$ $=N^2-1$.

As usual, the Lagrangian in (\ref{SYM action}) must be dimensionless. Recall that both derivatives and massless Yang-Mills fields have dimensions of length$^{-1}$ (mass). This implies that $F_{\plpl\mimi}^{\it k}$ has dimensions length$^{-2}$ (mass$^2$). Since $d^{2}x$ has dimensions of length$^2$ (mass$^{-2}$), a dimensionless Lagrangian will then imply that the bare gauge coupling $g$ in $d=1+1$ has dimensions of length$^{-1}$ or mass, i.e. it has canonical dimensions. This means that the effective $\it{dimensionless}$ gauge coupling in $d=1+1$, which we will henceforth denote as $g_{YM}$, must classically (i.e. without considering quantum loop corrections yet) be given by $g / u$, whereby $u$ is the energy scale of interest with dimensions of mass. Thus, we find that regardless of the amount of supersymmetry present, the effective gauge coupling of a supersymmetric Yang-Mills theory in $d=1+1$ can be seen to diverge in the IR (small $u$) even before considering the loop corrections, just as in ordinary Yang-Mills theory in $d=1+1$. This will in turn imply that the effective gauge coupling is small in the UV and that the theory is therefore asymptotically free. 

Notice that the action (\ref{SYM action}), for the supersymmetric $SU(N)$ theory in $d=1+1$, closely resembles that for QCD with a single flavour of massless quarks in $d=1+1$. Note that it has been shown that QCD with a single flavour of massless quarks in 2-dimensional spacetime can exist in two distinct phases, namely the 't Hooft phase (at large $N$) at weak gauge coupling, and the Higgs phase at strong gauge coupling \cite{thooft,pas,mitra}. In the 't Hooft phase at weak coupling, there is colour confinement and all gluons remain massless, thus resulting in an unbroken $SU(N)$ gauge symmetry. However, in the Higgs phase at strong coupling, it has been argued in \cite{pas} and further supported in \cite{mitra} and \cite{mitra1}, that there is a dynamical Higgs mechanism which breaks the non-abelian group down to its maximally abelian subgroup. The analysis in \cite{mitra} and \cite{mitra1} was defined for all solutions, physical or otherwise. Subsequently, an analysis to restrict the solutions in \cite{mitra} and \cite{mitra1} to physical ones was carried out in \cite{mitra3}, whereby it was demonstrated that even the maximally abelian subgroup is broken down, and that there remains a global $U(1) \times U(1)$ symmetry from a chiral and fermion $U(1)$ number carried by the relevant matter field. As expected, this breakdown of symmetry is shown to be accompanied by the emergence of multiple topological vacua that can be characterized by $2N$ distinct $\theta$-phases, whereby the factor of 2 arises from a trivial association with the relevant spinor fields' 2-component index $s=1,2$. In other words, the large $N$, non-abelian $SU(N)$ QCD theory in $d=1+1$, with a single flavour of massless quarks at weak gauge coupling in the UV, can undergo RG flow to an abelian $U(1) \times U(1)$ theory with strong gauge coupling in the IR, whence there is an emergence of multiple vacua. 

Note that the analysis in \cite{mitra}, and therefore that in \cite{mitra1} and \cite{mitra3}, although carried out with $N$ fermionic quark fields of the fundamental representation of $SU(N)$, is defined for an arbitrary number of fermionic matter fields, from which one can see that a dynamical Higgs mechanism which breaks the $SU(N)$ gauge group to an abelian group at strong coupling will persist provided the number of fermionic fields $\geq N$. Note also that the action in (\ref{SYM action}) has no classical chiral symmetry as it is not invariant under the chiral symmetry transformation $\chi_-^{\it k} \rightarrow e^{i \delta}\chi_-^{\it k}$, whereby $\delta \in \mathbb R$. Moreover, the gaugino $\chi_-^{\it k}$ are 1-component spinors, i.e. their spinor index spans a single value of $s=1$. Last but not least, as mentioned above, since the supersymmetry of the theory with action (\ref{SYM action}) dictates that the fermionic superpartners (i.e. gauginos) must come from the vector multiplets, their total number will be given by the dimension of the adjoint representation of $SU(N)$, which is $N^2 -1$ $> N$. Clearly, this means that we can expect to observe a similar phenomenon in the $\mathcal N = (1,0)$ supersymmetric pure $SU(N)$ theory in 1+1 dimensions at low energy, whereby the $SU(N)$ gauge group will be spontaneously broken down to a $U(1)$ abelian group, such that there is an emergence of $N$ inequivalent topological vacua, consequently characterized by $N$ distinct $\theta$-phases. In fact,  this phenomenon is consistent with our duality picture as we will see shortly.

\subsection{Gauge-theoretic interpretation and verification of the large $N$, type IIB, $Spin(7)$ duality} 

The effective theory in $d=1+1$ from a IIB compactification on $\mathcal Q$ with $N$ space-filling D5-branes wrapping the $\mathbb{CP}^2 \subset \mathcal Q$ is the $d=1+1$ worldvolume theory obtained from a dimensional reduction of the original $d=5+1$ worldvolume theory of the $N$ D5-branes along the tangent directions of the 4-dimensional $\mathbb{CP}^2$. Re-defining the bare gauge coupling constant in the $d=5+1$ worldvolume theory to be of order 1, the bare gauge coupling $g$ which appears in the action (\ref{SYM action}) of the $d=1+1$ theory will therefore be given by
\be
{1\over g^2} = vol (\mathbb {CP}^2_{bare}),
\label{bare}
\ee     
whereby $vol(\mathbb{CP}^2_{bare})$ is the volume of the $\mathbb {CP}^2 \subset \mathcal Q$ $\it{before}$ it starts to blow-down in a low energy geometric transition $\mathcal Q \rightarrow \mathcal X$. 

Recall that a IIB compactification on $\mathcal Q$ is equivalent to an initial compactification on $\widetilde Q \cong {\mathcal O(-1) \oplus \mathcal O(-1)\rightarrow {\mathbb {CP}^1}}$ to $d=3+1$ followed by a further compactification on the $\widetilde M_2$ fibre to $d=1+1$. Moreover, recall that since $\mathbb{CP}^2$ is given by $\widetilde M_2 \hookrightarrow \mathbb{CP}^2 \to \mathbb{CP}^1$, the $d=1+1$ worldvolume theory can be viewed as a dimensional reduction of the $d=3+1$ worldvolume theory from the $N$ space-filling D5-branes wrapping the $\mathbb{CP}^1 \subset \widetilde Q$, along the tangent directions of the $\widetilde M_2$ fibre. Note that the gauge coupling constant $g_{4d}(u)$ of the $\mathcal N =1$ supersymmetric pure $SU(N)$ worldvolume theory in $d=3+1$ from the initial IIB compactification on $\widetilde Q$ with $N$ space-filling D5-branes is dimensionless. Since $vol({\widetilde M_2})$ and energy $u$ have dimensions of mass$^{-2}$ (length$^2$) and mass respectively, the effective $\it{dimensionless}$ Yang-Mills gauge coupling of the $d=1+1$ theory at energy scale $u$ denoted as $g_{YM}(u)$ will be given by 
\be
{1\over {g^2_{YM}(u)}}\ = \ vol({\widetilde M_2}) \cdot ({u \over {g_{4d}(u)}})^2.
\label{gym}
\ee            
It is well known that up to 1-loop corrections, $g_{4d}(u)$ is given by the following relation \cite{QFT and strings 2}
\be
{1\over g^2_{4d}(u)} \ + \  i \theta_{4d}\ = \ {1\over {g^2_o}}\ + \ \alpha \ log\ ({u \over {\Lambda_{planck}}}).
\label{go}
\ee   
Here, $g_o$ is the bare gauge coupling of the $d=3+1$ worldvolume theory at the Planck scale and is thus given by a dimensional reduction of the gauge coupling (of order 1) of the $d=5+1$ worldvolume theory as ${1\over {g^2_o}} = vol (\mathbb{CP}^1_{bare})$, whereby $vol(\mathbb{CP}^1_{bare})$ is the volume of the $\mathbb {CP}^1 \subset \widetilde Q$ before it starts to blow-down in the low energy geometric transition $\widetilde Q \rightarrow \widetilde X$. $\alpha$ is a real positive constant, $\theta_{4d}$ is the theta angle in $d=3+1$ and $\Lambda_{planck}$ is a constant complex number such that $|\Lambda_{planck}|$ denotes the energy at the Planck scale. We therefore deduce from (\ref{go}) above that 
\be
{1\over g^2_{4d}(u)}\ = \ {1\over {g^2_o}} \ + \ \alpha \ log \ ({u \over {|\Lambda_{planck}|}}).
\label{g4d}
\ee
The bare coupling relation in (\ref{bare}) suggests that we can view the effective dimensionless Yang-Mills coupling $g_{YM}(u)$ of (\ref{gym}) in a similar way such that up to an irrelevant constant of proportionality with dimensions of mass$^2$, we have ${1 \over {g^2_{YM}(u)}} \sim vol (\mathbb{CP}^2_{eff})$, whereby $vol (\mathbb{CP}^2_{eff})$ is the $\it{effective}$ volume of the $\mathbb{CP}^2$ at the energy scale $u$. Notice also that $vol(\widetilde M_2)$ is a constant since $\widetilde M_2$ is a `spectator' space as explained previously. Hence, $vol (\mathbb {CP}^2_{bare}) = vol (\mathbb{CP}^1_{bare}) \cdot vol (\widetilde M_2)$. Substituting (\ref{g4d}) in (\ref{gym}), noting the relation in (\ref{bare}), we thus have for $g_{YM}(u)$ and $vol(\mathbb{CP}^2_{eff})$ the following result:
\be
vol(\mathbb{CP}^2_{eff}) \sim {1\over {g^2_{YM}(u)}} \ = \ {1\over {{(g/u)}^2}}\ + \ {\alpha}' \ u^2 log({u \over {|\Lambda_{planck}|}}), 
\label{RG flow}
\ee
whereby ${\alpha}'$ is a real positive constant. The first term on the RHS of (\ref{RG flow}) contains the dimensionless classical contribution of $(g/u)$ that we had anticipated earlier from dimensional considerations in the previous subsection while the second term is representative of the quantum correction. From (\ref{RG flow}), we see that $1\over {g^2_{YM}(u)}$ and thus $vol(\mathbb{CP}^2_{eff})$ gets smaller for decreasing $u$. In fact, for small enough values of $u$, we seem to get a $\it{negative}$ value for $1\over {g^2_{YM}(u)}$ and $vol(\mathbb{CP}^2_{eff})$ as the $d=1+1$ theory undergoes a RG flow to the IR, i.e.  $[vol(\mathbb{CP}^2_{eff}) > 0]$ $\rightarrow$ $[vol(\mathbb{CP}^2_{eff}) < 0]$. 

Before we proceed any further to make sense of a negative $vol(\mathbb{CP}^2_{eff})$ and $1\over {g^2_{YM}(u)}$, let us first ascertain if we will encounter any singularities in the $d=1+1$ theory as it undegoes a transition from $[vol(\mathbb{CP}^2_{eff}) > 0]$ $\rightarrow$ $[vol(\mathbb{CP}^2_{eff}) < 0]$ or equivalently, from $[{1\over {g^2_{YM}(u)}} > 0] \rightarrow [{1\over {g^2_{YM}(u)}} < 0]$. First, note that to ensure the absence of singularities in the ground state wave function of the quantum theory such that the wave function varies smoothly with the moduli, we must have a discrete spectrum if the space is compact and a potential that grows at infinity. For a non-compact space such as $\mathbb R^{1,1}$, this condition is satisfied by a non-zero vacuum energy density \cite{QFT and strings 2}. For instance if we have a non-compact space such as $X= \mathbb R$, whereby $V={1\over 2}k {x^2}$ (with $x$ a linear function on $\mathbb R$), then the ground state wave is a smooth function of $k$ as long as $k>0$ ($V>0$), but develops a singularity at $k=0$ ($V=0$). Next, note that the $\theta$-dependent vacuum energy density of $SU(N)$ gauge theories in 2-dimensions $E_{vac}(\theta)$ is given by \cite{QFT and strings 2}
\be
E_{vac}(\theta) \sim \ g^2 N \cdot min_{n \in \mathbb Z}\ (n- {\theta\over 2\pi}).
\label{vac}
\ee 
Hence, from (\ref{vac}) above, we conclude that regardless of the value of $vol (\mathbb{CP}^2_{eff})$ or $1\over {g^2_{YM}(u)}$ in the transition $[vol(\mathbb{CP}^2_{eff}) > 0]$ $\rightarrow$ $[vol(\mathbb{CP}^2_{eff}) < 0]$ or $[{1\over {g^2_{YM}(u)}} > 0] \rightarrow [{1\over {g^2_{YM}(u)}} < 0]$, there will not be any singularities in the $d=1+1$ theory as long as $E_{vac}(\theta) > 0$ or $\theta \neq 0$. Note also that the complexified gauge coupling $({1\over {g^2_{YM}(u)}} + i \theta)$ of the $d=1+1$ theory is a complexified moduli of the IIB theory on $\mathcal Q$ or $\mathcal X$. Thus, since ${1 \over {g^2_{YM}(u)}} \sim vol (\mathbb{CP}^2_{eff})$, the angle $\theta$ will be given by the vev of the self-dual RR 4-form $C^{+}_4$ along $\mathbb{CP}^2_{bare}$ because it is not relevant in perturbation theory and consequently does not undergo RG flow, i.e. 
\be
\theta  = \int_{\mathbb{CP}^2_{bare}} C^{+}_4. 
\label{theta}
\ee
Hence, we can conclude that if we at least turn on $C^+_4$ along the tangent directions of $\mathbb {CP}^2$, there will be no singularities in the transition from $[vol(\mathbb{CP}^2_{eff}) > 0]$ $\rightarrow$ $[vol(\mathbb{CP}^2_{eff}) < 0]$ or from $[{1\over {g^2_{YM}(u)}} > 0] \rightarrow [{1\over {g^2_{YM}(u)}} < 0]$ as the resulting $d=1+1$ theory flows to the IR. Notice also that turning on $C^+_4$ does not violate any anomaly-cancelling or supersymmetry-breaking conditions. Neither does it violate any of the duality arguments that were discussed and put forth earlier. Let us therefore assume that this condition is trivially satisfied in the rest of the paper. 

Notice at this point that in the $Spin(7)$ geometric transition $\mathcal Q \rightarrow \mathcal X$, there is a blow-down of a $\mathbb {CP}^2$ and a blow-up of an $S^5$ and consequently a $\mathbb {CP}^2$ since we have the Hopf fibration $S^1 \hookrightarrow S^5 \to \mathbb{CP}^2$. We can thus view the geometric transition $\mathcal Q \rightarrow \mathcal X$ as being induced by a $\mathbb {CP}^2$ flop within the 8-dimensional `resolved' and `deformed' $Spin(7)$ conifolds, such that the $\mathbb {CP}^2$ flop can be interpreted as a transition from $[vol(\mathbb{CP}^2) > 0]$ $\rightarrow$ $[vol(\mathbb{CP}^2) < 0]$, whereby the volume of the blown-up $\mathbb {CP}^2 \subset \ S^5 \subset \mathcal X$ is given by $|vol(\mathbb{CP}^2) < 0|$. Therefore, although the appearance of a negative $1\over {g^2_{YM}(u)}$ in the transition $[{1\over {g^2_{YM}(u)}} > 0] \rightarrow [{1\over {g^2_{YM}(u)}} < 0]$ seems meaningless, which occurs whenever the $SU(N)$ theory in $d=1+1$ undergoes a RG flow to the $\it{same}$ theory in the IR with strong coupling, we can make full physical sense of it by viewing the equivalent $\it{non}$-$\it{singular}$ transition from $[vol(\mathbb{CP}^2_{eff}) > 0]$ $\rightarrow$ $[vol(\mathbb{CP}^2_{eff}) < 0]$ as a $\mathbb{CP}^2$ flop that will induce a $\it{smooth}$ $Spin(7)$ geometric transition in the IIB background on $\mathcal Q$ at $\it{low}$ energy to the $\it{dual}$ background on $\mathcal X$ whereby the $N$ D5-branes wrapping the blown-down $\mathbb{CP}^2 \subset \mathcal Q$ vanish and are replaced by 3-form RR and NS fluxes through appropriate 3-cycles in $\mathcal X$, thus resulting in the $\it{same}$ albeit abelian $d=1+1$ theory in the IR with strong\footnote{Just after the geometric transition from $\mathcal Q$ to $\mathcal X$ at low energy, we have ${1\over {g^2_{YM}(u)}} \sim |vol(\mathbb{CP}^2_{eff}) < 0|$ and $|vol(\mathbb{CP}^2_{eff}) < 0| = vol (\mathbb {CP}^2 \subset S^5 \subset \mathcal X)$, whereby $vol (S^5) << 1$. This implies that $|vol(\mathbb{CP}^2_{eff}) < 0| << 1$ and $g_{YM} >> 1$ in the $\mathcal X$ background.} gauge coupling that is fully consistent with the purely gauge-theoretic discussion in $\S$3.2 on the low energy behaviour of an $\mathcal N = (1,0)$ supersymmetric pure $SU(N)$ theory in 1+1 dimensions, where it was noted that one can expect the non-abelian $SU(N)$ theory to be spontaneously broken down to an abelian one with strong gauge coupling due to a dynamical Higgs mechanism. Moreover, the 3-form RR and NS fluxes also give rise to an expected emergence of multiple vacua of the $d=1+1$ theory at low energy, which, as discussed in $\S$3.2, is characterized by $N$ distinct $\theta$-phases. This can be seen as follows: recall that the 3-form RR and NS fluxes, via (\ref{W}), contributes to an additional superpotential term in the resulting $d=3+1$ theory from an intermediate compactification of the IIB theory on $\widetilde X$ (which, according to the Gopakumar-Vafa superstring duality, is in turn equivalent to the low energy regime of the $d=3+1$ theory from a IIB compactification on $\widetilde Q$ with $N$ space-filling D5-branes and no 3-form fluxes). One is then led to (\ref{Nvacua}), the vev of the glueball condensate in $d=3+1$, which is given by $\langle S \rangle \sim e^{({i \theta'_{4d}} - {8 \pi^2 / g'^{2}_{o}})}$, where $\theta'_{4d} = {\theta_{4d} \over N}$ and $1/ g'^{2}_{o} = ({1\over N}) (1/ g^{2}_{o})$. Recall and note that since $1/ g^{2}_{o} = vol(\mathbb {CP}^1_{bare})$ and $\theta_{4d} = \int_{\mathbb {CP}^1_{bare}} C_2$, where $C_2$ is the RR 2-form, the $\it{effective}$ volume of $\mathbb {CP}^1_{bare}$ is now divided by a factor of $N$, i.e. $vol (\mathbb {CP}^{1'}_{bare}) = {vol (\mathbb {CP}^1_{bare}) \over N}$. Recall also that $vol (\mathbb {CP}^2_{bare})= vol (\mathbb {CP}^1_{bare}) \cdot vol (\widetilde M_2)$, which implies that $vol (\mathbb {CP}^{2'}_{bare}) = {vol (\mathbb {CP}^{2}_{bare}) \over N}$. Since $\theta = \int_{\mathbb {CP}^2_{bare}} C^+_4$, we then find that $\theta' = {\theta \over N}$. Therefore, the relevant phase factor $e^{i \theta}$ is effectively replaced by $e^{i\theta'}$ at low energy. A shift given by $\theta \rightarrow \theta + 2\pi q$, which leaves the phase factor $e^{i \theta}$ and thus, the vacuum of the theory in the high energy regime invariant, will then, at low energy, present one with a choice of $N$ inequivalent vacua labelled by distinct phases ${2\pi q} \over N$ via $e^{i 2\pi q \over N}e^{i\theta'}$, where $q=$0, 1, 2, ..., $N-1$.          

In short, we have arrived at a purely gauge-theoretic interpretation and verification of the large $N$, type IIB, $Spin(7)$ geometric transition duality; the $\it{smooth}$ large $N$, type IIB, $Spin(7)$ geometric transition from $\mathcal Q$ with $N$ D5-branes to the $\it{dual}$ theory on $\mathcal X$ at $\it{low}$ energy with no D-branes but with supersymmetry-breaking fluxes, is a consequence of a $\it{non}$-$\it{singular}$ RG flow of the resulting $\mathcal N = (1,0)$ supersymmetric pure $SU(N)$ theory in 1+1 dimensions to the $\it{same}$ (i.e. $\it{dual}$) albeit abelian $U(1)$ theory in the IR with $N$ inequivalent vacua characterized by $N$ distinct $\theta$-phases, such that the sizes of the blown-down $\mathbb{CP}^2 \subset \mathcal Q$ and blown-up $S^5 \subset \mathcal X$ before and after the geometric transition $\mathcal Q \rightarrow \mathcal X$ are governed by the (smoothly running) value of the effective dimensionless gauge coupling ${1\over {g^2_{YM}(u)}}$ of the $d=1+1$ theory at the observed energy scale $u$, which in turn determines the point in the geometric transition $\mathcal Q \rightarrow \mathcal X$!   

\section{Lifting the large $N$, $Spin(7)$, type IIB duality} 

In this section, we will systematically lift the large $N$, $Spin(7)$, type IIB duality to its equivalent F-theoretic $\mathbb{RP}^5$ flop description for small $\it{or}$ large string coupling via the following steps: we will first provide a geometric lift of the original IIB theory on $\mathcal Q$ or $\mathcal X$ with D5-branes or 3-form RR and NS fluxes to an equivalent background with $\it{no}$ D5-branes or 3-form fluxes via a IIB compactification on a suitable 8-dimensional manifold, eventually furnishing a gauge-theoretic interpretation of the IIB duality in this lifted background as a consistency check and for completeness. Next, we will discuss the corresponding results for large string coupling via an application of a IIB S-duality transformation. Afterwhich, we will review the F-theoretic description of a general IIB vacua or bacground. Finally, we will describe the equivalent F-theoretic description of this lifted background, which then allows us to demonstrate the F-theoretic $\mathbb{RP}^5$ flop for small or large string coupling $g_s$. Details of the solution of the metric on the flop manifold will also be furnished.           

\subsection{Geometric lift to a IIB background without D5-branes and 3-form fluxes} 
\vspace{-0.1cm}
In order to lift the IIB background on $\mathcal Q$ with $N$ space-filling D5-branes to an equivalent one without D5-branes but with an $\mathcal N=(1,0)$ supersymmetric theory in $d=1+1$ with $SU(N)$ gauge symmetry, we will need to consider a IIB compactification on a suitable 8-dimensional manifold with the following properties: firstly, it must possess the correct singularities which will introduce an $SU(N)$ gauge symmetry in the lower dimensional $d=1+1$ theory. Secondly, it must only preserve $1/32$ of the maximal supersymmetry from the original 10-dimensional theory. Lastly, It must be non-compact so that there can be a decoupling of stringy and gravitational effects as in the original background on $\mathcal Q$ and $\mathcal X$ with D5-branes and 3-form fluxes. 

\bigskip\noindent{\it  The $A_{N-1}$ Singularity and $SU(N)$ Gauge Theories}
\vspace{0.2cm}

The singular 4-dimensional subspace of the suitable non-compact 8-dimensional manifold required to introduce an $SU(N)$ gauge symmetry in $d=1+1$ is given by the singular $\mathbb R^4 / \mathbb Z_N$ space as follows: the McKay Correspondence \cite{mirrortext} defines an intricate mathematical relationship between finite subgroups of $SU(2)$ and a simply-laced ADE Lie algebra. The physical interpretation of this correspondence is readily manifest in the geometric engineering of 4-dimensional quantum field theories from string theory compactifications \cite{kkv,bershadsky}. In essence, the singular ALE space given by ${{\mathbb {C}^2} / \Lambda} = {{\mathbb {R}^4} / \Lambda}$, where $\Lambda = \mathbb Z_N$ is a finite subgroup of $SU(2)$, is said to have a singularity of $A_{N-1}$ type. Mathematically, it means that the resolution of this singular space will involve the blow-up of middle homology cycles (i.e.\ 2-cycles) which will intersect according to the corresponding $A_{N-1}$ Dynkin diagram; the 2-cycles ($S^2$s) correspond to nodes of the Dynkin diagram and lines joining adjacent nodes represent the intersections between the corresponding 2-cycles \cite{mayr}. Let the rank of the $A_{N-1} \equiv SU(N)$ algebra and therefore the number of nodes and hence $S^2$s be given by $r$. Then in IIB theory, one can have $r$ D3-branes that wrap around these $S^2$s whilst the remaining string-like degrees of freedom wind around the 4-dimensional subspace normal to the 4-dimensional resolution of the singular $\mathbb R^4 / \mathbb Z_N$ manifold in the 8-dimensional space, such that there will be an enhanced gauge symmetry as the resolved subspace becomes singular again due to the appearance of extra massless point-like states in $d=1+1$ that come from the D3-branes when the $S^2$s shrink to zero volume in which the winding orientation determines their various charges. In other words, there will be an enhanced non-abelian $SU(N)$ gauge symmetry in the lower-dimensional field theory in $d=1+1$, furnished by the massless gauge field states in the adjoint representation, whenever one has a string compactification on a manifold containing a singular $\mathbb R^4/ \mathbb Z_N$ subspace or a locus of $A_{N-1}$ singularity. From the hyperk\"ahler moment map of a $U(1)$ action in $\mathbb R^4$ \cite{aw}
\be
{{\mathbb R^4}/ U(1)} \cong {\mathbb R^3},
\label{hyper}
\ee 
we can obtain an $\mathbb R^4/ \mathbb Z_N$ space via identifying the points along the embedded $U(1)$ fibre in $\mathbb R^4$ which are connected by a $\it{fixed}$-acting $\mathbb{Z}_N \subset U(1)$ action.

\bigskip\noindent{\it  The Joyce Construction}
\vspace{0.2cm}

In order to derive a new 8-dimensional manifold such that it will only retain $1/2$ of the maximal supersymmetry preserved by the original one, one can adopt a similar construction by Joyce; a smooth non-simply-connected 8-manifold $\mathcal Z$, with $SU(4) \odot \mathbb Z_2$ holonomy, which preserves $1/16$ of the maximal supersymmetry, can be obtained from a desingularization of a singular manifold $Z$, whereby $Z$ can be constructed from a Calabi-Yau 4-orbifold $Y$ with $SU(4)$ holonomy (which preserves $1/8$ of the maximal supersymmetry) via the relation $Z=Y/\langle \gamma \rangle$, in which $\gamma$ is a freely-acting\footnote{Note that as opposed to a fixed-acting action, a freely-acting one will not result in fixed points which give rise to additional singularities $\it{not}$ already present in a manifold before the action is imposed.} antiholomorphic isometric involution on $Y$, i.e. $\gamma: Y \to Y$ is a diffeomorphism satisfying ${\gamma}^2 = $ id and $\gamma^{*}(J)=-J$, where $J$ is the complex structure on $Y$ \cite{joyce}. This suggests that one can consider a new, non-simply-connected, 8-manifold $\widehat {\mathcal M} = \mathcal M /\langle{\sigma}\rangle$, whereby $\sigma$ is a freely-acting isometric involution on $\mathcal M$, such that if $\mathcal M$ is either $\mathcal Q$ or $\mathcal X$, it will preserve $1/32$ of the maximal supersymmetry as required. 

Let us now specialize to the case when $\mathcal M$ is a fibration of a certain base manifold and discuss the corresponding action of $\sigma$ on it. Note  that from the general definition of a fibration, we see that points on a (base) manifold which are being identified under the action of a freely-acting isometric involution such as $\sigma$ will have identical and hence identified fibres defined over them upon its fibration, i.e. the action of $\sigma$ can be equivalently lifted from the base space to the entire fibre bundle and vice-versa. Hence, when $\mathcal M$ is a fibration over a base manifold with fibre $F$ and base space $B$ as given by $F \hookrightarrow {\mathcal M} \to B$, the action of $\sigma$ on $\mathcal M$ will descend onto the base space $B$ such that $\widehat {\mathcal M} = {\mathcal M} /\langle{\sigma}\rangle$ will be given by the fibration $F \hookrightarrow {\widehat {\mathcal M}} \to B/ {\langle \sigma \rangle}$. 

\bigskip\noindent{\it The Cayley 4-form and $Spin(7)$ Structures}
\vspace{0.2cm}
     
It is known that an 8-manifold with $Spin(7)$ holonomy $\mathcal M$ is endowed with a torsion-free $Spin(7)$ structure defined by the existence of a Cayley 4-form $\Psi$ given by \cite{sparks,gst}  
\be
\Psi = dt \wedge *\rho + \rho,
\label{spin7}
\ee 
where $d \Psi = 0$ and $\rho \in \Omega_{exact}(L)$ is a $U(1)$-invariant 4-form such that the principal orbits of $\mathcal M$ are copies of the 7-manifold $L$. $t$ is a scalar which parameterizes $\rho$. Note that $\Psi$ is a constant tensor, i.e. $\nabla \Psi = 0$, where $\nabla$ is the Levi-Civita connection of the associated $Spin(7)$ metric. It is also known that there is a 1-1 correspondence between constant tensors and the holonomy of a manifold \cite{joyce}. This means that if $\Psi$ is unchanged under an arbitrary action on the manifold $X$ or on its subspace thereof, the holonomy and thus the amount of maximal supersymmetry it will preserve must remain invariant. 

\bigskip\noindent{\it  The Physically Equivalent 8-manifolds $\widehat{\mathcal Q}$ and $\widehat{\mathcal X}$}
\vspace{0.2cm}

Now, from all of the preceding discussions above, it is clear that we can geometrically lift the IIB background on $\mathcal Q \cong \mathbb{R}^4 \times \mathbb{CP}^2$, with $N$ space-filling D5-branes wrapping the $\mathbb{CP}^2 \subset \mathcal Q$, to an equivalent background on the non-compact $\it{singular}$ 8-manifold $\widehat{\mathcal Q} \cong ({\mathbb{R}^4 / \mathbb Z_N) \times (\mathbb{CP}^2 / \langle \sigma \rangle})$ $\it{without}$ any D5 branes. This can be justified as follows: in anticipation of an $SU(N)$ gauge symmetry, we identify points along the $U(1)$ fibre embedded in the $\mathbb R^4$ bundle of $\mathcal Q \cong \mathbb{R}^4 \times \mathbb{CP}^2$ connected by a fixed $\mathbb Z_N$ action so as to obtain a singular space with an $A_{N-1}$ singularity. Since the action is given by $\mathbb Z_N  \subset U(1) \subset U(3)$, whereby $U(3)$ is the isometry group of $\mathcal Q$, $\rho$ and therefore the constant tensor or Cayley 4-form $\Psi$ remains invariant from eqn.(\ref{spin7}), i.e. quotienting the $\mathbb R^4$ bundle of $\mathcal Q$ by  $\mathbb Z_N$ does not change its $Spin(7)$ structure and $(\mathbb R^4 /\mathbb Z_N) \times \mathbb {CP}^2$ continues to be a (singular) $Spin(7)$ 8-manifold that preserves $1/16$ of the maximal supersymmetry. Thus, from the earlier discussions, a new 8-manifold that should preserve $1/32$ of the maximal supersymmetry will be isomorphic to the trivial fibre bundle $[(\mathbb R^4 /\mathbb Z_N) \times \mathbb {CP}^2] / \langle \sigma \rangle$. As explained before, since the action of $\sigma$ will descend onto the $\mathbb {CP}^2$ base of the trivial $\mathbb R^4 /\mathbb Z_N$ bundle with an $A_{N-1}$ singularity over $\mathbb {CP}^2$ denoted by $(\mathbb R^4 /\mathbb Z_N) \times \mathbb {CP}^2$, the new non-compact 8-manifold is effectively given by $\widehat{\mathcal Q} \cong ({\mathbb{R}^4 / \mathbb Z_N) \times (\mathbb{CP}^2 / \langle \sigma \rangle})$. Notice that there is a $({\mathbb {CP}^2 / \langle \sigma \rangle}) \times \mathbb R^{1,1} $ locus of an $A_{N-1}$ singularity. This ensures that there will be an $SU(N)$ gauge symmetry in $d=1+1$.   

Likewise, we can geometrically lift the IIB background on $\mathcal X \cong \mathbb R^3 \times S^5$, with $N$ units of $H_{RR}$ and a unit of $H_{NS}$ flux through the compact and non-compact 3-cycles of $\mathcal X$ respectively, to an equivalent background $\it{without}$ any 3-form fluxes on the non-compact and $\it{smooth}$ 8-manifold $\widehat {\mathcal X} \cong {\mathbb R^3} \times ({S^5}'/{\langle \sigma \rangle})$, whereby $vol({S^5}')={{vol (S^5)} \over N}$. This can be justified as follows: since there are no space-filling D5-branes in the background on $\mathcal X$, there is an absence of a non-abelian $SU(N)$ gauge symmetry i.e. the equivalent manifold must be non-singular. Since the presence of supersymmetry-breaking 3-form RR and NS fluxes effectively results in $\mathcal N = (1,0)$ supersymmetry in $d=1+1$ as argued earlier, the equivalent manifold must preserve $1/32$ of the maximal supersymmetry. Moreover, as explained in $\S$3.3, the presence of 3-form RR and NS fluxes results in the change $\theta \rightarrow {\theta\over N}$ of the effective $\theta$-angle, thereby accounting for the $N$ inequivalent vacua (due to $N$ distinct $\theta$-phases) expected of an $\mathcal N=(1,0)$ supersymmetric pure $SU(N)$ theory in $d=1+1$. As pointed out in $\S$3.3, since $\int_{\mathbb {CP}^2_{bare}} C^+_4 = \theta$, one will also have $vol(\mathbb {CP}^2_{bare}) \rightarrow {vol (\mathbb {CP}^2_{bare}) \over N}$, which in turn implies that $vol(S^5) \rightarrow {vol (S^5)\over N}$, as $S^5$ is a Hopf fibration of $\mathbb {CP}^2$. In order to lift to a background with no 3-form RR and NS fluxes, one will need a 5-sphere base space whose volume is divided by a factor of $N$. To do so, one can divide a 5-sphere by a group of order $N$ such as $\mathbb Z_N$ along its $U(1)$ subspace. Since $U(1) / \mathbb Z_N \cong U(1)$, one will still have a 5-sphere, albeit with a volume that is divided by $N$, as required. Let us denote this new 5-sphere by $S^{5'}$. Since $\mathbb Z_N \subset U(1) \subset U(3)$, where $U(3)$ is the isometry group of the $Spin(7)$-manifold isomorphic to $\mathbb R^3 \times S^5$, it means that $\mathbb R^3 \times S^{5'}$ is again a $Spin(7)$-manifold. Hence, from the earlier discussions, the new, non-compact and smooth 8-manifold that should preserve $1/32$ of the maximal supersymmetry and also result in the emergence of multiple vacua in the $d=1+1$ theory, will be given by $\widehat {\mathcal X} \cong {[\mathbb R^3 \times {S^5}'] / \langle \sigma \rangle} \equiv {\mathbb R^3 \times ({S^5}' / \langle \sigma \rangle)}$.

\bigskip\noindent{\it A Closer Look at $\widehat {\mathcal Q}$ and $\widehat {\mathcal X}$}
\vspace{0.2cm}

Now recall that $\mathbb {CP}^2$ is given by the fibration $\widetilde M_2 \hookrightarrow \mathbb {CP}^2 \to \mathbb {CP}^1$ and that $S^{5'}$ can be viewed as the fibration $\widetilde M_2 \hookrightarrow S^{5'} \to S^{3'}$. Since the action of $\sigma$ on a fibre bundle descends onto its base space, we will have the fibrations $\widetilde M_2 \hookrightarrow {\mathbb {CP}^2 / {\langle \sigma \rangle}} \to {\mathbb {CP}^1/ {\langle \sigma \rangle}}$ and $\widetilde M_2 \hookrightarrow {S^{5'} / {\langle \sigma \rangle}} \to {S^{3'} / {\langle \sigma \rangle}}$. Consequently, $\widehat {\mathcal Q} \cong (\mathbb R^4 / \mathbb Z_N) \times (\mathbb {CP}^2 / \langle \sigma \rangle)$ and $\widehat {\mathcal X} \cong \mathbb R^3 \times (S^{5'} / \langle \sigma \rangle)$ are given by the fibrations $\widetilde M_2 \hookrightarrow {\widehat {\mathcal Q}} \to {\widehat {\widetilde Q}}$ and $\widetilde M_2 \hookrightarrow {\widehat {\mathcal X}} \to {\widehat {\widetilde X}}$ respectively, whereby ${\widehat {\widetilde Q}}\cong (\mathbb R^4 / \mathbb Z_N) \times (\mathbb{CP}^1/{\langle \sigma \rangle})$ and $\widehat {\widetilde X} \cong \mathbb R^3 \times (S^{3'} / \langle \sigma \rangle)$. In other words, the `lifted' geometric transition $\widehat {\mathcal Q} \rightarrow \widehat {\mathcal X}$ is also solely effected by the geometric transition of its 6-dimensional base space $\widehat {\widetilde Q} \rightarrow \widehat {\widetilde X}$. Just as in the case of the original geometric transition $\mathcal Q \rightarrow \mathcal X$ with D5-branes and 3-form fluxes, the $\widetilde M_2$ fibre is again a `spectator' subspace. Notice also that $\widehat {\widetilde Q} = {\widetilde Q'/ {\langle \sigma \rangle}}$ and $\widehat {\widetilde X} = {\widetilde X' / {\langle \sigma \rangle}}$ whereby $\widetilde Q' \cong (\mathbb R^4 / \mathbb Z_N) \times \mathbb {CP}^1$ and $\widetilde X' \cong \mathbb R^3 \times S^{3'}$, such that $\widetilde Q'$ and $\widetilde X'$ are isomorphic to CY 3-folds that preserve $1/4$ of the maximal supersymmetry.\footnote{Note here that the fixed $Z_N \subset U(1)$ action along the $U(1)$ fibre of the $\mathbb R^4$ bundle of the CY 3-fold $\widetilde Q \cong \mathbb R^4 \times \mathbb {CP}^1\cong \mathcal O(-1) \oplus \mathcal O(-1) \rightarrow \mathbb {CP}^1$ is an isometry of the $\mathbb R^4$ space. Therefore, it leaves the (constant tensor) holomorphic 3-form $\Omega$ and hence holonomy invariant. Thus, $\widetilde Q' \cong (\mathbb R^4 / \mathbb Z_N) \times \mathbb {CP}^1$ continues to preserve $1/4$ of the maximal supersymmetry since it is still isomorphic to a CY. The $S^{3'}$ in $\widetilde X' \cong \mathbb R^3 \times S^{3'}$ is a 3-sphere divided by $\mathbb Z_N$ along its $U(1)$ fibre. Since $\mathbb Z_N \subset U(1)$ is an isometry subgroup of the trivial bundle $\mathbb R^3 \times S^3 \cong T^*S^3$, $\widetilde X'$ is also CY, thus preserving $1/4$ of the maximal supersymmetry.} Thus, just like $\gamma$ on the CY 4-orbifold $Y$, $\sigma$ is a freely-acting $\it{antiholomorphic}$ isometric involution on $\widetilde Q'$ and $\widetilde X'$. Hence, from a lower, 6-dimensional extension of the construction of the 8-manifold $\mathcal Z$ (which preserves $1/16$ of the maximal supersymmetry) from the CY 4-orbifold $Y$ (which preserves $1/8$ of the maximal supersymmetry) via the action of $\gamma$ discussed earlier, we can expect $\widehat {\widetilde Q}$ and $\widehat {\widetilde X}$ to preserve $1/8$ of the maximal supersymmetry from $d=9+1$ in $d=3+1$. Let us look at $\widehat {\widetilde Q}$ and $\widehat {\widetilde X}$ in greater detail so as to verify this statement and consequently, the fact that $\widehat {\mathcal Q}$ and $\widehat {\mathcal X}$ will preserve $1/32$ of the maximal supersymmetry as required.

The lifted background on $\widehat {\widetilde Q}$ or $\widehat{\widetilde X}$ is without 3-form RR and NS fluxes. Thus, from the discussion surrounding (\ref{flux}), the $(1,0)$-form $W_5$ representing the torsion class of the $SU(3)$ structure of $\widetilde X'$, in the construction of $\widehat {\widetilde X} = {\widetilde X' / {\langle \sigma \rangle}}$, is zero. This means that one has a torsion-free $SU(3)$ structure, i.e. $\nabla \Omega = 0$ and $d\Omega = 0$, where $\Omega$ is the holomorphic volume $(3,0)$-form on $\widetilde X'$ and $\nabla$ is the Levi-Civita connection of the metric on $\widetilde X'$. The $SU(3)$ structure of $\widetilde Q'$ is always torsion-free since there are no calibrated 3-cycles in $\widetilde Q'$ with varying volume during the geometric transition $\widehat {\widetilde Q} \rightarrow \widehat {\widetilde X}$. Let the metric, K\"ahler $(1,1)$-form and holomorphic volume $(3,0)$-form on $\widetilde Q'$ and $\widetilde X'$ with torsion-free $SU(3)$ structure be given by $\widetilde {\it{g}}$, $\omega$ and $\Omega$ respectively. One can then define a free antiholomorphic isometric involution $\sigma$ which acts on $\widetilde Q'$ and $\widetilde X'$, satisfying $\sigma: Y \rightarrow Y$ for any manifold $Y$ and ${\sigma}^2 =$ id, such that \cite{joyce}
\be
\sigma^{*}(\widetilde{\it{g}}) = {\widetilde {\it{g}}}, \qquad \sigma^{*}(\omega) = - {\omega}, \qquad \sigma^{*} (\Omega) = \bar {\Omega}.
\label{sigmaaction}
\ee    
Since ${\omega_{ab}} = {J^{b}_a {\widetilde {\it{g}}}_{bc}}$, from the action of $\sigma$ on $\omega$ in (\ref{sigmaaction}), we thus have $\sigma^{*}(J)=-J$, where $J$ is the complex structure on $\widetilde Q'$ or $\widetilde X'$. 

To see that there is indeed a well-defined $\sigma$ action on $\widetilde Q'$ which satisfies the above conditions, first recall that $\widetilde Q \cong \mathbb R^4 \times \mathbb {CP}^1 \cong  \mathcal O(-1) \oplus \mathcal O(-1) \rightarrow \mathbb {CP}^1$. Note that we can describe $\widetilde Q$ as a subspace in $\mathbb C^4$ as follows \cite{mirrortext}: let $(z_1, z_2, z_3, z_4) \in \mathbb C^4$. Then, $\widetilde Q$ will be given by the hypersurface 
\be
|z_1|^2 + |z_2|^2 - |z_3|^2 -|z_4|^2 = {\zeta},
\label{Q}
\ee 
subject to the $C^*$ identification
\be
(z_1, z_2, z_3, z_4) \sim (e^{i\theta} z_1, e^{i\theta} z_2, e^{-i\theta} z_3, e^{-i\theta}z_4),
\label{equivalence}
\ee
where $\theta \in \mathbb R$. $(z_1, z_2)$ and $(z_3, z_4)$ parameterize the $\mathbb {CP}^1$ base (with radius$^2$ $=\zeta$) and $\mathbb R^4$ fibre respectively. Consequently, $\widetilde Q' \cong \mathbb R^4/ \mathbb Z_N \times \mathbb {CP}^1$, in addition to being characterized by $(\ref{Q})$ and (\ref{equivalence}), must also be subjected to the following identification in the $\mathbb R^4$ fibre due to the $Z_N$ action:    
\be
(z_1, z_2, z_3, z_4) \sim (z_1, z_2, e^{ {2\pi i n} \over N} z_3, e^{ { 2\pi i n} \over N}z_4), \qquad 0 \leq n \leq N-1,
\label{znaction}
\ee
whereby $(z_3, z_4) = (0,0)$, the zero section of the fibre bundle $\widetilde Q'$, is a locus of fixed points under (\ref{znaction}) with an $A_{N-1}$ singularity.\footnote{From the perspective of the initial 6-dimensional IIB compactification, there is an $\mathbb R^{3,1} \times \mathbb {CP}^1$ locus of an $A_{N-1}$ singularity, resulting in an $SU(N)$ gauge symmetry in $d=3+1$, consistent with the expected $\mathcal N= 1$ pure $SU(N)$ theory. Hence, we see that $\widetilde Q'$ is the appropriate manifold to use in defining $\widehat {\widetilde Q} = \widetilde Q' / \langle \sigma \rangle$.} Define an action $\sigma$ which doesn't coincide with the existing identifications (\ref{equivalence}) and (\ref{znaction}), such that 
\be
\sigma:(z_1, z_2, z_3, z_4) \mapsto (z_3, z_4, z_1, z_2).
\label{sigmamapQ'}
\ee
The action of $\sigma$ on $\widetilde Q'$ will then result in a flop to a $\mathbb {CP}^1$ with the same albeit `negative' volume. Since the K\"ahler form $\omega$ evaluates the volume of the $\mathbb {CP}^1$ base via $\int_{\mathbb {CP}^1} \omega$, we find that $\sigma^*(\omega)= - \omega$, from which we can obtain $\sigma^*(J) =-J$ as discussed above. The flop due to the action of $\sigma$ just exchanges the base and the fibre spaces, which means that we still have the same manifold $\widetilde Q'$, i.e. $\sigma^{*}(\widetilde{\it{g}}) = {\widetilde {\it{g}}}$ and $\sigma: \widetilde Q' \rightarrow \widetilde Q'$. $\sigma^*(\Omega)$ is then proportional to $\bar \Omega$, and by multiplying $\Omega$ by $e^{i \psi} \in C^*$, one can arrange for $\sigma^* (\Omega) = {\bar \Omega}$ \cite{joyce}.     

Likewise, one can also find a well-defined $\sigma$ action on $\widetilde X'$ which will satisfy the required conditions above. To see this, first recall that $\widetilde X' \cong \mathbb R^3 \times S^{3'}$. We can therefore describe $\widetilde X'$ as a subspace in $\mathbb C^4$ as follows: let $z_j = x_j + i p_j$, where $x_j, p_j$ are real and $(z_1, z_2, z_3, z_4) \in \mathbb C^4$. Then, $\widetilde X'$ is given by the hypersurface
\be
\sum_{j=1}^4 {z_j}^2 = \mu, 
\ee 
or equivalently,   
\be
\sum_{j=1}^4 ({x_j}^2 - {p_j}^2) = \mu, \qquad \sum_{j=1}^4 x_j p_j = 0,
\label{X}
\ee  
subject to the additional equivalence relation 
\be
(x_1, x_2, x_3, x_4) \sim ( e^{{2 \pi i n} \over N} x_1,  e^{{2 \pi i n} \over N} x_2,  e^{{2 \pi i n} \over N} x_3,  e^{{2 \pi i n} \over N} x_4), \qquad 0 \leq n \leq N-1, 
\label{s3'}
\ee
due to the identification of the 3-sphere under a $\it{free}$ $\mathbb Z_N$ action along its $U(1)$ subspace.\footnote{Note that the identification in (\ref{s3'}) implies that $vol (S^{3'}) = {vol (S^3) / N}$ and since a 3-sphere is a Hopf fibration of $\mathbb {CP}^1$, this therefore implies that $vol (\mathbb {CP}^{1'}_{bare}) = {vol (\mathbb {CP}^1_{bare}) / N}$. Recall from the relevant discussions in $\S$3.1 and $\S$3.3 that this will result in the $N$ inequivalent vacua characteristic of $\mathcal N=1$ pure $SU(N)$ theory in $d=3+1$ at low energy. Thus, from the perspective of the $d=3+1$ theory from the initial 6-dimensional compactification of the IIB theory, one can see that if one wants to lift to a background with no 3-form RR and NS fluxes, $\widetilde X'$ is indeed the appropriate manifold to use in $\widehat{\widetilde X} = \widetilde X' / \langle \sigma \rangle$.} From (\ref{X}) and (\ref{s3'}), we find that the $x_j$s parameterize the $S^{3'}$ base space while the $p_j$s parameterize the $\mathbb R^3$ fibre space $\it{normal}$ to it. $\sqrt {\mu}$ is the radius of the $S^{3'}$ base, which sits at the zero section of the fibre bundle $\widetilde X'$ given by $p_j =0$, $\forall j$. The K\"ahler form $\omega$ is given by 
\be  
\omega = \sum_{j=1}^4 dp_j \wedge dx_j.
\label{omegaX} 
\ee
Define the action of $\sigma$ such that 
\be
\sigma:(z_1, z_2, z_3, z_4) \mapsto - (\bar z_1, \bar z_2, \bar z_3, \bar z_4).
\label {sigmamapX}
\ee 
Notice that since $z_j = x_j + i p_j$, we effectively have $\sigma :(x_1, x_2, x_3, x_4) \mapsto (-x_1, -x_2, -x_3, -x_4)$. Thus, we see that (\ref{X}) defining $\widetilde X'$ is invariant under the action of $\sigma$, i.e. $\sigma^{*}(\widetilde{\it{g}}) = {\widetilde {\it{g}}}$ and $\sigma: \widetilde X' \rightarrow \widetilde X'$. In addition, from (\ref{omegaX}), we find that $\sigma^*(\omega) = -\omega$, which gives $\sigma^*(J) = -J$ as usual. $\sigma^* (\Omega) = \bar {\Omega}$ follows from a similar argument made in the discussion on $\widetilde Q'$.   

Note that $d \Omega =0$ implies that $\nabla \Omega = 0$ \cite{joyce}, i.e. $\Omega$ is a constant tensor. As can be seen from (\ref{sigmaaction}), only $Re (\Omega)$ is invariant under the action of $\sigma$. This means that only $Re(\Omega)$ survives on $\widehat {\widetilde Q} = \widetilde Q'/ {\langle \sigma \rangle}$ and $\widehat {\widetilde X} = \widetilde X'/ {\langle \sigma \rangle}$; since points on the manifold $\widetilde Q'$ or $\widetilde X'$ which are connected by the action of $\sigma$ are identified, it means that a well-defined tensor will be the same over these points. Thus, due to the action of $\sigma$, the constant tensor on $\widehat {\widetilde Q}$ and $\widehat {\widetilde X}$ is now given by $Re(\Omega)$. Due to the 1-1 correspondence between constant tensors and the holonomy of a manifold, which therefore determines the fraction of the maximal supersymmetry preserved or the number of constant spinors it possesses, we find that since only $Re(\Omega)$ from $\Omega = Re(\Omega) + i Im(\Omega)$ survives, the number of constant spinors on $\widehat{\widetilde Q}$ and $\widehat {\widetilde X}$ will be half that on $\widetilde Q'$ and $\widetilde X'$. Alternatively, note that $\Omega$ on $\widetilde Q'$ and $\widetilde X'$ can be expressed in terms of the constant spinor $\eta$ and the affine connection $\Gamma$ as \cite{gsw}
\be
\Omega_{ijk} = \eta^T \Gamma_{ijk}\eta.
\label{omegaijk}
\ee 
Recall that $\sigma^*(\widetilde {\it{g}}) = {\widetilde {\it{g}}}$. Since $\Gamma$ is constructed from $\widetilde {\it{g}}$, it means that $\Gamma$ is invariant under the action of $\sigma$. Since $\sigma^*(\Omega) = \bar{\Omega}$, (\ref{omegaijk}) will then imply that $\Gamma$ is real while $\eta$ must be a complex spinor, such that 
\be
\bar \Omega_{ijk} = {\bar {\eta}}^T \Gamma_{ijk} {\bar {\eta}}.
\label{baromegaijk}
\ee
This means that the surviving constant tensor $Re(\Omega)$ on $\widehat {\widetilde Q}$ and $\widehat {\widetilde X}$ can be written as
\be
Re(\Omega_{ijk}) = \hat {\eta}^T \Gamma_{ijk}\hat{\eta},
\label{reomegaijk}
\ee
whereby $\hat{\eta}$ is the surviving constant spinor on $\widehat {\widetilde Q}$ and $\widehat {\widetilde X}$, such that its degrees of freedom will be half that of $\eta$ on ${\widetilde Q'}$ and ${\widetilde X'}$, i.e., the number of conserved supercharges on $\widehat {\widetilde Q}$ and $\widehat {\widetilde X}$ is half that on $\widetilde Q'$ and $\widetilde X'$. In other words, since $\widetilde Q'$ and $\widetilde X'$ will preserve $1/4$ of the maximal supersymmetry from $d=9+1$ in $d=3+1$, it means that $\widehat{\widetilde Q}$ and $\widehat{\widetilde X}$ will then preserve $1/8$ of the maximal supersymmetry from $d=9+1$ in $d=3+1$, as required. 

Next, notice that $\sigma^2=$id implies that $\langle \sigma \rangle \cong \mathbb Z_2$. As such, one will have the fundamental groups $\pi_{1}(\widehat {\widetilde Q})=\pi_{1}(\widehat {\widetilde X}) = \mathbb Z_2$, which in turn translates to the fact that $\widehat {\widetilde Q}$ and $\widehat {\widetilde X}$ are non-simply-connected. Hence, the holonomy of $\widehat{\widetilde Q}$ and $\widehat {\widetilde X}$ will be given by $SU(3)\odot \mathbb Z_2$ \cite{joyce}. Thus, the construction of $\widehat{\widetilde Q}$ and $\widehat{\widetilde X}$ are indeed lower dimensional extensions of the construction of $\mathcal Z$. A further compactification on the 2-dimensional `spectator' $\widetilde M_2$ fibre after an initial compatification on $\widehat {\widetilde Q}$ or $\widehat {\widetilde X}$ should then preserve another $1/4$ of the remaining supersymmetry from $d=3+1$ in $d=1+1$, such that the compactification on $\widehat{\mathcal Q}$ or $\widehat {\mathcal X}$ will effectively preserve $1/32$ of the maximal supersymmetry from $d=9+1$ in $d=1+1$ as desired.   

\bigskip\noindent{\it A Note  Concerning $\widehat {\mathcal Q}$ and $\widehat {\mathcal X}$}
\vspace{0.2cm}

Before we proceed any further, we would like to bring to the reader's attention a subtle but important point concerning the above non-simply-connected 8-manifolds $\widehat {\mathcal Q}$ and $\widehat {\mathcal X}$; if and only if $\widehat {\mathcal Q}$ and $\widehat {\mathcal X}$ undergo the  geometric transition $\widehat {\mathcal Q} \rightarrow \widehat {\mathcal X}$ considered in this paper will the above derivation in which  the 8-manifolds are shown to preserve $1/32$ of the maximal supersymmetry from $d=9+1$ in $d=1+1$ be mathematically consistent. This can be explained as follows: notice that one has $\widehat {\mathcal Q} \cong \mathcal Q' / {\langle \sigma \rangle}$ and $\widehat{\mathcal X} \cong \mathcal X'/{\langle \sigma \rangle}$, whereby $\mathcal Q' \cong (\mathbb R^4 /\mathbb Z_N) \times \mathbb {CP}^2$ and $\mathcal X' \cong \mathbb R^3 \times S^{5'}$ are $Spin(7)$-manifolds. In a situation when the $\mathbb {CP}^2$ or $\mathbb {CP}^{2'}$ 4-cycle of $\mathcal Q'$ or $\mathcal X'$ has non-varying volume, whereby one is considering a IIB compactifcation on the fixed, non-transitioning 8-manifold $\mathcal Q'$ or $\mathcal X'$ with $Spin(7)$ holonomy, the Cayley 4-form $\Psi$ is closed, i.e. $d\Psi =0$, and $(\Psi,\it{g})$, where $\it{g}$ is the manifold metric, defines a torsion-free $Spin(7)$ structure, which implies $\nabla \Psi =0$, so that $\Psi$ is a constant tensor.\footnote{See Proposition 10.5.3 of \cite{joyce}.} It is then known that $\Psi$ is $\sigma$-invariant such that $\sigma^*(\Psi) = \Psi$ \cite{joyce}. This means that on a new manifold, constructed as the quotient of the fixed-volume $\mathcal Q'$ or $\mathcal X'$ by a freely-acting isometric involution such as $\sigma$, one does not obtain a different constant tensor from $\Psi$. From the 1-1 correspondence between the constant tensors, holonomy and hence covariantly constant spinors on a manifold \cite{joyce}, this in turn implies that one cannot arrive at a new manifold which preserves $1/2$ of the maximal supersymmetry that $\mathcal Q'$ or $\mathcal X'$ does. In fact, the new manifold will preserve the same amount of supersymmetry as $\mathcal Q'$ or $\mathcal X'$ since the constant tensor $\Psi$ remains invariant. However, in a geometric transition, since the volume of $\mathbb{CP}^2 \subset \mathcal Q'$ and $\mathbb{CP}^{2'} \subset S^{5'} \subset \mathcal X'$ is varying, one will have $\delta \int_{\mathbb {CP}^2} \Psi = \int_{\delta \mathbb {CP}^2} \Psi \neq 0$ and $\delta \int_{\mathbb {CP}^{2'}} \Psi= \int_{\delta \mathbb {CP}^{2'} } \Psi \neq 0$. Equivalently, we can view $\mathcal Q'$ and $\mathcal X'$ to be non-varying, i.e. $\delta \mathbb {CP}^2 = \delta \mathbb {CP}^{2'} =0$, such that $\delta \int_{\mathbb {CP}^2} \Psi = \int_{\mathbb {CP}^2} \delta \Psi \neq 0$ and $\delta \int_{\mathbb {CP}^{2'}} \Psi= \int_{\mathbb {CP}^{2'} } \delta \Psi \neq 0$. This means that $\delta \Psi \neq 0$, which in turn implies that $d\Psi \neq 0$, i.e., the expression for $\Psi$ is modified. In other words, since $d\Psi \neq 0$ implies $\nabla \Psi \neq 0$, we effectively have a $Spin(7)$-structure with torsion on $\mathcal Q'$ and $\mathcal X'$. Let us denote this $Spin(7)$-structure with torsion by $(\Psi^t, \it{g}^t)$. Then, one can show, via proposition 10.5.10 of \cite{joyce}, that by expressing $d\Psi^t$ in terms of a suitable 4-form $\phi \in C^{\infty}(\Lambda^4 T^*M)$, whereby $M= \mathcal Q'$ or $\mathcal X'$, such that $d\Psi^t = - d\phi$ and $(\Psi^t + \phi) \neq 0$, one can alternatively view the varying $Spin(7)$-manifold $\mathcal Q'$ or $\mathcal X'$ as a fixed $Spin(7)$-manifold with a torsion-free $Spin(7)$-structure $(\Psi', \it {g}')$, in which $d\Psi'= 0$, $\nabla' \Psi' =0$ and $\Psi'=\Psi^t + \phi$, thus implying that one can now have $\sigma^*(\Psi') \neq \Psi'$. To demonstrate this, first note that from proposition 10.5.10 of \cite{joyce}, one has the following statement: there exists positive non-zero constants $\epsilon_1$, $\epsilon_4$ and $\epsilon_5$ such that the following holds. Let $(\Psi^t, g^t)$ be a $Spin(7)$-structure on an 8-manifold M, and suppose $\phi \in C^{\infty}(\Lambda^4 T^*M)$ satisfies $|\phi| \leq \epsilon_1$ and $d\Psi^t + d\phi =0$. Let $\mathcal A M$ be a subbundle of $\Lambda^4 T^*M$ with fibre $GL_+(8, \mathbb R) / Spin(7)$ such that a 4-form $\Psi$ on $M$ is $\it{admissable}$ if $\Psi \in \mathcal {A}M$. Let the neighbourhood of $\mathcal A M$ in $\Lambda^4 T^*M$ be given by $\mathcal T M$. Define the smooth and surjective map of bundles $\Theta: {\mathcal T M} \rightarrow {\mathcal A M}$. Since $\Psi^t$ is an admissable 4-form, i.e. $\Psi^t \in C^{\infty}(\mathcal A M)$, one will have $({\Psi^t + \phi}) \in C^{\infty} (\mathcal {T} M)$, such that $\Psi' = \Theta (\Psi^t + \phi)$ lies in $C^{\infty}(\mathcal {A} M)$. Define $(\Psi', \it{g}')$ to be the induced $Spin(7)$-structure, and set $\phi ' = \Psi^t + \phi - \Psi'$. Then $d\Psi' + d\phi' =0$, and 
\be
|\phi'|_{\it{g}'} \leq \epsilon_4 (|\pi_{27}(\phi)|_{\it{g}^t} + |\phi|_{\it{g}^t}^2) \qquad and \qquad |\nabla' \phi'|_{\it{g}'} \leq \epsilon_5 |\nabla \phi|_{\it{g}}, 
\label{prop}
\ee
whereby $\pi_{27}(\phi)$ is just an orthogonal projection from the 4-form $\phi$ to an irreducible representation of $Spin(7)$ of dimension 27. It is clear from (\ref{prop}) that one is free to choose $|\phi'| = |\nabla' \phi'| =0$. In doing so, we find that $\phi' = \Psi^t + \phi - \Psi' = 0$ or $\Psi' = \Psi^t + \phi$, and $d \Psi' = 0$. Since $\Psi' \in C^{\infty}(\mathcal A M)$, via proposition 10.5.3 of \cite{joyce}, $d \Psi'$ will then imply $\nabla' \Psi' =0$ and a torsion-free $Spin(7)$-structure $(\Psi', \it{g}')$ as mentioned. Only then can one expect to find the appropriate combination of complex 4-forms $\Psi^t$ and $\phi$ such that $\sigma^*(\Psi') =\bar \Psi'$, i.e. only $Re(\Psi')$ is $\sigma$-invariant, hence implying that only half of the original degrees of freedom of ${\Psi}'$ survive on the new manifold. Since $\nabla' \Psi' = 0$, it is a constant tensor. Thus, from the 1-1 correspondence between constant tensors and covariantly constant spinors on a manifold, one finds that by quotienting the $\it{varying}$ $Spin(7)$-manifold $\mathcal Q'$ or $\mathcal X'$ by a suitable freely-acting isometric involution $\sigma$, one can arrive at a new, non-simply-connected 8-manifold $\widehat {\mathcal Q} \cong \mathcal Q' / \langle \sigma \rangle$ or $\widehat{\mathcal X} \cong {\mathcal X' / \langle \sigma \rangle}$ which will, as desired, preserve $1/2$ of the maximal supersymmetry that $\mathcal Q'$ or $\mathcal X'$ would. The careful reader might have noticed earlier that the overall $\widetilde M_2$ fibre spaces of $\widehat {\mathcal Q}$ and $\widehat{\mathcal X}$ must differ from the overall $\widetilde M_2$ fibre spaces of $\mathcal Q'$ and $\mathcal X'$ due to the action of $\sigma$ on the base of the former. However, from the above analysis relevant to the case at hand, we find that a further compactification on the overall $\widetilde M_2$ fibre spaces of $\widehat {\mathcal Q}$  and $\widehat{\mathcal X}$ will continue to preserve $1/4$ of the remaining supersymmetry from $d=3+1$ in $d=1+1$ such that $\widehat{\mathcal Q}$ and $\widehat{\mathcal X}$ will preserve $1/32$ of the maximal supersymmetry from $d=9+1$ in $d=1+1$ as required.       

\bigskip\noindent{\it Support Of Constant Spinors on $\widehat {\mathcal Q}$ and $\widehat {\mathcal X}$}
\vspace{0.2cm}

Unlike for the case of a simply-connected, orientable manifold, the non simple connectedness and thus non-trivial topology of $\widehat {\widetilde Q} \subset \widehat {\mathcal Q}$ and $\widehat {\widetilde X} \subset \widehat {\mathcal X}$ renders their support of spinor fields and hence, covariantly constant spinors neccessary in a physically consistent supersymmetric compactification, questionable. Fortunately, one can show that both $\widehat {\widetilde Q}$ and $\widehat {\widetilde X}$, and consequently $\widehat {\mathcal Q}$ and $\widehat {\mathcal X}$, neccessarily support spinor fields, thus verifying their physical validity as the geometric lifts of $\mathcal Q$ and $\mathcal X$ to a background without D5-branes and 3-form RR and NS fluxes respectively. Let us look at this claim in greater detail.       

Note that one has the relation $S^n / \mathbb Z_2 = \mathbb{RP}^n$ \cite{hou}. Since $\langle \sigma \rangle \cong \mathbb Z_2$, we will have $\widehat {\widetilde Q} \cong (\mathbb R^4/ \mathbb Z_N) \times (\mathbb {CP}^1/ \mathbb Z_2) \equiv (\mathbb R^4/ \mathbb Z_N) \times \mathbb {RP}^2$. Similarly, one will have $\widehat {\widetilde X} \cong \mathbb R^3 \times (S^{3'} / \mathbb Z_2) \equiv \mathbb R^3 \times \mathbb {RP}^{3'}$. Note also that one has the result $W_3(M) = 0$, if $M \equiv \mathbb{RP}^i$ for $i \leq 4$, or $S^5$ \cite{baryons}, where one recalls from $\S$3.1 that $W_3(M) \in H^3(M, \mathbb Z)$ such that $W_3(M) = \beta(w_2(M))$, in which $w_2(M) \in H^2 (M, \mathbb Z_2)$ is the second Stiefel-Whitney class of $M$ and $\beta$ is given by the ``Bockstein" map $\beta : H^2(M; {\mathbb Z}_2) \rightarrow H^3(M;\mathbb Z)$. A manifold $M$ is $spin^c$ if and only if $W_3(M) =0$ \cite{michel}. Hence, $\mathbb {RP}^2 \subset \widehat {\widetilde Q}$ and $\mathbb {RP}^{3'} \subset \widehat {\widetilde X}$ are $spin^c$ submanifolds. Last but not least, note that every oriented manifold of dimension $\leq$ 4 is $spin^c$ \cite{freed-witten}. It is clear that the $\mathbb R^3$ fibre of $\widehat {\widetilde X}$ is an orientable manifold of dimension $\leq$ 4. Thus, it is $spin^c$. As for the $\mathbb R^4 / \mathbb Z_N$ fibre of $\widehat{\widetilde Q}$, since the $\mathbb Z_N \subset U(1)$ action along the $U(1)$ fibre of $\mathbb R^4$ is orientation preserving, $\mathbb R^4 / \mathbb Z_N$, like $\mathbb R^4$, is an orientable (albeit singular) manifold. Because it is of dimension $\leq$ 4, it is also $spin^c$. Therefore, both $\widehat{\widetilde Q}$ and $\widehat {\widetilde X}$ are $spin^c$ manifolds. 

Next recall that we have the fibration $\widetilde M_2 \hookrightarrow \mathbb {CP}^2 \to \mathbb {CP}^1$. Note that since $w_1(S^n) =0$ for any $n$, whereby $w_1(M)$ is the first Stiefel-Whitney class of $M$, $\mathbb {CP}^1= S^2$ is orientable. Being of dimensions $\leq$ 4, it is thus $spin^c$. Recall from $\S$3.1 that $\mathbb {CP}^2$ is also $spin^c$. This implies that the $\widetilde M_2$ fibre must be $spin^c$. This property of the $\widetilde M_2$ fibre can also be derived from an alternative viewpoint as follows: recall that we can view $S^5$ as the fibration $\widetilde M_2 \hookrightarrow S^5 \to S^3$. Since $w_1(S^3) = 0$, $S^3$ is oriented. Since its dimension $\leq$ 4, it is thus $spin^c$. Because $W_3(S^5) =0$ as mentioned, $S^5$ is also $spin^c$. Hence, $\widetilde M_2$ is indeed $spin^c$. Therefore, since we have the fibrations $\widetilde M_2 \hookrightarrow {\widehat {\mathcal Q}} \to {\widehat{\widetilde Q}}$ and $\widetilde M_2 \hookrightarrow {\widehat {\mathcal X}} \to {\widehat{\widetilde X}}$, it will mean that both $\widehat {\mathcal Q}$ and $\widehat {\mathcal X}$ are also $spin^c$. Note that $spin^c$ manifolds support $spin^c$ bundles over them. Thus, as required, this in turn means that $\widehat{\widetilde Q}$ and $\widehat {\widetilde X}$, as well as $\widehat {\mathcal Q}$ and $\widehat {\mathcal X}$, do support spinor fields, which are, in these cases, sections of the corresponding $spin^c$ bundles with fibres being the spinor representations of $spin^c(6)$ and $spin^c(8)$ respectively. Note also that since $\mathbb {CP}^2 / \mathbb Z_2 = S^4$ \cite{aw}, $\widehat {\mathcal Q}\cong (\mathbb R^4/ \mathbb Z_N) \times S^4$ and  $\widehat {\mathcal X}\cong \mathbb R^3 \times \mathbb {RP}^{5'}$.      

\bigskip\noindent{\it The Metrics on $\widehat {\mathcal Q}$ and $\widehat {\mathcal X}$}
\vspace{0.2cm}
                           
Since quotienting an old manifold by a freely-acting isometric involution to construct a new manifold simply involves the identifying of points on the old manifold connected by the action of the involution, the metric on the new manifold is the same as that on the old manifold. Recall that since the $\mathbb Z_N \subset U(1)$ action in the construction of $\mathcal Q' \cong (\mathbb R^4/ \mathbb Z_N) \times \mathbb {CP}^2$ and $\mathcal X' \cong \mathbb R^3 \times S^{5'}$ is a subgroup of the $U(3)$ isometry group of $\mathcal Q \cong \mathbb R^4 \times \mathbb {CP}^2$ and $\mathcal X \cong \mathbb R^3 \times S^5$, the holonomy on $\mathcal Q'$ ($\mathcal X'$) and $\mathcal Q$ ($\mathcal X$) are the same. Hence, the metric (but not holonomy) on $\widehat {\mathcal Q}\cong \mathcal Q' / \langle \sigma \rangle$ and $\widehat {\mathcal X}\cong \mathcal X' /\langle \sigma \rangle$ will be of type $Spin(7)$, and it is given by the metric on $\mathcal Q$ and $\mathcal X$, modded out by a $\mathbb Z_N$ group along the relevant $\mathbb R^4$ bundle and $S^5$ directions respectively. Since the Riemann curvature and hence Ricci tensor is derived purely from the metric, this will imply that $\widehat {\mathcal Q}$ and $\widehat {\mathcal X}$, like $\mathcal Q$ and $\mathcal X$, are both Ricci-flat and are thus physically valid as compactification manifolds since the IIB worldsheet theory remains conformally invariant up to lowest order corrections. We refer the reader to the appendix of \cite{gst} for explicit details of these Ricci-flat $Spin(7)$ metrics on $\mathcal Q$ and $\mathcal X$.

\subsection{Gauge-theoretic interpretation of the IIB duality in the lifted background}    

Let the bare gauge coupling of the $d=3+1$ theory (from the initial IIB compactification on $\widehat {\widetilde Q}$ with no D5-branes) at the Planck scale be given by $\widehat{g}_o$, and the corresponding effective dimensionless coupling at the observed energy scale $u$ be given by $\widehat {g}_{4d}(u)$. Due to the singular $\mathbb R^4 / \mathbb Z_N$ bundle of $\widehat {\widetilde Q}$, there will be a locus of an $A_{N-1}$ singularity along $\mathbb {RP}^2 \times \mathbb R^{3,1}$. In other words, we have a (5+1)-dimensional $SU(N)$ gauge theory on $\mathbb {RP}^2 \times \mathbb R^{3,1}$. By a usual redefinition of the bare $SU(N)$ gauge coupling constant in $d=5+1$ to be of order 1, we have, via a dimensional reduction (along $\mathbb {RP}^2 $) from $d=5+1$ to $d=3+1$, the following relation:       
\be
{1\over {{\widehat{g}_o}^2}} = vol (\mathbb {RP}^2_{bare}),
\label{gocap}
\ee
whereby $vol(\mathbb {RP}^2_{bare})$ is the volume of the smooth 2-cycle $\mathbb {RP}^2_{bare}$ $\it{before}$ it starts to blow-down in the `lifted' geometric transition $\widehat {\widetilde Q} \rightarrow \widehat {\widetilde X}$ at low energy. Note also that from the arguments in the previous subsection, an initial IIB compactification on the $\it{non}$-$\it{compact}$ 6-manifold $\widehat {\widetilde Q} \cong \mathbb (R^4 /{\mathbb Z_N}) \times \mathbb {RP}^2$, in a background with no space-filling D5-branes, will result in an $\mathcal N=1$ supersymmetric pure $SU(N)$ theory in $d=3+1$. Hence, as in (\ref{g4d}), due to the non-abelian $SU(N)$ gauge symmetry, which thus ensures that one will have asymptotic freedom in $d=3+1$ \cite{cheng}, we will have the following relation:
\be
{1\over {\widehat{g}^2_{4d}(u)}} \ = \ {1\over {\widehat {g}^2_o}}\ + \ \alpha \ log\ ({u \over {|\Lambda_{planck}}|}),
\label{g4dcap}
\ee   
where $\alpha$ is a positive constant. Note that the $log$ term on the RHS of (\ref{g4dcap}) is due to a 1-loop quantum correction, while the $1/ {\widehat{g}^2_o}$ term is purely classical. Hence, in view of the relation (\ref{gocap}), we see that the running of the effective dimensionless gauge coupling $1/ \widehat {g}^2_{4d}(u)$ as reflected in (\ref{g4dcap}), can be interpreted as being induced by quantum effects in the presence of the $\mathbb Z_N$ singularity in measuring the volume of the 2-cycle $\mathbb {RP}^2 \subset \widehat {\widetilde Q}$, which is at the singular locus $\mathbb {RP}^2 \times \mathbb R^{3,1}$.      

Let the bare dimensionful gauge coupling of the $d=1+1$ theory (which results from the IIB compactification on $\widehat {\mathcal Q}$) be given by $\widehat g$ at the Planck scale. Also, let the corresponding effective dimensionless gauge coupling at the observed energy scale $u$ be given by ${\widehat g}_{YM}(u)$. Recall that $\widehat {\mathcal Q}$ is given by a fibration structure such that $\widetilde M_2 \hookrightarrow \widehat {\mathcal Q} \to \widehat {\widetilde Q}$. Hence, a full compactification on $\widehat {\mathcal Q}$ can viewed as an initial compactification on $\widehat {\widetilde Q}$ down to $d=3+1$, followed by a further compactification on the $\widetilde M_2$ fibre down to $d=1+1$. By a simple dimensional reduction, we find that $({1/ {\widehat {g}^2}}) = vol (\widetilde M_2) \cdot ({1/ {{\widehat{g}_o}^2}})$. From (\ref{gocap}), one has the relation $({1/ {{\widehat{g}_o}^2}}) = vol(\mathbb {RP}^2_{bare})$. Moreover, one also has the fibration $\widetilde M_2 \hookrightarrow \mathbb {CP}^2 /\mathbb Z_2 \to \mathbb {RP}^2$, whereby $\mathbb {CP}^2 /\mathbb Z_2 = S^4$, thus implying that ${vol(\mathbb{RP}^2_{bare}) \cdot vol(\widetilde M_2)}={vol(S^4_{bare})}$. Therefore, we have  
\be
{1\over {\widehat {g}^2}} = vol (S^4_{bare}),
\label{barecap}
\ee     
where $vol (S^4_{bare})$ is the volume of $S^4 \subset \widehat {\mathcal Q}$ before it starts to blow-down in the `lifted' geometric transition $\widehat {\mathcal Q} \rightarrow \widehat {\mathcal X}$ at low energy. By similar dimensional reduction arguments, since $vol (\widetilde M_2)$ has dimensions of mass$^{-2}$ (length$^2$) while $\widehat {g}_{4d}(u)$ is dimensionless, we have the following relation for the effective dimensionless gauge coupling in $d=1+1$: 
\be
{1\over {\widehat {g}^2_{YM}(u)}}\ = \ vol({\widetilde M_2}) \cdot ({u \over {\widehat{g}_{4d}(u)}})^2.
\label{gymcap}
\ee            
In light of the bare relation given in (\ref{barecap}), by substituting (\ref{g4dcap}) and (\ref{gocap}) into (\ref{gymcap}), we have   
\be
vol(S^4_{eff}) \sim {1\over {\widehat {g}^2_{YM}(u)}} \ = \ {1\over {{({\widehat g}/u)}^2}}\ + \ {\alpha}' \ u^2 log({u \over {|\Lambda_{planck}|}}), 
\label{RG flow,cap}
\ee    
whereby $vol(S^4_{eff})$ is the $\it{effective}$ volume of the 4-cycle $S^4$ at the energy scale $u$. ${\alpha}'$ is a real positive constant. Since $vol (\mathbb {RP}^2) = {vol (\mathbb {CP}^1) \over 2}$, we have $(1/ {g^2_o}) \sim (1/ {\widehat {g}^2_o})$ from discussion below (\ref{go}) and (\ref{gocap}) itself. Likewise, since $vol (S^4)= {vol (\mathbb {CP}^2) \over 2}$, we have $({1/ {{g}^2}}) \sim ({1/ {\widehat {g}^2}})$ from (\ref{bare}) and (\ref{barecap}). Thus, as expected of a geometrically lifted and hence physically equivalent background, up to a redefinition of the bare value of the gauge coupling at the Planck scale, which can be trivially reabsorbed by defining the `bare'\footnote{i.e. before it undergoes a geometric transition $\widehat {\mathcal Q} \rightarrow \widehat {\mathcal X}$ at low energy.} volume of $\widehat {\mathcal Q}$  to be twice  the `bare' volume of $\mathcal Q$, the expression for the Yang-Mills gauge coupling at energy $u$, of the $d=1+1$ theory from the IIB compactification on the manifold $\widehat {\mathcal Q}$ as given by (\ref{RG flow,cap}), is identical to that of the $d=1+1$ theory obtained from the IIB compactification on the original manifold $\mathcal Q$, as given by (\ref{RG flow}).

From (\ref{RG flow,cap}), $vol(S^4_{eff})$ seems to get smaller for decreasing $u$, such that for small enough values of $u$, there will be an $S^4_{eff}$  flop characterized by $[vol(S^4_{eff}) > 0]$ $\rightarrow$ $[vol(S^4_{eff}) < 0]$ as the $d=1+1$ theory undergoes a RG flow to the IR. In addition, via the arguments surrounding (\ref{vac}) and (\ref{theta}) of $\S$3.3, we find that the $S^4_{eff}$ flop will be smooth as long as we turn on the self-dual RR 4-form $C^+_4$ along the tangent directions of the 4-cycle $S^4$. We will henceforth assume this to be true. Now, note that $S^{5'}/ \mathbb Z_2 = \mathbb {RP}^{5'}$ in $\widehat{\mathcal X}\cong \mathbb R^3 \times \mathbb {RP}^{5'}$ is given by the fibration $S^1 \hookrightarrow \mathbb {RP}^{5'} \to \mathbb {CP}^{2'} /\mathbb Z_2$, whereby $\mathbb {CP}^{2'} /\mathbb Z_2 = S^{4'}$ as usual. Thus, since $\widehat {\mathcal Q} \cong (\mathbb R^4 / \mathbb Z_N) \times S^4$, one can view the `lifted' geometric transition ${\widehat {\mathcal Q}} \rightarrow {\widehat {\mathcal X}}$, as being induced by an $S^4$ flop within $\widehat {\mathcal Q}$ and $\widehat {\mathcal X}$, characterized by $[vol(S^4) > 0]$ $\rightarrow$ $[vol(S^4) < 0]$, whereby the volume of the blown-up $S^{4'} \subset \mathbb {RP}^{5'} \subset \widehat {\mathcal X}$ is proportional to $|vol(S^4) < 0|$. Also note here that $\theta' = \int_{S^{4'}_{bare}} C^+_4$, and since $vol (S^{4'}) = vol (S^4) / N$ implies that $vol (S^{4'}_{bare}) = vol (S^4_{bare}) / N$, it means that $\theta' = \theta /N$. This will result in the $N$ inequivalent vacua of the effective $d=1+1$ theory due to $N$ distinct $\theta$-phases. Therefore, consistent with an (equivalent) geometric lift of the large $N$, type IIB, $Spin(7)$ duality involving the geometric transition $\mathcal Q \rightarrow \mathcal X$, it is clear that the $\it{smooth}$ 8-dimensional geometric transition ${\widehat {\mathcal Q}} \rightarrow {\widehat {\mathcal X}}$, which now occurs in a background with $\it{no}$ space-filling D5-branes or 3-form RR and NS fluxes at low energy, is a consequence of a $\it{non}$-$\it{singular}$ RG flow of the resulting $\mathcal N = (1,0)$ supersymmetric pure $SU(N)$ theory in 1+1 dimensions to the $\it{same}$ (i.e. $\it{dual}$) albeit abelian $U(1)$ theory in the IR with $N$ inequivalent vacua characterized by $N$ distinct $\theta$-phases, such that the sizes of the blown-down $S^4 \subset \widehat {\mathcal Q}$ and blown-up $\mathbb {RP}^{5'} \subset \widehat {\mathcal X}$ before and after the geometric transition $\widehat{\mathcal Q} \rightarrow \widehat {\mathcal X}$, are governed by the (smoothly running) value of the effective dimensionless gauge coupling $1/ {\widehat{g}^2_{YM}(u)}$ of the $d=1+1$ theory at the observed energy scale $u$, which in turn determines the point in the geometric transition $\widehat {\mathcal Q} \rightarrow \widehat {\mathcal X}$. 

\subsection{IIB S-duality and the lifted background at large $g_s$}

The bosonic field content of the type IIB theory coming from the NS-NS sector consists of the metric $\it{g}$, the 2-form potential $B_2$, and the scalar dilaton $\Phi$. On the other hand, the bosonic field content coming from the R-R sector consists of the axion field or the 0-form potential $C_0$, the 2-form potential $C_2$, and the self-dual 4-form potential $C^+_4$. Under the exact $SL(2,\mathbb Z)$ symmetry of the IIB theory, the fields transform as follows:
\be
\tau \rightarrow {{a\tau + b} \over {c \tau +d}},
\label{tau}
\ee  
where
\be
\tau = C_0 + ie^{-\Phi},
\ee
and 
$a$, $b$, $c$, $d$ are real $\it{integers}$ obeying $ad-bc=1$. In addition, one must simultaneously make a transformation on the 2-form fields such that 
\be
\left(\begin {array}{c}
B_2 \\
C_2
\end{array}\right)
\rightarrow
\left(\begin {array}{cc}
\hspace{0.2cm}a & \hspace{0.1cm}{-c} \\
\hspace{-0.1cm}{-b} &\hspace{0.3cm} d \\
\end{array}\right)
\left(\begin {array}{c}
B_2 \\
C_2
\end{array}\right).
\ee
The self-dual 4-form RR potential $C^+_4$ is invariant under the transformation because one has 
\be
C^+_4 \rightarrow C^+_4.
\ee
The $SL(2,\mathbb Z)$ symmetry transformation can alternatively be expressed in terms of the matrices
\be
\mathbb M = e^{\Phi}
\left(\begin {array}{cc}
|\tau|^2\ &\ C_0 \\
C_0 \ & \ 1 \\
\end{array}\right),
\qquad
\Lambda =
\left(\begin {array}{cc}
a \ & \ b \\
c \ & \ d \\
\end{array}\right),
\ee
such that 
\be
\mathbb M \rightarrow \Lambda \mathbb M {\Lambda}^T, 
\ee
whereby for $H_{NS}=dB_2$ and $H_{RR}=dC_2$, one must also have 
\be
\tilde H=
\left(\begin {array}{c}
H_{NS} \\
H_{RR}
\end{array}\right),
\qquad
\tilde H \rightarrow (\Lambda^T)^{-1} \tilde H.
\label{flux tx}
\ee 

It is clear that for $a=d=0$, such that one must have $b=-c=1$ since $ad-bc=1$, one will obtain the weak-strong S-duality subgroup transformation $\tau \rightarrow {-1/\tau}$, whereby $\tau = C_0 +  {i\over g_s}$. Under this S-duality transformation, one will also have $B_2 \rightarrow C_2$ and $C_2 \rightarrow -B_2$, or $H_{NS} \rightarrow H_{RR}$ and $H_{RR} \rightarrow -H_{NS}$. The self-dual 4-form $C^+_4$ remains invariant as mentioned, and as discussed before, we will not set it to zero. Note also that since the F-string (fundamental string) and D-string (D1-brane) couple electrically to $B_2$ and $C_2$ respectively, because $B_2$ and $C_2$ are exchanged under an S-duality transformation, one will have F-string $\leftrightarrow$ D-string. Moreover, since the NS5-branes and D5-branes couple magnetically to $B_2$ and $C_2$ respectively, one will also have NS5-branes $\leftrightarrow$ D5-branes. The absence of D5-branes in the lifted background thus implies that there must be an absence of NS5-branes. However, since there must be F-strings, there will be D-strings. The presence of D-strings should not affect our results thus found as they are a typical component of the non-perturbative IIB spectrum. However, because open-string worldsheets can end on D-strings, one must check for a Freed-Witten anomaly as follows: let the worldvolume of the D-string be given by $\mathbb R^{0,1} \times \mathcal S$. From an extension of (\ref{zeta}) and the discussion thereafter, one has 
\be
H_{NS}|_{\mathbb R^{0,1}\times \mathcal S} = 2\pi  W_3(\mathcal S),
\label{HNS}
\ee
whereby one recalls that $W_3(M) = 0$ if and only if $M$ is $spin^c$. Since $H_{NS} = 0$ in the lifted background, the following condition must be satisfied for the theory to be anomaly-free:
\be      
W_3 (\mathcal S)= 0.
\label{anomaly-free}
\ee      
Notice that the D-string must extend over or wrap a 1-dimensional subspace of $\mathbb R^{1,1} \times \widehat {\mathcal Q}$ and $\mathbb R^{1,1} \times \widehat {\mathcal X}$. Recall that since $\widehat {\mathcal Q} \cong (\mathbb  R^4/\mathbb Z_N) \times S^4$ and $\widehat {\mathcal X} \cong \mathbb  R^3 \times \mathbb {RP}^{5'}$, $\mathcal S \cong \mathcal I$, where $\mathcal I$ is an open interval representing an unwrapped string, or\footnote{Recall that $\mathbb {RP}^{5'} = S^{5'}/\mathbb Z_2$ is given by the fibration $S^1 \hookrightarrow \mathbb {RP}^{5'} \to \mathbb {CP}^{2'} /\mathbb Z_2$, whereby $\mathbb {CP}^{2'} /\mathbb Z_2 = S^{4'}$. Thus, the D-string can wrap on the $S^1$ fibre of $\mathbb {RP}^{5'} \subset \widehat {\mathcal X}$.} $\mathcal S \cong S^1$. As mentioned before, any oriented manifold of dimension $\leq$ 4 is $spin^c$ \cite{freed-witten}. Since $\mathcal I$ is topologically trivial and orientable, it is $spin^c$. Since $w_1(S^n)=0$ for any $n$, whereby $w_1$ is the first Stiefel-Whitney class of $S^n$, one sees that $S^1$ is also orientable and thus, $spin^c$. Hence, (\ref{anomaly-free}) is always satisfied. Therefore, one finds that the theory continues to be anomaly-free even in the presence of D-strings.  

Now, recall that the lifted background on $\widehat {\mathcal Q}$ and $\widehat {\mathcal X}$ are such that $H_{NS}= dB_2=0$ and $H_{RR}=dC_2=0$. This implies that $B_2$ and $C_2$ are constant fields. One is then free to set $B_2=C_2=0$. Hence, only $C_0$ and $g_s$ are modified under an S-duality transformation. Via this $\it{dual}$ transformation, one can express the IIB background on $\widehat {\mathcal Q}$ and $\widehat {\mathcal X}$ at strong string coupling ($g_s \gg 1$) in terms of an $\it{equivalent}$ IIB background on $\widehat {\mathcal Q}$ and $\widehat {\mathcal X}$ at weak string coupling ($g_s \ll 1$), whereby only $C_0$ is different. In other words, type IIB on $\mathcal Q$ with $N$ space-filling D5-branes wrapping $\mathbb {CP}^2 \subset \mathcal Q$ at low energy and $g_s \gg 1$, is equivalent, via a geometric lift and the above S-duality transform, to type IIB on $\widehat{\mathcal Q}$ without  D5-branes at low energy and $g_s \ll 1$. Since $C_0$ is not a moduli controlling the geometric transition $\widehat {\mathcal Q} \rightarrow \widehat {\mathcal X}$, i.e. its value does not modify our results, and because $C^+_4$ (whose non-zero vev along the 4-cycle $S^4_{bare}$ is essential in avoiding singularities) remains invariant under an S-duality transformation, as argued in the preceding discussions thus far, the IIB background on $\widehat{\mathcal Q}$ without  D5-branes at low energy and $g_s \ll 1$ will undergo a smooth geometric transition to a $\it{dual}$ background on $\widehat {\mathcal X}$ at low energy and $g_s \ll 1$. Via an S-duality transformation and $\widehat {\mathcal X}$ 's equivalence as a geometrically lifted configuration, this dual background on $\widehat {\mathcal X}$ is then equivalent to type IIB on $\mathcal X$ with no D5-branes but with $H_{NS}$ and $H_{RR}$ fluxes through 3-cycles in $\mathcal X$ at $g_s \gg 1$. In other words, the large $N$, type IIB duality via a $Spin(7)$ geometric transition $\mathcal Q \rightarrow \mathcal X$, is observed to hold for large values of string coupling $g_s$ as well! Since one has the 't Hooft coupling $t=g_s N$, it means that the duality holds for small and large values of $t$ too. This is consistent with the conjecture by Vafa in \cite{v} that the large $N$ duality should hold for $\it{all}$ values of 't Hooft coupling. The chain of dualities is depicted as follows:
\be
\def\mapdown#1{\Big\downarrow\rlap{$\vcenter{\hbox{$\scriptstyle#1$}}$}}
\def\mapup#1{\Big\uparrow\rlap{$\vcenter{\hbox{$\scriptstyle#1$}}$}}
\begin{array}{rccc}
& \hbox{\scriptsize $N$ D5-branes}& & \hbox{\scriptsize 3-form RR and NS flux}\\
\noalign{\smallskip}
\mbox{$g_s \gg 1$} :~~ &{\mathcal Q}\cong {\mathbb R^4 \times \mathbb {CP}^2} & {\longrightarrow {\rm geometric} \hspace{0.3cm} {\rm transition} \longrightarrow}  & {\mathcal X \cong \mathbb R^3 \times S^5}  \\
\noalign{\medskip}
& \mapdown{{\rm lift}}&  &\mapup{{\rm unlift}}\\
\noalign{\medskip}
&{\widehat {\mathcal Q}\cong (\mathbb R^4 /\mathbb Z_N) \times S^4} &  & {\widehat {\mathcal X} \cong \mathbb R^3 \times \mathbb {RP}^{5'}} \\
\noalign{\bigskip}
& \mapdown{\rm S-duality}& &\mapup{\rm S-duality}\\
\noalign{\medskip}
&\widehat {\mathcal Q}\cong (\mathbb R^4 /\mathbb Z_N) \times S^4 & & \widehat {\mathcal X} \cong \mathbb R^3 \times \mathbb {RP}^{5'} \\
\noalign{\bigskip}
& \mapdown{{\rm unlift}}&  &\mapup{{\rm lift}}\\
\noalign{\medskip}
\mbox{$g_s \ll 1$} :~~ &{\mathcal Q}\cong {\mathbb R^4 \times \mathbb {CP}^2} & {\longrightarrow {\rm geometric} \hspace{0.3cm} {\rm transition} \longrightarrow}  & {\mathcal X \cong \mathbb R^3 \times S^5} \\
\noalign{\smallskip}
& \hbox{\scriptsize $N$ D5-branes}& & \hbox{\scriptsize 3-form RR and NS flux}
\end{array}
\label{diagram}
\ee
Notice that this result is only readily manifest when we geometrically lift to the background on $\widehat {\mathcal Q}$ and $\widehat {\mathcal X}$ $\it{withhout}$ D5-branes and 3-form fluxes. This is because from (\ref{flux tx}), one can see that $H_{NS}$ and $H_{RR}$ are not invariant under an S-duality transformation. Moreover, the transformation will also result in the exchange D5-branes $\leftrightarrow$ NS5-branes. Hence, the $N$ space-filling D5-branes wrapping the $\mathbb {CP}^2 \subset \mathcal Q$ will be replaced by $N$ NS5-branes instead, of which the large $N$ duality does not involve. Note also that $B_2= C_2=0$ is the condition considered in \cite{ftheory} in constructing an equivalent F-theory background. This suggests that the lifted IIB backgrounds on $\widehat {\mathcal Q}$ and $\widehat {\mathcal X}$ naturally lend themselves to an equivalent F-theoretic description. We will pursue this next in the following subsections.   
                
\subsection{Equivalent F-theoretic description of a general IIB background} 

Let us now review the equivalent F-theoretic description of a general IIB background. 12-dimensional F-theory was first constructed in \cite{ftheory} as a geometrically economical way to describe the physics of type IIB vacua. Let us review the relevant aspects of the construction in \cite{ftheory}: the existence of the 12-dimensional F-theory of signature (10,2) is derived from a consideration of the strong/weak duality of the IIB theory that it is supposed to describe. BRST invariance of the corresponding worldsheet theory reveals that the resulting physical states are identical to those of the 10-dimensional IIB theory, i.e. F-theory and IIB strings have an equivalent spectrum up to a BRST quantization. Furthermore, via a proposal made in the context of $\mathcal N =2$ strings in \cite{ftheoryref8}, an object with a 4-dimensional worldvolume of signature (2,2) can be considered such that its compactification on the compact subspace with signature (1,1) will result in a worldsheet of signature (1,1) in the effective 10-dimensional spacetime of signature (9,1). This can be identified as the worldsheet of the fundamental IIB string. It can also be shown that an F-theory compactification on this same 2-dimensional compact subspace of signature (1,1), which results in the 10-dimensional IIB theory with conventional signature (9,1), is actually equivalent to a $\it{Euclidean}$ $T^2$ compactification, such that the only physical moduli coming from this compactification is the complex structure of the $T^2$ (henceforth referred to as the F-torus). The K\"ahler class of the $T^2$ is therefore frozen (i.e. its size does not vary). The IIB $SL(2,\mathbb Z)$ torus symmetry is thus naturally encoded in the geometry of an F-torus in the equivalent F-theoretic description. Recall that the $SL(2,\mathbb Z)$ action on the relevant fields which leaves the IIB theory invariant includes a modular transformation of the complexified IIB coupling. Consequently, the physical moduli related to the complex structure of the F-torus is identified with the complexified coupling of the IIB theory $\tau=C_{0} + i e^{-\Phi}$, where $C_{0}$ is the RR axion field, $\Phi$ is the dilaton field, $e^{-\Phi}=1/g_{s}$ and $g_{s}$ is the IIB string coupling. By the application of fibrewise duality \cite{witten/vafa,lectures}, we arrive at the following statement: IIB on a manifold $\mathcal M$ is equivalent to F-theory on an elliptic ($T^2$) fibration of $\mathcal M$. However, do note that this is a specific example of a more general equivalence statement which we will now discuss.   

Let us consider an application of this IIB/F-theory equivalence. In particular, consider the IIB BPS vacua in which there are twenty-four space-filling D7-branes on an $S^2$ compactification manifold with a complexified coupling that varies over it \cite{ftheory,shapere}. The complexified IIB coupling is thus given by $\tau(z)=C_{0}(z) + i e^{-\Phi(z)}$, whereby the complex parameter $z$ is the coordinate on $S^2$. Recall that $\tau(z)$ is also the complex structure of the F-torus defined at the point $z$. By the specific equivalence statement above, compactification of F-theory down to eight dimensions is achieved on a 4-manifold $M$ given by an elliptic fibration of $S^2$. Since $\tau(z)$ varies along $S^2$ in the IIB vacua considered, there is a different F-torus (due to an different complex structure) at every point on the $z$-parameterised $S^2$. This translates to the fact that $M$ is a $\it{non}$-$\it{trivial}$ elliptic fibration. Note also that because of the presence of the `magnetic' dual $C_{0}$ charges of the D7-branes,\footnote{The D7-brane has a magnetically dual D(-1)-brane which carries the magnetic charge of $C_{0}$.} as we travel along non-trivial closed cycles around the points where the branes sit on $S^2$, the complexified IIB coupling and hence complex structure of the elliptic fibre can undergo a non-trivial $SL(2,\mathbb Z)$ monodromy:\footnote{The RR `electric' charge of the D7-brane $Q_e$ in ten spacetime dimensions is given by the appropriate flux through the $S^1$ that surrounds it, i.e. $Q_e = {\int_{S^1} *dC_{8}} = {\int_{S^1} dC_{0}}$. In order to register a unit of RR charge, we must have $C_{0} \rightarrow C_{0} + 1$ each time we circle the point where it sits on $S^2$.} $\tau(z) \rightarrow \tau(z) + 1$. Mathematically, this implies that the fibre degenerates (pinches off to zero size) over these points. This happens at twenty-four positions on $S^2$ for each of the twenty-four D7-branes. A non-trivial elliptic fibration of $S^2$ with 24 degenerate points happens to be a particular representation of the 4-dimensional $K3$ manifold. In other words, $M$ is $K3$. Indeed, $K3$ preserves $1/2$ of the maximal supersymmetry in the F-theory, consistent with the BPS vacua of the IIB theory in eight dimensions. Moreover, it has been shown that the perturbative $\it{and}$ non-perturbative aspects of this particular IIB vacua are being encoded in this $K3$ geometry \cite{sen}. Hence, this IIB background can be equivalently described in terms of an F-theory on $K3$ \cite{ftheory,sen}.  

Likewise, via a fibrewise duality, F-theory on an elliptic $K3$ fibration of a base $B$ should correctly describe the physics of the $\it{above}$ IIB background on an $S^2$ fibration of this same base. Eventually, one is led to the following equivalence result for an arbitrary IIB vacua which generalizes the earlier statement: the physics of a IIB theory on a manifold $\mathcal M$ is equivalently described by the corresponding F-theory on an elliptic fibration of $\mathcal M$, such that the fibration degenerates at certain points when there are D7-branes sitting on the base or on its even-dimensional hypersurface thereof.

\subsection{Equivalent F-theoretic description of the IIB background without D5-branes and fluxes} 

The superstring extension of the large $N$ duality conjecture of Gopakumar-Vafa \cite{gv} found in \cite{v} was shown to hold in a resolved/deformed conifold background $\it{without}$ D7-branes. Since the large $N$ type IIB $Spin(7)$ geometric transition duality considered in this paper is just an 8-dimensional extension of this original duality, there are $\it{no}$ D7-branes in this case either. We will henceforth be looking at the equivalent F-theoretic description of this IIB $Spin(7)$ background.

At this point, the reader might have realised that  an F-theoretic description of a IIB background without D7-branes is unconventional since typical F-theoretic descriptions of IIB vacua usually involve some D7-branes. This simply means that the IIB theory on either side of the geometric transition of our large $N$, $Spin(7)$ duality $\it{cannot}$ be related to a heterotic or type I string (via an F-theory/het-type I duality);\footnote{Note that a IIB background with twenty-four D7-branes on an $S^2$, which is equivalent to F-theory on elliptically fibred $K3$, has been conjectured and established to be dual to the heterotic string on $T^2$ in \cite{ftheory} and \cite{sen}. Moreover, a IIB orientifold background with D7-branes and O7-planes, which is equivalent to F-theory on a $K3$ orbifold, is related to type I theory on a $T^2$ via T-duality \cite{sen}. We thus see the need to involve D7-branes/O7-planes if we wish to extend our duality relation to include the heterotic or type I string theory via an F-theory/het-type I duality.} indeed it has been shown in \cite{das1,das2,das3}, that to relate the Gopakumar-Vafa large $N$ duality conjecture involving the resolved/deformed conifold to heterotic or type I theory via the F-theory/het-type I duality established in \cite{sen}, one must include D7-branes/O7-planes in the background before a lift to F-theory is performed. Therefore, we wish to emphasize that since we are considering a background without D7-branes, we will not be claiming any equivalence between the vacua of our large $N$, type IIB, $Spin(7)$ theory and the vacua of a heterotic or type I string. Neither will we claim that the duality can be extended to involve the heterotic or type I theory in this paper. 

From the discussion in $\S$4.2, the absence of D7 branes (i.e. RR charge $Q_e = \int_{S^1} dC_{0} = 0$) on the 8-dimensional IIB compactification manifolds ${\widehat {\mathcal Q}}$ and ${\widehat {\mathcal X}}$ will then imply the absence of monodromy in the complexified coupling and hence degeneracy in the elliptic fibres of the F-theory. In other words, the 10-dimensional elliptic fibration of $\widehat {\mathcal Q}$ and $\widehat {\mathcal X}$ to be considered in our F-theoretic description is $\it{non}$-$\it{degenerate}$. Notice that the introduction of non-degenerate elliptic fibres over $\widehat {\mathcal Q}$ and $\widehat {\mathcal X}$ do not result in additional enhanced gauge symmetries (which arise due to coincident D7-branes in the IIB picture) in the resulting lower dimensional theory beyond those that are already present due to the singular and non-singular geometry of the $\widehat {\mathcal Q}$ and $\widehat {\mathcal X}$ bases alone, i.e. the gauge symmetries of the effective theory from IIB on $\widehat {\mathcal Q}$ ($\widehat {\mathcal X}$) and F-theory on the elliptic fibration of $\widehat {\mathcal Q}$ ($\widehat {\mathcal X}$) are the same, consistent with the IIB/F-theory equivalence in a background without D7-branes. The complex structure and geometry of the F-theory elliptic fibre merely function as means to encode the complexified coupling and $SL(2, \mathbb Z)$ symmetry of the IIB theory respectively. 

The $\it{constant}$ 't Hooft coupling is given by $t = g_s N$, where $g_s$ is the string coupling and $N$ is large. Notice then that $g_s$ is also constant since $N$ is fixed, i.e. $g_s \sim e^{\Phi}$ does not vary along $\widehat {\mathcal Q}$ and $\widehat {\mathcal X}$. Recall that we will be considering a non-trivial albeit finite and smoothly varying axion field background $C_{0} = C_{0}(\vec{z})$, where $\vec{z}$ is a position vector on  $\widehat {\mathcal Q}$ or $\widehat {\mathcal X}$ and $C_{0}(\vec{z})$ is periodic, i.e. $C_{0}(\vec{z}) \sim C_{0}(\vec{z}) +1$ (due to the IIB $SL(2, \mathbb Z)$ symmetry $\tau(\vec{z}) \sim \tau(\vec{z}) +1$). Note that this background choice is not in conflict with the condition $\int_{S^1} dC_{0}(\vec{z}) = 0$ from an absence of D7-branes since $dC_{0}(\vec{z})$ is only required to vanish up to an integral along a circle, i.e. $dC_{0}(\vec{z})$ can be non-zero locally and thus $C_0(\vec{z})$ can be allowed to vary along $\widehat {\mathcal Q}$ and $\widehat {\mathcal X}$. Therefore, the complexified IIB coupling or the complex structure of the F-theory elliptic fibre $\tau(\vec {z})=C_{0}(\vec{z}) + i e^{-\Phi}$ will vary along $\widehat {\mathcal Q}$ and $\widehat {\mathcal X}$. In other words, the 10-dimensional F-theory manifolds which we will consider are $\it{non}$-$\it{trivial}$ elliptic fibrations of $\widehat {\mathcal Q}$ and $\widehat {\mathcal X}$. 

\subsection{Large $N$, type IIB, $Spin(7)$ duality as an F-theoretic $\mathbb {RP}^5$ flop} 

Let us denote the two 10-dimensional F-theory compactification manifolds, which are non-degenerate and non-trivial elliptic fibrations of $\widehat {\mathcal Q}$ and $\widehat {\mathcal X}$, as $\widehat{\mathcal Q}_F$ and $\widehat {\mathcal X}_F$ respectively. Recall that $\widehat {\mathcal Q} = {\mathcal Q}' / \mathbb Z_2$ and $\widehat {\mathcal X} = {\mathcal X}'/ \mathbb Z_2$, where ${\mathcal Q}' \cong (\mathbb R^4 /\mathbb Z_N) \times \mathbb {CP}^2$ and $\mathcal X' \cong \mathbb R^3 \times S^{5'}$. Let the 10-dimensional, non-degenerate and non-trivial elliptic fibrations of ${\mathcal Q}'$ and $\mathcal X'$ be ${\mathcal Q}'_F$ and ${\mathcal X}'_F$ accordingly. As explained earlier in $\S$4.1, since the $\mathbb Z_2$ action on $\mathcal Q'$ and $\mathcal X'$ lifts to act on the elliptic fibres defined over them as well, one will then have $\widehat{\mathcal Q}_F = {\mathcal Q}'_F / \mathbb Z_2$ and $\widehat {\mathcal X}_F = {\mathcal X}'_F / \mathbb Z_2$. 

As elucidated in $\S$4.4, F-theory compactified on a Euclidean $T^2$ results in a type IIB theory. Since the $T^2$ compactification manifold does not break any supersymmetry, it also means that like in IIB, F-theory is defined with 32 supercharges.\footnote{This feature of  F-theory is also related to the fact that unlike M-theory, it has no low-energy interpretation as a Lorentz-covariant 12-dimensional SUGRA theory with signature (11,1), in which the minimal spinor dimension is 64 and not 32. However, it can be related to a theory with signature (10,2), in which the minimum dimension of a Majorana-Weyl spinor is 32.} This further implies that $\widehat{\mathcal Q}_F$ and $\widehat{\mathcal X}_F$ must preserve the same amount of maximal supersymmetry that $\widehat {\mathcal Q}$ and $\widehat {\mathcal X}$ would, i.e., they must also preserve $1/32$ of the maximal supersymmetry. Moreover, in order to be a physically consistent compactification such that the worldsheet theory is conformally invariant (up to lowest order corrections), $\widehat{\mathcal Q}_F$ and $\widehat{\mathcal X}_F$, like $\widehat {\mathcal Q}$ and $\widehat {\mathcal X}$, must also be Ricci-flat. Since quotienting an old manifold by a freely-acting isometric involution (such as a $\mathbb Z_2$ action) to construct a new manifold simply involves the identifying of points on the old manifold connected by the action of the involution, the metric on the new manifold is the same as that on the old manifold. This means that the metric on $\widehat{\mathcal Q}_F$ and $\widehat {\mathcal X}_F$ will be the same as that on ${\mathcal Q}'_F$ and ${\mathcal X}'_F$ respectively. Since the Ricci curvature tensor is determined purely from the metric, the requirement of Ricci-flatness in $\widehat{\mathcal Q}_F$ and $\widehat {\mathcal X}_F$ will in turn translate to the requirement of Ricci-flatness in ${\mathcal Q}'_F$ and ${\mathcal X}'_F$. Moreover, via a 10-dimensional extension of the Joyce construction discussed in $\S$4.1, one can see that if $\widehat{\mathcal Q}_F$ and $\widehat {\mathcal X}_F$ are to preserve $1/32$ of the maximal supersymmetry, $\widehat {\widetilde Q}$ and $\widehat {\widetilde X}$ must then preserve $1/16$ of it. The only non-trivial choice for a physically consistent, 10-dimensional, Ricci-flat, Riemannian manifold that will preserve $1/16$ of the maximal supersymetry is one that possesses an $SU(5)$ holonomy group \cite{joyce}, i.e. ${\mathcal Q}'_F$ and ${\mathcal X}'_F$ must be CY 5-folds. Let us now take a closer look at things.       

\bigskip\noindent{\it Derivation of ${\mathcal Q}'_F$ and ${\mathcal X}'_F$}
\vspace{0.2cm}

As shown in $\S$4.3, the results for the background with $g_s \ll 1$ and $g_s \gg 1$ are similar; they are S-dual to each other, and this is implicit in the equivalent F-theoretic interpretation since $B_2= C_2=0$ and the elliptic fibre possesses an $SL(2,\mathbb Z)$ modular symmetry of which the S-duality is a subgroup thereof. Hence, a discussion of the background in which $g_s \ll 1$ would suffice. Now notice that the $\it{algebraic}$ F-torus fibre has a complex structure $\tau=C_{0} + {i\over g_s}$ such that its height is given by $1\over g_s$. In the limit where $g_s \ll 1$, it will also mean that ${1\over g_s} \gg 1$, i.e., one of the two $S^1$ cycles of the F-torus fibre, whose radius is represented by the height $1\over g_s$, decompactifies. In other words, the F-torus fibre will be given by $(S^1 \times \mathbb R)$. Consequently, ${\mathcal Q}'_F$ and ${\mathcal X}'_F$ will be given by the non-degenerate and non-trivial $(S^1 \times \mathbb R)$ fibrations of ${\mathcal Q}'$ and $\mathcal X'$ respectively.  

Notice that  an $\mathbb R^4$ space is radially contractible, and this still holds even when one divides the space by $\mathbb Z_N$, a group of order $N$, i.e., the $\mathbb R^4/ \mathbb Z_N$ bundle of ${\mathcal Q}' \cong (\mathbb R^4 /\mathbb Z_N) \times \mathbb {CP}^2$ is a contractible space. In view of the non-trivial Hopf fibration structure of $S^5$ given by $S^1 \hookrightarrow S^5 \to \mathbb {CP}^2$, and the fact that all bundles over a contractible space must be trivial \cite{hou}, it is clear that the $(S^1 \times \mathbb R)$ space can only be non-trivially fibred over the compact and non-contractible $\mathbb{CP}^2$ base of ${\mathcal Q}' \cong (\mathbb R^4 /\mathbb Z_N) \times \mathbb {CP}^2$. This non-trivial $(S^1\times \mathbb R)$ fibration of $\mathcal Q'$ results in a trivial $\mathbb R^4 /\mathbb Z_N$ bundle over a 6-dimensional base given by a $\it{non}$-${trivial}$ real line bundle over an $S^5$. Because of the triviality of the $\mathbb R^4 /\mathbb Z_N$ bundle over this 6-dimensional base, the geometry of the base and hence that of its real line bundle will not vary along the $\mathbb R^4 /\mathbb Z_N$ directions. Thus, one can trivially combine the non-trivial real line bundle of the 6-dimensional base with the $\mathbb R^4 /\mathbb Z_N$ fibre over it into an $[\mathbb R \times (\mathbb R^4 /\mathbb Z_N)]$ space, from which one can see that $\mathcal Q'_F$ is actually isomorphic to a $\it{non}$-$\it{trivial}$ $[\mathbb R \times (\mathbb R^4 /\mathbb Z_N)]$ bundle over $S^5$. 

On the other hand, recall that one has the hyperk\"ahler moment map of a $U(1)$ action in $\mathbb R^4$ given by ${{\mathbb R^4}/ U(1)} \cong {\mathbb R^3}$ \cite{aw}. Thus, it is clear that the $S^1= U(1)$ fibre would be naturally embedded in an $\mathbb R^4$ bundle, of which the $\mathbb R^3$ bundle of $\mathcal X' \cong \mathbb R^3 \times S^{5'}$ is now a subspace thereof. Since the resulting $\mathbb R^4$ bundle is contractible, the remaining real line $\mathbb R$ can only be non-trivially fibred over the compact and non-contractible $S^{5'}$ base of $\mathcal X'$. This non-trivial $(S^1 \times \mathbb R)$ fibration of $\mathcal X'$ will therefore result in a trivial $\mathbb R^4$ bundle over a 6-dimensional base given by a $\it{non}$-$\it{trivial}$ real line bundle over $S^{5'}$. Likewise, because of the triviality of the $\mathbb R^4$ bundle over this 6-dimensional base, the geometry of the base and hence that of its real line bundle will not vary along the $\mathbb R^4$ directions. Thus, one can trivially combine the non-trivial real line bundle of the 6-dimensional base with the $\mathbb R^4$ fibre over it into an $\mathbb R^5$ space, from which it is clear that $\mathcal X'_F$ is actually isomorphic to a $\it{non}$-$\it{trivial}$ $\mathbb R^5$ bundle over $S^{5'}$. 

There exists a non-compact CY 5-fold, with an $SU(5)$ holonomy group, given by a $\it{non}$-$\it{trivial}$ $\mathbb R^5$ bundle over $S^5$. It is isomorphic to $T^*S^5$ as a symplectic manifold and isomorphic to the hypersurface $z_1^2 + z_2^2 + z_3^2 + z_4^2 + z_5^2 + z_6^2 = 1$ in $\mathbb C^6$, for $z_i \in \mathbb C$, as a complex manifold \cite{private comms}. It is $SO(6)$ invariant while its generic orbits are copies of a 9-dimensional manifold, i.e. it is of cohomogeneity one. Notice that the orientation-preserving $\mathbb Z_N \subset U(1)$ action along the non-trivial $[\mathbb R \times (\mathbb R^4 /\mathbb Z_N)]$ bundle of $\mathcal Q'_F$, is an isometry subgroup of the $\mathbb R^4$ space, thus implying that the holonomy of $\mathcal Q'_F$, will be the same as that of a non-trivial $\mathbb R^5$ bundle over $S^5$. In other words, the holonomy group of $\mathcal Q'_F$ is $SU(5)$ and $\mathcal Q'_F$ is thus a CY 5-fold. The orientation-preserving $\mathbb Z_N \subset U(1)$ action is also an isometry subgroup of a 5-sphere. This means that the holonomy of $\mathcal X'_F$, will be the same as that of a non-trivial $\mathbb R^5$ bundle over $S^5$. In other words, the holonomy group of $\mathcal X'_F$ is $SU(5)$ and $\mathcal X'_F$ is also a CY 5-fold. As had been done for $\widehat {\widetilde Q}$ and $\widehat {\widetilde X}$ in $\S$4.1, it remains for us to find explicit examples of freely-acting, antiholomorphic, isometric involutions on $\mathcal Q'_F$ and $\mathcal X'_F$ so that we can define $\widehat{\mathcal Q}_F = {\mathcal Q}'_F / \mathbb Z_2$ and $\widehat {\mathcal X}_F = {\mathcal X}'_F / \mathbb Z_2$. 

\bigskip\noindent{\it The 10-manifolds $\widehat{\mathcal Q}_F$ and $\widehat{\mathcal X}_F$}
\vspace{0.2cm}

First notice that we can describe the CY 5-fold ${\mathcal Q}'_F $ as a subspace in $\mathbb C^6$ as follows: let $z_j = x_j + i p_j$, where $x_j, p_j$ are real and $(z_1, z_2, z_3, z_4, z_5, z_6) \in \mathbb C^6$. Then, ${\mathcal Q}'_F $ is given by the hypersurface
\be
\sum_{j=1}^6 {z_j}^2 = \mu,
\ee 
or equivalently,
\be
\sum_{j=1}^6 ({x_j}^2 - {p_j}^2) = \mu, \qquad \sum_{j=1}^6 x_j p_j = 0,
\label{Q'}
\ee
subject to the additional equivalence relation
\be
(p_1, p_2, p_3, p_4, p_5, p_6) \sim (e^{{2\pi i n} \over {N}} p_1, e^{{2\pi i n} \over {N}} p_2, e^{{2\pi i n} \over {N}} p_3, e^{{2\pi i n} \over {N}} p_4, p_5, p_6), \hspace {0.5cm} 0 \leq n \leq N-1,
\label{singularaction}
\ee
because of the identification of the $\mathbb R^4$ subspace of the bundle under a $\it{fixed}$ $\mathbb Z_N$ action. It is clear from (\ref{Q'}) and (\ref{singularaction}) that the $x_j$s parameterize the $S^5$ base of radius $\sqrt \mu$, whilst the $p_j$s parameterize the $[\mathbb R \times (\mathbb R^4 / \mathbb Z_N)]$ bundle $\it{normal}$ to this 5-sphere base. Let the manifold metric be $\widetilde{\it{g}}_5$, the complex structure be $J_5$, the K\"ahler (1,1)-form be $\omega_5$, and the holomorphic (5,0)-form be $\Omega_5$. The K\"ahler form $\omega_5$ will then be given by 
\be  
\omega_5 = \sum_{j=1}^6 dp_j \wedge dx_j.
\label{omegaQ'} 
\ee
One can then define the action of $\sigma = \mathbb Z_2$ on ${\mathcal Q}'_F$ such that 
\be
\begin{array}{cc}
\sigma: & (x_1, x_2, x_3, x_4,x_5,x_6) \mapsto i (p_1, p_2, p_3, p_4, p_5, p_6) \\
& \\
& \hspace {0.5cm}{ (p_1, p_2, p_3, p_4,p_5,p_6) \mapsto -i (x_1, x_2, x_3, x_4, x_5, x_6)}. \\ 
\end{array}
\label {sigmamapQ'}
\ee 
Thus, we see that (\ref{Q'}) defining ${\mathcal Q}'_F$ is invariant under the action of $\sigma$, i.e. $\sigma^{*}(\widetilde{\it{g}}_5) = {\widetilde {\it{g}}_5}$ and $\sigma: {\mathcal Q}'_F \rightarrow {\mathcal Q}'_F$. In addition, from (\ref{omegaQ'}), we find that $\sigma^*(\omega_5) = -\omega_5$, which gives $\sigma^*(J_5) = -J_5$ since $\omega_5 = J_5 \otimes {\widetilde{\it {g}}_5}$. $\sigma^* (\Omega_5) = {\bar {\Omega}}_5$ follows from a similar argument made in $\S$4.1 on the discussion of $\widetilde {Q}'$.   

Next, notice that the CY 5-fold ${\mathcal X}'_F$, which is isomorphic to a non-trivial $\mathbb R^5$ bundle over $S^{5'}$, can also be described as the following hypersurface in $\mathbb C^6$:
\be
\sum_{j=1}^6 ({p_j}^2 - {x_j}^2) = \mu, \qquad \sum_{j=1}^6 p_j x_j = 0,
\label{X'}
\ee
subject to the additional equivalence relation
\be
(p_1, p_2, p_3, p_4, p_5, p_6) \sim (e^{{2\pi i n} \over {N}} p_1, e^{{2\pi i n} \over {N}} p_2, e^{{2\pi i n} \over {N}} p_3, e^{{2\pi i n} \over {N}} p_4, p_5, p_6), \hspace {0.5cm} 0 \leq n \leq N-1,
\label{actionS5'}
\ee
because of the identification on the 5-sphere under a $\it{free}$ $\mathbb Z_N$ action along the $U(1)$ subspace of its 3-sphere base. It is clear from (\ref{X'}) and (\ref{actionS5'}) that the $p_j$s parameterize the $S^{5'}$ base, whilst the $x_j$s parameterize the $\mathbb R^5$ bundle $\it{normal}$ to this base. As usual, let the metric on  ${\mathcal X}'_F$ be $\widetilde{\it{g}}_5$, the complex structure be $J_5$, the K\"ahler (1,1)-form be $\omega_5$, and the holomorphic (5,0)-form be $\Omega_5$. The K\"ahler (1,0)-form will be given by
\be  
\omega_5 = \sum_{j=1}^6 dp_j \wedge dx_j.
\label{omegaX'} 
\ee
One can then define the action of $\sigma =\mathbb Z_2$ on ${\mathcal X}'_F$ such that 
\be
\begin{array}{cc}
\sigma: & (p_1, p_2, p_3, p_4,p_5,p_6) \mapsto -i (x_1, x_2, x_3, x_4, x_5, x_6) \\
& \\
& \hspace {0.3cm}{ (x_1, x_2, x_3, x_4,x_5,x_6) \mapsto i (p_1, p_2, p_3, p_4, p_5, p_6) }. \\ 
\end{array}
\label {sigmamapX'}
\ee 
Thus, we see that (\ref{X'}) defining ${\mathcal X}'_F$ is invariant under the action of $\sigma$, i.e. $\sigma^{*}(\widetilde{\it{g}}_5) = {\widetilde {\it{g}}_5}$ and $\sigma: {\mathcal X}'_F \rightarrow {\mathcal X}'_F$. In addition, from (\ref{omegaX'}), we find that $\sigma^*(\omega_5) = -\omega_5$, which gives $\sigma^*(J_5) = -J_5$. $\sigma^* (\Omega_5) = {\bar {\Omega}}_5$ follows as usual. Hence, via a trivial 10-dimensional extension of the discussion surrounding $\widehat {\widetilde Q}$ and $\widehat {\widetilde X}$ in $\S$4.1, on the 1-1 correspondence between the constant tensors, holonomy, and ultimately constant spinors on a CY manifold quotiented by a freely-acting, antiholomorphic, isometric involution, we thus find that $\widehat{\mathcal Q}_F$ and $\widehat{\mathcal X}_F$ will preserve $1/32$ of the maximal supersymmetry whilst possessing an extended $SU(5) \odot \mathbb Z_2$ holonomy group as required.

\bigskip\noindent{\it Support of Constant Spinors on $\widehat{\mathcal Q}_F$ and $\widehat{\mathcal X}_F$}
\vspace{0.2cm}

Notice that the action of a freely-acting isometric involution on a fibre bundle descends onto its base space as explained before. Hence, since $\widehat {\mathcal Q}_F = {\mathcal Q}'_F / \mathbb Z_2$ and $S^5/\mathbb Z_2 = \mathbb {RP}^5$, we find that $\widehat {\mathcal Q}_F$ is isomorphic to a $\it{non}$-$\it{trivial}$ $[\mathbb R \times (\mathbb R^4 /\mathbb Z_N)]$ bundle over $\mathbb {RP}^5$. Similarly, since $\widehat {\mathcal X}_F = {\mathcal X}'_F / \mathbb Z_2$, we find that $\widehat {\mathcal X}_F$ is isomorphic to a $\it{non}$-$\it{trivial}$ $\mathbb R^5$ bundle over $\mathbb {RP}^{5'}$, where one recalls that $\mathbb {RP}^{5'}$ is a 5-dimensional real projective space whose volume is divided by a factor of $N$. We would like to highlight to the reader at this point the non-trivial topological structure of $\mathbb {RP}^5$ ($\mathbb {RP}^{5'}$) that will motivate one to investigate its ability to support spinor fields as required of a physically consistent, supersymmetry preserving compactification. To this end, note that we have the result that real projective spaces $\mathbb {RP}^m = S^m / \mathbb Z_2$ are orientable iff $m$ is odd; for $m>1$ there are either two inequivalent $(s)pin$ structures or none \cite{trautman}:
\be
\begin{array}{rccccc}
\mbox {$m$}: & \ 0 & \ 1 & \ 2 & \ 3 & \ mod \ 4 \\
\mbox{structure}: & \ pin_{m,0} & \ \rm{no} & \ pin_{0,m} & \ spin_m &  \\
\end{array}
\label{m}
\ee
It is therefore clear from (\ref{m}) above that the $\it{orientable}$ $\mathbb {RP}^5$ ($\mathbb {RP}^{5'}$) base manifold does not possess any $spin$ or $pin$ structures, i.e. it is neither a $spin$ nor $pin$ manifold. However, it has been shown in $\S$5 of \cite{spin spaces}, that one can still construct two inequivalent spinor bundles that are not associated with any $spin$ or $pin$ structures over $\mathbb {RP}^5$, i.e. $\mathbb {RP}^5$ ($\mathbb {RP}^{5'}$) is nevertheless able to support spinor fields as required. In fact, $\mathbb {RP}^5$ is $spin^c$ as follows: recall that we have the fibration $S^1 \hookrightarrow \mathbb {RP}^5 \to \mathbb {CP}^2 / \mathbb Z_2$, whereby $\mathbb {CP}^2 / \mathbb Z_2 = S^4$. Also recall that an oriented manifold with dim $\leq 4$ is $spin^c$ \cite{freed-witten}. Since $S^1$ and $S^4$ are orientable and with dim $\leq 4$, it means that $\mathbb {RP}^5$ is $spin^c$. This is consistent with the findings of $\S$4.1, whereby $\widehat {\mathcal X} \cong \mathbb R^3 \times \mathbb {RP}^{5'}$ is found to be a $spin^c$ manifold. This, together with the fact that the construction of arbitrary spinor bundles over the fibre spaces of $\widehat{\mathcal Q}_F$ and $\widehat{\mathcal X}_F$ is trivial, implies that the orientable, 10-dimensional manifolds $\widehat{\mathcal Q}_F$ and $\widehat{\mathcal X}_F$ are therefore $spin^c$. Thus, they have the ability to support spinor fields, hence rendering them physically consistent as supersymmetric compactifications.                

\bigskip\noindent{\it The Metrics on $\widehat{\mathcal Q}_F$ and $\widehat{\mathcal X}_F$}
\vspace{0.2cm}

As mentioned, the metrics on $\widehat {\mathcal Q}_F$ and $\widehat {\mathcal X}_F$ are the same as those on ${\mathcal Q}'_F$ and ${\mathcal X}'_F$. Hence, from the discussions surrounding the derivation of ${\mathcal Q}'_F$ and ${\mathcal X}'_F$, one can see that the metric on the 10-dimensional, orientable, Ricci-flat manifolds $\widehat {\mathcal Q}_F$ and $\widehat {\mathcal X}_F$, will be given by the metric on a non-trivial $\mathbb R^5$ bundle over an $S^5$ or simply $T^*S^5$, the cotangent bundle of $S^5$, modded out by a $\mathbb Z_N$ subgroup along the direction of the $\mathbb R^5$ bundle and $S^5$ base respectively. It is known that the K\"ahler potential for the Ricci-flat K\"ahler metric on $T^*S^5$, is given by the solution of the following O.D.E. \cite{stenzel}:
\be
{d \over {dw}} (f'(w))^5 = 5c (sinh \ w)^4,
\ee
where $w=cosh^{-1} r$, $r$ is the radial coordinate of an $S^5$, and $c$ is a non-zero positive constant. Let $(z_1,...,z_5, {\bar z}_1,...,{\bar z}_5)$ be complex coordinates on $\widehat {\mathcal Q}_F$ and $\widehat {\mathcal X}_F$. The Ricci-flat K\"ahler metric on $\widehat {\mathcal Q}_F$ and $\widehat {\mathcal X}_F$ will then be given by $g_F/ \mathbb Z_N$, whereby   
\be
(g_F)_{i \bar j} = \partial_i \partial_{\bar j} \ f (z_i, {\bar z}_j), \qquad 0 \leq \ i,j \ \leq 5,
\ee
and $\mathbb Z_N \subset U(1) \subset \mathbb R^5$ and $\mathbb Z_N \subset U(1) \subset S^5$ respectively.

\bigskip\noindent{\it At Last, an F-theoretic $\mathbb {RP}^5$ Flop}
\vspace{0.2cm}

Finally, recall that in the IIB geometric transition of the lifted background $\widehat {\mathcal Q} \rightarrow \widehat {\mathcal X}$, where $\widehat {\mathcal Q} \cong (\mathbb R^4/\mathbb Z_N) \times S^4$ and $\widehat {\mathcal X} \cong \mathbb R^3 \times \mathbb {RP}^{5'}$, there is a blow-down of the 4-cycle $S^4 \subset \widehat {\mathcal Q}$, and a blow-up of the 5-cycle $ \mathbb {RP}^{5'} \subset \widehat {\mathcal X}$. Notice that the blown-down $S^4$ 4-cycle in the IIB theory is a subspace of the $\mathbb {RP}^5$ base of $\widehat {\mathcal Q}_F$ in the F-theoretic picture, i.e. $[\mathbb R \times (\mathbb R^4 /\mathbb Z_N)] \hookrightarrow \widehat {\mathcal Q}_F \to \mathbb {RP}^5$ where $S^1 \hookrightarrow \mathbb {RP}^5 \to S^4$, in which the $S^1$ fibre is the 1-cycle of the F-torus $(S^1 \times \mathbb R)$. On the other hand, the blown-up $\mathbb {RP}^{5'}$ 5-cycle in the IIB theory is itself a base manifold in the F-theoretic picture, i.e. $\mathbb R^5 \hookrightarrow \widehat {\mathcal X}_F \to \mathbb {RP}^{5'}$. From the fibration structure of the 5-cycle $\mathbb {RP}^5 \subset \widehat {\mathcal Q}_F$, one can see that $vol (\mathbb {RP}^5 \subset \widehat {\mathcal Q}_F) = vol (S^1) \cdot vol (S^4)$, whereby $S^4 \subset \widehat {\mathcal Q}$ in the IIB picture. Since the F-torus fibre has no K\"ahler moduli, i.e. the size of the $S^1$ fibre is fixed, it is clear that there will be an F-theoretic $\mathbb {RP}^5$ flop, effected by a blow-down and blow-up of an $\mathbb {RP}^5$ and $\mathbb {RP}^{5'}$ 5-cycle respectively, as one goes from $\widehat {\mathcal Q} \rightarrow \widehat {\mathcal X}$ in the equivalent IIB theory! Recall that $C^+_4$ has been  turned on along the appropriate 4-cycles. Hence, the resulting $\mathcal N=(1,0)$ supersymmetric $SU(N)$ theory in $d=1+1$, obtained via a IIB compactification on the 8-dimensional manifolds $\widehat {\mathcal Q}$ and $\widehat {\mathcal X}$, or via an F-theory compactification on the 10-dimensional manifolds $\widehat {\mathcal Q}_F$ and $\widehat {\mathcal X}_F$, will undergo a non-singular RG flow to the IR, which in turn effects the smooth geometric transition $\widehat {\mathcal Q} \rightarrow \widehat {\mathcal X}$ in the IIB theory, and a smooth $\mathbb {RP}^5$ flop $\widehat {\mathcal Q}_F \rightarrow \widehat {\mathcal X}_F$ in the equivalent F-theory. In short, as claimed, the large $N$ type IIB duality via a $Spin(7)$ geometric transition $\mathcal Q \rightarrow \mathcal X$, with $N$ space-filling D5-branes and 3-form RR and NS fluxes, can be consistently lifted, for either small $\it{or}$ large string coupling, to a purely geometric and $\it{smooth}$ $\mathbb {RP}^5$ flop in the equivalent F-theoretic description! Moreover, one can see, via the equations (\ref{Q'})-(\ref{sigmamapX'}) which characterize $\widehat {\mathcal Q}_F$ and $\widehat {\mathcal X}_F$, that this large $N$ type IIB duality via an F-theoretic $\mathbb {RP}^5$ flop, can in fact be expressed as a simple and elegant mathematical equivalence; from (\ref{Q'})-(\ref{sigmamapX'}), one can see that the background on $\widehat {\mathcal Q}_F$, with a $\it{negative}$-volumed (i.e. $\mu < 0$) $\mathbb {RP}^5$, is mathematically equivalent to a background on $\widehat {\mathcal X}_F$, with a $\it{positive}$-volumed (i.e. $\mu > 0$) $\mathbb {RP}^{5'}$! This observation therefore verifies the mathematical consistency of our physical interpretation of a negative-volumed $S^4 \subset \widehat {\mathcal Q} \subset \widehat {\mathcal Q}_F$ as a positive-volumed (i.e. blown-up) $S^{4'} \subset \widehat {\mathcal X} \subset \widehat {\mathcal X}_F$ in $\S$4.2 earlier. Hence, one can conclude that the large $N$ type IIB duality via an F-theoretic $\mathbb {RP}^5$ flop, is indeed physically $\it{and}$ mathematically consistent.

\section{Summary}   

The two $Spin(7)$ manifolds resulting from distinct resolutions of the cone over an $SU(3)/U(1)$ base, namely the trivial $Spin^c$  or $\mathbb R^4$ bundle over $\mathbb{CP}^2$ and the trivial $\mathbb R^3$ bundle over $S^5$, termed respectively as the `resolved' and `deformed' $Spin(7)$ conifold, have been considered in this paper. An investigation into their structures via their toric and geometric descriptions reveals that they possess  6-dimensional subspaces which are isomorphic to the resolved and deformed Calabi-Yau 3-folds respectively. Consequently, it is observed that the geometric transition between these two distinct $Spin(7)$ resolutions as considered by Gukov et al. in \cite{gst}, is simply effected by the geometric transition between the 6-dimensional resolved and deformed Calabi-Yau subspaces. 

Building upon the above insight and a direct application of the Gopakumar-Vafa large $N$ superstring duality of \cite{gv, v} in the intermediate $d=3+1$ theory that results from the initial string compactification on the 6-dimensional resolved/deformed Calabi-Yau subspaces, a physically consistent and valid large $N$, type IIB duality via a geometric transition from the `resolved' to the `deformed' $Spin(7)$ conifold is derived, thus furnishing a natural $Spin(7)$ extension of the large $N$ superstring duality; IIB superstring theory compactified on the `resolved' $Spin(7)$ conifold with $N$ space-filling D5-branes wrapping the even-dimensional $\mathbb {CP}^2$ supersymmetric cycle, is observed to undergo an anomaly-free large $N$ geometric transition at low energy to the $\it dual$ `deformed' $Spin(7)$ conifold geometry with no branes but with $N$ units of 3-form RR flux and a unit of NS flux through the 3-cycle isomorphic to a 3-sphere and 3-ball respectively. Moreover, it can be shown via purely gauge theoretic arguments, that the large $N$ duality via this $\it{smooth}$ $Spin(7)$ geometric transition involving vanishing D5-branes and resultant 3-form fluxes, is a consequence of a $\it{non}$-$\it{singular}$ RG flow of the resulting $\mathcal N = (1,0)$ supersymmetric pure $SU(N)$ theory in 1+1 dimensions to the $\it{same}$ (i.e. $\it{dual}$) albeit abelian $U(1)$ theory in the IR with $N$ inequivalent vacua characterized by $N$ distinct $\theta$-phases, such that the sizes of the blown-down $\mathbb{CP}^2$ and blown-up $S^5$ of the `resolved' and `deformed' $Spin(7)$ conifold before and after the geometric transition respectively, are governed by the (smoothly running) value of the effective dimensionless gauge coupling of the $d=1+1$ theory at the observed energy scale $u$, which in turn determines the point in the geometric transition.  

The $Spin(7)$ compactification above with D5-branes or 3-form fluxes, can be geometrically lifted to an equivalent background without D5-branes or 3-form fluxes via the singular or smooth non-simply connected compactification 8-manifold constructed from the `resolved' or `deformed' $Spin(7)$ conifold by defining quotients by a $\mathbb Z_N$ action along the appropriate subspaces, together with an identification of the underlying manifold by a freely-acting isometric involution. The large $N$, $Spin(7)$, type IIB duality can be shown to hold in this lifted background via purely gauge theoretic arguments as before, whereby the sizes of the blow-down $S^4$ and and blow-up $\mathbb {RP}^5$ in the lifted geometric transition, are governed by the (smoothly running) value of the effective dimensionless gauge coupling of the $d=1+1$ theory at the observed energy scale $u$.  

In a non-trivial albeit finite and smoothly varying axion field background, such that one is to consider a non-trivial elliptic fibration over the above-mentioned equivalent 8-manifolds in the F-theoretic description, it is eventually shown that for small $\it{or}$ large values of string coupling, the large $N$, $Spin(7)$, type IIB duality can be lifted to a purely geometric $\mathbb {RP}^5$ flop without D5-branes and 3-form fluxes in the corresponding F-theory. The non-compact, 10-dimensional Ricci-flat manifold undergoing the smooth $\mathbb {RP}^5$ flop has an extended $SU(5) \odot {\mathbb Z_2}$ holonomy group, which preserves $1/32$ of the maximal supersymmetry, in agreement with the effective $\mathcal N =(1,0)$ supersymmetric pure $SU(N)$ theory in $d=1+1$ from the IIB compactification on the $Spin(7)$ with D5-branes/fluxes. Although the flop manifold lacks the usual $spin$ or $pin$ structures, it can be argued to possess a $spin^c$ structure, which in turn implies that it is able to support a covariantly constant spinor as required of a supersymmetric compactification. As expected, the large $N$ type IIB duality via an F-theoretic $\mathbb {RP}^5$ flop, is eventually observed to be $\it{both}$ physically and mathematically consistent. 
\newline

\hspace{-1.0cm}{\Large \bf Acknowledgements:}\\
I would like to take this opportunity to thank M. Atiyah, B. Baaquie, A.P. Balachandran, S.J. Gates Jr., D. Joyce, Miao Li, P. Minkowski, P. Mitra, E. Teo, G. 't Hooft, A. Trautman and E. Witten for directing me to the relevant references and/or for useful discussions. 
\newline
 
\hspace{-1.0cm}{\Large \bf {Appendix:}}
\appendix {\section{$\mathcal N = (1,0)$ supersymmetric pure $SU(N)$ in $d=1+1$}}
\label{sec:2dsym}

In this appendix, a supersymmetric gauge invariant action of the $\mathcal N = (1,0)$ pure $SU(N)$ theory in 1+1 dimensions will be presented. The action is constructed using $(1,0)$ superspace methods. To this end, the $\mathcal N=(1,0)$ supersymmetry algebra and the super Yang-Mills multiplet in 1+1 dimensions will first be reviewed and a consistent construction of the requisite action is given thereafter. 

We begin by stating the conventions adopted herein for describing $d=2$, (1,0) superspace. The uncompactifed part of the $d=5+1$ worldvolume of the N coincident D5-branes which wrap the $\mathbb {CP}^2$ 4-cycle of  $\mathcal Q$ corresponds to a flat Minkowski manifold with coordinates $x^\mu=(x,t)$ and metric $\eta_{\mu\nu}$. The worldvolume coordinates $x$ and $t$ can be assembled into a pair of bosonic light-cone/null coordinates given by $(x^\plpl,x^\mimi)=(x+t,x-t)$. Together with the single real fermionic coordinate denoted by $\theta^+$, the $d=2$, (1,0) superspace is then a supermanifold parameterised by $(x^\plpl,x^\mimi, \theta^+)$. Note that $\theta^+$ is a real
Grassmann-odd coordinate, which is anticommuting and in particular $(\theta^+)^2=0$.
\\
\subsection{$\mathcal N = (1,0)$ supersymmetry algebra}

The (1,0) supersymmetry algebra is given by 
\begin{align}
\label{eq:10susyalgebra}
  \{ Q_+, Q_+ \} = 2iP_\plpl ~.
\end{align}
As usual in superspace theories, we can construct supersymmetry covariant derivatives $D_M \equiv (D_+, \partial_{\plpl}, \partial_{\mimi})$. The supercovariant derivative is given by 
\begin{align}
D_+ = \partial_{\theta^+} + i \theta ^+ \partial_{\plpl},
\label{D+}
\end{align}
where $\partial_{\theta^+}=\partial/\partial\theta^+$ such that $\partial_{\theta^+} \theta^+ = 1$, $\partial_{\plpl}=\partial/\partial x^{\plpl}$ and $\partial_{\mimi}=\partial/\partial x^{\mimi}$. We also have 
\begin{align}
\{D_+, D_+\} = 2i \partial_{\plpl}.
\label{D+D+}
\end{align}
This gives ${D_+}^2 = i\partial_{\plpl}$. Note that we also have the following property
\begin{align}
 \{Q_+,D_+\}=0 ~.
\end{align}
This allows one to define $Q_+$ such that 
\begin{align}
\label{eq:10susycharge}
Q_+ &= \frac{\partial}{\partial\theta^+} -
  i\theta^+\frac{\partial}{\partial x^\plpl}  ~
\end{align}
in component form. We also have the realisation $P=(P_{\plpl}, P_{\mimi})=(-\partial_{\plpl}, -\partial_{\mimi})$. It can be verifed that $P_{\plpl}$ and $Q_+$ satisfy the supersymmetry algebra given by \eqref{eq:10susyalgebra}.
 
\subsection{(1,0) super Yang-Mills multiplet}

The (1,0)-supersymmetric Yang-Mills multiplet with arbitrary gauge group G is described by a super connection $\mathcal A=(\mathcal A_{\plpl}, \mathcal A_{\mimi}, \mathcal A_{+})$. The superderivatives are made covariant with respect to this super connection via minimal coupling. They are given by

\begin{gather}
   \tilde {\nabla}_+ \equiv D_+ - ig\mathcal A_+,
\mspace{100mu} \tilde{\nabla}_\plpl \equiv \partial_{\plpl}-ig\mathcal A_{\plpl},     
\nonumber\\
\nonumber\\
\tilde{\nabla}_\mimi \equiv \partial_{\mimi}-ig\mathcal A_{\mimi},
\end{gather}
where $g$ is the gauge coupling constant. It has dimensions of mass in $d=2$ spacetime. These covariant superderivatives obey a certain set of constraints given by \cite{gates} 

\begin{align}
 \{ \tilde{\nabla}_+, \tilde{\nabla}_+ \} = 2i\tilde{\nabla}_\plpl,
\label{constart}
\end{align}
\begin{align}
[ \tilde{\nabla}_+, \tilde{\nabla}_{\plpl} ] = 0,
\end{align}
\begin{align}
[ \tilde{\nabla}_+, \tilde{\nabla}_{\mimi}] = -ig\mathcal W_-,
\end{align}
\begin{align}
[ \tilde{\nabla}_{\plpl}, \tilde{\nabla}_{\mimi}] = -ig\mathcal F,
\label{conend}
\end{align}
where $\mathcal W_-$ and $\mathcal F$ are the components of the supercurvature of the superconnection $\mathcal A$. 

The constraints of \eqref{constart}-\eqref{conend} are easily solvable in terms of the superspace potentials $\mathcal A_+$ and $\mathcal A_{\mimi}$. The corresponding Taylor expansions of these potentials quickly terminate due to ${\theta^+}^2 = 0$. They are given by \cite{gates}
\begin{align}
\mathcal A_+ = {1\over g} \rho_+(x,t) +i\theta^+ A_{\plpl}(x,t),
\end{align}
\begin{align}
\mathcal A_{\mimi} = A_{\mimi}(x,t) + \theta^+[ \chi_- (x,t) + {1\over g} \nabla_{\mimi} \rho_+(x,t)],
\label{nabla}
\end{align}
where $\nabla_{\mimi}$ is a $\it spacetime$ covariant derivative and $A_{\plpl(\mimi)}$ are the components of the $\it spacetime$ gauge potential $A(x^{\plpl},x^{\mimi})$. The $\pm$ indices on the spacetime spinors $\rho_+(x,t)$ and $\chi_-(x,t)$ indicate their transformation property under an internal (spin) Lorentz rotation such that for any given arbitrary spacetime spinor $\psi_{\pm}$ and an infinitesimal Lorentz group generator $M$, we have
\begin{align}
[M,\psi_{\pm}]=\pm{1\over2}\psi_{\pm}.
 \end{align}
Via constraint \eqref{constart}, the remaining potential $\mathcal A_{\plpl}$ is determined in terms of $\mathcal A_+$ as \cite{gates} 
\begin{align}
\mathcal A_{\plpl} = -i{D_+} \mathcal A_+ - {1\over2}g \{ {\mathcal A_+}, {\mathcal A_+} \}.
\end{align}
Note that all the relevant gauge indices and generators have been suppressed for simplicity. 

The Bianchi identities on $\tilde{\nabla}_+$, $\tilde{\nabla}_{\plpl}$ and $\tilde{\nabla}_{\mimi}$ imply that 
\begin{align}
\tilde{\nabla}_+ \mathcal W_- = i \mathcal F,
\mspace{100mu}
\tilde{\nabla}_+ \mathcal F=\tilde{\nabla}_{\plpl}\mathcal W_-.
\end{align}
Therefore, the independent components of the gauge multiplet are 
\begin{align}
\chi_-(x,t) = \mathcal W_-|_{\theta^+ = 0},
\mspace{100mu}
F_{\plpl\mimi}(x,t) = -i \tilde{\nabla}_+\mathcal W_-|_{\theta^+ = 0},
\end{align}
whereby $\chi_-(x,t)$ is the spacetime gaugino field and $F_{\plpl\mimi}(x,t)$ is the spacetime two-form gauge field strength.  
  
\subsection{The supersymmetric action}
The most general gauge invariant supersymmetric action is given by \cite{gates,machin}
\begin{align}
S_{SYM} &=- \intd{^2x\rd\theta^+}Tr \bigl( \it u \mathcal W_- \tilde{\nabla}_+ \mathcal W_-  - i \it z  \mathcal W_-  \bigr).
\label{action}
\end{align}
Note here that $\it u$ and $\it z$ are real, constant scalar superfields whose lowest components are proportional to the spacetime gauge coupling $1\over {g^2}$ and $\theta$ term respectively. 

The action of the (1,0) supersymmetric pure Yang-Mills theory described by the action \eqref{action} can be expressed in component form by using the
definition of the various component fields of the (1,0)-multiplets described earlier and integrating out the $\theta^+$ coordinate via the Berezin relation ${\int d\theta^+} = \partial_{\theta^+}$. In doing so, we obtain 
\begin{align}
S_{SYM}= {{a} \over {g^2}} \intd{^2x}Tr(F_{\plpl\mimi}^{\it k} F_{\plpl\mimi}^{\it k}   - i  \chi_-^{\it k} {\nabla}_{\plpl}\chi_-^{\it k}) + \ b \ \theta \intd{^2x} Tr(F_{\plpl\mimi}^{\it k}),  
\label{finalaction} 
\end{align}
where $a$ and $b$ are dimensionless constants of proportionality and the trace has been taken over the matrix components of the generators (labelled by $k$) in the corresponding representation of the gauge group, which have been suppressed here for simplicity in expression. Like $\nabla_{\mimi}$ in \eqref{nabla}, ${\nabla}_{\plpl}$ here is a $\it spacetime$ gauge covariant derivative compatible with the spinor gaugino fields $\chi^{\it k}_-(x,t)$. As the Yang-Mills multiplets are in the adjoint representation of the gauge group, the sum over $\it k$ in \eqref{finalaction} is taken over the values ${\it k}=1,2,3$.....dim $\mathfrak g$. In our case of interest, $\mathfrak g=\mathfrak {su(n)}$, where $\mathfrak {su(n)}$ is the Lie algebra of $SU(N)$ and dim $\mathfrak {su(n)}$ $=N^2 -1$.

\bigbreak\bigskip

{\renewcommand{\Large}{\normalsize}
}
\end{document}